\documentclass[hyper,12pt,letterpaper]{JHEP3}

\usepackage{amsmath,amssymb,amsfonts,bm}
\usepackage{graphicx}
\usepackage{multirow}
\usepackage{verbatim}
\usepackage{appendix}
\usepackage{epstopdf}
\usepackage{verbatim}
\usepackage{ulem}
\usepackage[small,nohug,heads=littlevee]{diagrams}
\diagramstyle[labelstyle=\scriptstyle]

\numberwithin{equation}{section}

\newtheorem{proposition}{Proposition}
\newtheorem{conjecture}{Conjecture}

\newcommand{\mb}{\mathbf}
\newcommand{\eg}{\emph{e.g.}}
\newcommand{\ie}{\emph{i.e.}}
\newcommand{\cf}{\emph{cf.}}
\newcommand{\ol}{\overline}

\newcommand{\be}{\begin{equation}}
\newcommand{\ee}{\end{equation}}
\newcommand{\ben}{\begin{equation*}}
\newcommand{\een}{\end{equation*}}
\newcommand{\bea}{\begin{eqnarray}}
\newcommand{\eea}{\end{eqnarray}}
\newcommand{\bean}{\begin{eqnarray*}}
\newcommand{\eean}{\end{eqnarray*}}
\newcommand{\bse}{\begin{subequations}}
\newcommand{\ese}{\end{subequations}}
\newcommand{\ds}{\displaystyle}
\newcommand{\nno}{\nonumber}
\newcommand{\imp}{\Rightarrow}

\newcommand{\bTi}{\begin{itemize} \setlength{\itemsep}{-.1cm}}
\newcommand{\eTi}{\end{itemize}}

\newcommand{\pd}{\partial}
\newcommand{\Tr}{{\rm Tr}}

\renewcommand{\Im}{{\rm Im}}
\renewcommand{\Re}{{\rm Re}}

\newcommand{\Vol}{{\rm Vol}}

\newcommand{\Li}{{\rm Li}}

\newcommand{\bs}{\backslash}

\newcommand{\mtt}[4]{{\left(\begin{array}{cc} #1 & #2 \\ #3 & #4 \end{array}\right)}}

\newcommand{\R}{{\mathbb{R}}}
\newcommand{\C}{{\mathbb{C}}}
\newcommand{\Z}{{\mathbb{Z}}}
\newcommand{\Q}{{\mathbb{Q}}}

\renewcommand{\H}{{\mathbb{H}}}

\newcommand{\fg}{{\frak{g}}}

\newcommand{\Hom}{{\rm Hom}}

\newcommand{\CA}{\mathcal{A}}
\newcommand{\CB}{\mathcal{B}}
\newcommand{\CC}{\mathcal{C}}
\newcommand{\CD}{\mathcal{D}}

\newcommand{\CH}{\mathcal{H}}
\newcommand{\CI}{\mathcal{I}}
\newcommand{\CJ}{\mathcal{J}}

\newcommand{\CL}{\mathcal{L}}

\newcommand{\CN}{\mathcal{N}}

\newcommand{\CP}{\mathcal{P}}

\newcommand{\CR}{\mathcal{R}}

\newcommand{\CT}{\mathcal{T}}

\newcommand{\CX}{\mathcal{X}}

\newcommand{\CZ}{\mathcal{Z}}

\title{Quantum Riemann Surfaces in Chern-Simons Theory}

\author{Tudor Dimofte\\
Institute for Advanced Study, Einstein Dr., Princeton, NJ 08540, USA\\
\email{tdd@ias.edu} \vspace{.3cm} \\
Trinity College, Cambridge CB2 1TQ, UK \vspace{.3cm} \\
Max-Planck-Institut f\"ur Mathematik, Vivatsgasse 7, 53111 Bonn, Germany \vspace{.2cm}
}

\abstract{We construct from first principles the operators $\hat A_M$ that annihilate the partition functions (or wavefunctions) of three-dimensional Chern-Simons theory with gauge groups $SU(2)$, $SL(2,\R)$, or $SL(2,\C)$ on knot complements $M$. The operator $\hat A_M$ is a quantization of a knot complement's classical A-polynomial $A_M(\ell,m)$. The construction proceeds by decomposing three-manifolds into ideal tetrahedra, and invoking a new, more global understanding of gluing in TQFT to put them back together. We advocate in particular that, properly interpreted, ``gluing $=$ symplectic reduction.'' We also arrive at a new finite-dimensional state integral model for computing the analytically continued ``holomorphic blocks'' that compose any physical Chern-Simons partition function.}

\begin{document}
\normalem

\begin{flushright} \emph{You can leave your hat on.}\\
--- Randy Newman
\end{flushright}

\section{Introduction}
\label{sec:intro}

This paper is in part about quantizing Riemann surfaces. The surfaces in question are algebraic ones, defined as the zero-locus of some polynomial function on a semi-classical phase space. For example, we can consider the surface
\be f(x,y) = 0\,, \label{fcl} \ee
thought of as a subset of the phase space $\CP = \{(x,y)\} \simeq \C^*\times\C^*$ with a symplectic structure $\omega = (i\hbar)^{-1}(dx/x)\wedge(dy/y)$. We have included a factor of $\hbar$ in the symplectic form, where $\hbar$ is to be thought of as a small, formal quantization parameter. The goal, then, is to promote $f(x,y)$ to a quantum operator $\hat f(\hat x,\hat y;q)$, where
\be \hat x \hat y = q \hat y \hat x, \label{ficomm} \ee
with $\boxed{q = e^{\hbar}}$, as dictated by the semi-classical Poisson bracket. The operators $\hat x$, $\hat y$, and $\hat f$ itself should act on an appropriate quantum Hilbert space $\CH$, typically obtained from (a real slice of) $\CP$ by geometric quantization.

Unfortunately, the choice of a polynomial operator $\hat f(\hat x,\hat y;q)$ that reduces to $f(x,y)$ in the classical limit $q\to 1$ is far from unique. As usual, one encounters ``ordering ambiguities'' when attempting to quantize. These ambiguities are aggravated by the fact that $f(x,y)$ is \emph{not} a polynomial function in the canonical linear coordinates on $\CP$, which would be $X=\log x$ and $Y= \log y$. Therefore, well understood mathematical quantization methods, such as deformation quantization, do not immediately apply. Indeed, in a few known examples where the quantization of $f(x,y)$ has a precise physical interpretation and the correct answer for $\hat{f}(\hat x,\hat y)$ is known (by various indirect methods), the actual resolution of ordering ambiguities appears wildly complicated.

In general, one might also consider phase spaces of higher dimension. Instead of a Riemann surface, the relevant variety to quantize would then be a higher-dimensional Lagrangian submanifold --- describing a semi-classical state. Just as in \eqref{fcl} above, we would be interested in the case where the defining equations $f_i=0$ for this submanifold were polynomials in the \emph{exponentiated} canonical coordinates on phase space. Again, we would like to promote the equations to quantum operators $\hat f_i$.

\subsection*{Chern-Simons theory}

We will describe a solution to the quantization of certain functions like $f(x,y)=0$ above in the context of Chern-Simons theory. In particular, we consider an analytically continued version of three-dimensional Chern-Simons theory with rank-one gauge group $SU(2)$, or $SL(2,\R)$, or $SL(2,\C)$, and we put this theory on an oriented three-manifold $M$ that is the complement of a (thickened) knot or link in some other compact manifold $\ol{M}$. Let's suppose that $M$ is a knot complement,
\be M = \ol{M}\bs K\,.\ee
There is a classical phase space $\CP_{T^2}$ associated to the boundary of $M$, which is a torus, $\pd M \simeq T^2$. The coordinates of $\CP_{T^2}$ can be taken as the eigenvalues $\ell$ and $m$ of the holonomies of a flat connection (\ie\ a classical solution to Chern-Simons theory) on the two one-cycles of $T^2$. Chern-Simons theory further induces a symplectic structure $\omega_{T^2} = (i\hbar/2)^{-1}(d\ell/\ell)\wedge(dm/m)$ on $\CP_{T^2}$, where $\hbar$ is the coupling constant, or inverse level, of the theory.

In an analytically continued setting, as developed in \cite{gukov-2003} (and later in \cite{DGLZ, Wit-anal}), one is interested in complexified classical solutions to Chern-Simons theory, \ie\ in the set of flat $SL(2,\C)$ connections on $M$ that extend from the boundary $T^2$ to the entire bulk. These are characterized by a single polynomial condition $A_M(\ell,m)=0$, where $A_M(\ell,m)$ is the so-called A-polynomial of $M$ \cite{cooper-1994}. This condition cuts out a Lagrangian submanifold
\be \CL_M=\{A_M(\ell,m) = 0\} \quad\subset \quad \CP_{T^2}=\{(\ell,m)\}\approx \C^*\times\C^*\,,\ee
or a semi-classical state in Chern-Simons theory \cite{gukov-2003}. We would like to quantize the A-polynomial, promoting it to an operator $\hat A_M(\hat \ell,\hat m;q)$ that annihilates the \emph{quantum} wavefunction or partition function of Chern-Simons theory on $M$. More precisely, the operator $\hat A_M(\hat\ell,\hat m;q)$ will annihilate the ``holomorphic blocks'' of Chern-Simons theory on $M$. The holomorphic blocks are universal, locally holomorphic functions $Z_M^\alpha(m)$, which can be summed to form any analytically continued $SU(2)$, $SL(2,\R)$, or $SL(2,\C)$ partition function. According to the symplectic structure $\omega_{T^2}$, $\hat \ell$ and $\hat m$ should act on $Z_M^\alpha(m)$ as
\be \hat \ell\, Z_M^\alpha(m) = Z_M^\alpha(q^{1/2}m)\,,\qquad \hat m Z_M^\alpha(m)=mZ_M^\alpha(m)\,,\ee
so that%
\footnote{We choose $q^{1/2}$ to appear in \eqref{lmicomm}, as opposed to $q$ in our ``basic'' example \eqref{ficomm}, in order to agree with conventions in later sections. The reason is related to the fact that $A(\ell,m)$ is typically a polynomial in $m^2$ rather than just $m$; in terms of $m^2$, the $q$-commutation would be $\hat \ell\hat m^2 = q\hat m^2\hat\ell\,.$}
\be \hat \ell\hat m = q^{1/2}\hat m\hat\ell\,, \label{lmicomm} \ee
and we expect, following \cite{gukov-2003}, that $\hat A_M(\hat \ell,\hat m;q)\,Z_M^\alpha(m)=0$.

The quantum A-polynomial $\hat A(\hat\ell,\hat m;q)$ has made previous appearances in the guise of a recursion relation for colored Jones polynomials \cite{Gar-Le, garoufalidis-2004} (see also \cite{Frohman-Gelca}). Famously, colored Jones polynomials are equivalent to Chern-Simons partition functions with gauge group $SU(2)$ \cite{Wit-Jones, Resh-Tur, Kir-Mel}. Deferring further details to Section \ref{sec:CS}, we note that this connection (so far) has provided the only known tool for finding the properly quantized $\hat A_M(\hat\ell,\hat m;q)$ in various geometries. As an example, consider the complement of the figure-eight knot in the three-sphere, $M=S^3\bs \mb{4_1}$. The classical A-polynomial is easily calculated%
\footnote{There is a universal factor of $(\ell-1)$ in the classical A-polynomials of knot complements in $S^3$ that was removed here. We will be discussing this factor in detail in Section \ref{sec:Apoly}, as well as Section \ref{sec:classrem}.} as \cite{cooper-1994}
\be A_{\mb{4_1}}(\ell,m) = m^4\ell^2-(1-m^2-2m^4-m^6-m^8)\ell+m^4\,. \label{A41eg} \ee
The quantum version was obtained in \cite{garoufalidis-2004} by searching for a recursion relation for the colored Jones polynomials of the figure-eight knot, and found to be
\begin{align} \hat{A}(\hat{\ell},\hat{m};q)_{\mb{4_1}}
 = &q^{5/2}(1-q\hat{m}^4)\hat{m}^4\hat{\ell}^2-(1-q^2\hat{m}^4)(1-q\hat{m}^2-(q+q^3)\hat{m}^4-q^3\hat{m}^6+q^4\hat{m}^8)\hat{\ell} \nno \\
 &\quad +q^{3/2}(1-q^3\hat{m}^4)\hat{m}^4\,.  \label{Ah41eg}
\end{align}
This example explicitly illustrates just how severe ordering ambiguities can be! We observe that in addition to an extra factor of the form $(1-q^{\#}\hat m^4)$, which has no meaning in the classical A-polynomial, monomials like $2m^4$ in $A(\ell,m)$ split into expressions like $(q+q^3)\hat m^4$ in $\hat A(\hat \ell,\hat m;q)$; thus the quantization is not even linear.

We attempt in this paper to provide an intrinsic, three-dimensional construction of quantum $\hat A$-polynomials for knot and link complements. Our method utilizes ideal triangulations of three-manifolds, along with a convenient relation between flat $SL(2,\C)$ connections and hyperbolic structures in three dimensions (\cf\ \cite{Witten-gravCS, gukov-2003}). This relation allows us to use many well-developed tools of hyperbolic geometry and decompositions into ideal \emph{hyperbolic} tetrahedra \cite{thurston-1980, NZ}.%
\footnote{Nevertheless, it should be entirely possible to use appropriately decorated ideal (topological) tetrahedra to describe flat connections of any complex gauge group, not just $SL(2,\C)$.} %
We make (and justify) the assumption that quantization at the level of a single tetrahedron is \emph{simple}. As we will sketch out momentarily, a tetrahedron $\Delta$ has its own boundary phase space and its own version of a constraint ``$\CL_\Delta=0$'' (or a Lagrangian submanifold) that should be quantized to an operator $\hat\CL_\Delta$ that annihilates the tetrahedron's partition function. The trick, then, is to glue tetrahedra together in an appropriate way, while also preserving information about the operators $\hat \CL_\Delta$ --- and to somehow use this extra information to find an operator $\hat A$ that annihilates the Chern-Simons partition function on an entire glued manifold $M$.

\subsection*{Symplectic gluing}

This brings us to our second major focus: a new perspective on gluing in topological quantum field theory (TQFT). According to the standard rules of TQFT, or QFT, the gluing of two manifolds along a common boundary should correspond to multiplying together component wavefunctions or partition functions and then integrating over all possible boundary conditions at the gluing. This is an exceedingly useful prescription for computing partition functions, but it tells us very little about the operators that annihilate them.

We reformulate the notion of ``integrating over boundary conditions'' in terms of symplectic geometry. Semi-classically, we find that gluing corresponds to forming a product of the phase spaces associated to two identified boundaries, and then taking a symplectic quotient, or reduction, of this product. In the reduction, we use as moment maps the functions that would relate boundary conditions at the two boundaries. For example, suppose that we glue together $M$ and $N$ along a common boundary $\Sigma$, and that the phase space $\CP_{\Sigma}$ is two-dimensional. There must be two functions $C_1,C_2$ on $\CP_{\Sigma}\times \CP_{-\Sigma}$ that identify the boundary conditions of $M$ to those of $N$ by requiring $C_1=C_2=0$. In this case, the resulting phase space of $M\cup N$ is a symplectic reduction of a four-dimensional space ($\CP_{\Sigma}\times \CP_{-\Sigma}$) by two moment maps ($C_1$ and $C_2$), and hence zero-dimensional (empty). This is, trivially, as expected for a closed manifold $M\cup N$. However, when a gluing happens to be \emph{incomplete}, so that (say) $M\cup N$ still has some boundary left over, the prescription still works and the result is no longer so tautological. The case of gluing together ideal tetrahedra to form a manifold $M$ with a left-over torus boundary is precisely such a situation.

The notion of gluing by forming products of phase spaces and then symplectically reducing via gluing functions has immediate implications both for semi-classical states (\emph{a.k.a.} Lagrangian submanifolds) and for quantum states and the operators that annihilate them. Roughly speaking, Lagrangian submanifolds can be ``pulled through'' symplectic reductions by projecting perpendicular to flows and then intersecting with moment maps. One can use this to construct a semi-classical state on a glued manifold from the states of its pieces. The analogous procedure for quantum operators will be discussed in great detail in Section \ref{sec:opglue}.

In terms of partition functions, our new understanding of gluing essentially replaces the rule ``multiply and integrate over boundary values'' with an equivalent rule, ``multiply and Fourier transform.'' Applying this to a three-manifold with an ideal triangulation leads immediately to a new state integral model for the holomorphic blocks of Chern-Simons theory.

\subsection*{Some detail}

In order to whet our appetites a bit further, let us actually consider rank-one, analytically continued Chern-Simons theory on an ideal tetrahedron. We will discover in Section \ref{sec:hyp} that the phase space of flat $SL(2,\C)$ connections on the surface $\pd\Delta$ of an ideal tetrahedron (which could alternatively be viewed as a four-punctured sphere) is two-dimensional, parameterized as
\be \CP_{\pd \Delta} = \{(z,z',z'')\in (\C\bs\{0,1,\infty\})^3\;|\;zz'z''=-1\}\,, \label{Pintro} \ee
with symplectic structure
\be \omega_{\pd \Delta} = (i\hbar)^{-1}\frac{dz}{z}\wedge\frac{dz'}{z'}\,.  \label{Tsi} \ee
The complex variables $z,z',z''$ might be recognized as the hyperbolic shape parameters of the tetrahedron, while \eqref{Tsi} is one tetrahedron's worth of the Neumann-Zagier symplectic form \cite{NZ}. (Alternatively, if the $z$'s were real, \eqref{Tsi} would be the Weil-Petersson form on the Teichm\"uller space of the four-punctured sphere \cite{Fock-Teich}.)

The condition that a flat connection on the boundary of a tetrahedron extend through its bulk is given by the Lagrangian submanifold
\be \{\CL_\Delta:=z + z'{}^{-1}-1 = 0\} \quad\subset\quad\CP_{\pd\Delta}\,. \label{Lintro} \ee
(This is also a well-known equation from hyperbolic geometry, relating classically equivalent shape parameters $z$ and $z'$!)
Let us use the condition $zz'z''=-1$ in \eqref{Pintro} to eliminate $z''$ from the parametrization of the phase space. We will argue in Section \ref{sec:quant} that $\CL_\Delta$ has the almost trivial quantization
\be \hat\CL_\Delta = \hat z+ \hat z'{}^{-1}-1\,, \ee
where, according to \eqref{Tsi},
\be \hat z\hat z' = q\hat z'\hat z\qquad (q=e^{\hbar})\,. \ee
If we denote by $\psi(z')$ the Chern-Simons holomorphic block of an ideal tetrahedron, then we should require that $\hat\CL_\Delta\, \psi(z') = 0$, or
\be \psi(qz') = (1-z'{}^{-1})\psi(z')\,. \label{qdleqi} \ee
The formal solution to \eqref{qdleqi} is a quantum dilogarithm function \cite{Fad-Kash},
\begin{align} \psi(z) &= \prod_{r=1}^\infty (1-q^rz'{}^{-1}) \;\;\underset{\hbar\to 0}{\sim}\;\; e^{ \frac{1}{2\hbar}\Li_2(z'{}^{-1})}\,, \label{QDLi}
\end{align}
whose leading asymptotic in the classical $\hbar\to0$ limit reproduces a (holomorphic version of) the volume of an ideal tetrahedron, given by the classical dilogarithm $\Li_2(z'{}^{-1})$ \cite{milnor-1982, thurston-1980}. As explained in \cite{gukov-2003, DGLZ, VCReview} (also \cf\ \cite{kashaev-1997}), this is exactly what one would expect for analytically continued rank-one Chern-Simons theory on an ideal tetrahedron.

Now suppose that a knot or link complement $M$ has an ideal triangulation $\{\Delta_i\}_{i=1}^N$. In Section \ref{sec:quant}, our perspective on gluing will identify the quantum $\hat A$-polynomial $\hat A(\hat \ell,\hat m;q)$ of $M$ as a distinguished element in the left ideal $\hat\CI_{M,\Delta}$ generated by the operators $\CL_{\Delta_i}=\hat z_i+\hat z_i'{}^{-1}-1$ for $i=1,...,N$. (If $M$ is the complement of a link with $\nu$ components, there would actually be $\nu$ classical equations $A_a=0$ characterizing flat connections, and a corresponding distinguished left sub-ideal of $\hat\CI_{M,\Delta}$ generated by at least $\nu$ quantum operators.) We will show, under certain assumptions, that the quantum polynomials $\hat A(\hat\ell,\hat m;q)$ so constructed are in fact independent of the precise choice of triangulation for $M$.

The state integral model predicted by our gluing construction will be explored in Section \ref{sec:wf}. We find that the holomorphic blocks $Z^\alpha(m)$ of Chern-Simons theory on a manifold $M$ with triangulation $\{\Delta_i\}_{i=1}^N$ can be expressed (roughly) as certain multiple integrals of a product of $N$ tetrahedron blocks $\psi(z_i')$,%
\footnote{The tetrahedron blocks actually needed for the state integral model will be nonperturbative completions of \eqref{QDLi}, constructed from ``noncompact'' quantum dilogarithm functions \cite{Fad-modular}.}
\be Z^\alpha \sim \int\!\!\!\int_{\CC^\alpha} \psi(z_1')\cdots\psi(z_N')\,. \label{ZSIMi} \ee
The label `$\alpha$' of the block, corresponding to a choice of complex ($SL(2,\C)$) flat connection $\CA^\alpha$ on $M$, determines the choice of integration cycle $C^\alpha$ used on the right hand side. This is highly reminiscent of the state integral model for analytically continued Chern-Simons theory presented in \cite{DGLZ} (based in turn on \cite{hikami-2006}), as well as of the structure of \emph{infinite}-dimensional integration cycles for the Chern-Simons path integral that define holomorphic blocks in \cite{Wit-anal, Witten-path, Witten-HK}. We believe that the present state integral model is equivalent to that of \cite{DGLZ}, although we have not yet attempted to show this directly. In principle, both state integral models should thought of as finite-dimensional versions of the infinite-dimensional path integrals in \cite{Wit-anal, Witten-path, Witten-HK}.

\subsection*{Topological strings} 

As a final relevant topic in this introduction, let us mention a rather different place in physics where quantum Riemann surfaces arise: open B-model topological string amplitudes. The precise context involves the B-model on a noncompact Calabi-Yau manifold $X$ that is described by an equation
\be X= \{\xi\xi' = H(x,y)\} \quad\subset\quad \C^2\times \C^*{}^2\,.\ee
Such a geometry is typically mirror to a noncompact toric Calabi-Yau in the A-model. It is a fibration of the $(x,y)$ plane by complex hyperbolas, with the hyperbolas degenerating to a reducible union of lines on the Riemann surface
\be \Sigma = \{H(x,y)=0\}\quad\subset\quad \C^*\times \C^*. \ee
After placing a noncompact B-brane at a point $x$ on $\Sigma$ and extending in either the $\xi$ or $\xi'$ fiber directions, the open topological string amplitude becomes (locally) a function of the open string modulus $x$,
\be Z^{\rm open} = Z^{\rm open}(x)\,. \ee

It is argued in \cite{adkmv} that in fact $Z^{\rm open}(x)$ should be treated as a wavefunction that is annihilated by a quantized version of the Riemann surface $\Sigma$, \ie
\be \hat H(\hat x,\hat y;q) Z^{\rm open}(x) = 0\,, \ee
with $\hat x\hat y=q\hat y\hat x$, where now $q = e^{ig_s}$.
The known methods for quantizing $H(x,y)$ involve matrix models \cite{remodelingBmodel, EO}, and express $\hat H(\hat x,\hat y;q)$ not as a finite polynomial in its three arguments (\cf\ \eqref{Ah41eg}) but as an infinite series in $\hbar$, the terms of which must be computed one by one, with increasing difficulty. It is tempting to hope that the quantization of $H(x,y)$ in topological string theory might be related to the quantization of $A(\ell,m)$ in Chern-Simons theory --- or, more generally, that quantization of ``Riemann surfaces'' is context-independent. Some promising experiments to test this idea were conducted by \cite{dijkgraaf-2009, DFM}.

In terms of our present gluing methods, it is very interesting to note that an ideal hyperbolic tetrahedra behaves very much like a pair of pants in a pants decomposition of the Riemann surface $\Sigma$. Namely, the algebraic equation for a pair of pants is just
\be x+y^{-1}-1 = 0 \quad\subset\quad \C^*\times\C^*\,. \ee
Moreover, the wavefunction $Z^{\rm open}(x)$ for a B-brane on a pair of pants obeys the equation
\be (\hat x+\hat y^{-1}-1)\,Z^{\rm open}(x)\,,\ee
and is given precisely by the quantum dilogarithm \eqref{QDLi}. This is the B-model mirror of a toric A-brane in $\C^3$, otherwise known to be computed by a one-legged topological vertex. One might hope that the gluing of pairs of pants to form a complete Riemann surface $\Sigma$ proceeds much along the same lines as the gluing of tetrahedra to form a complete three-manifold. These ideas --- leading in essence to a B-model mirror of the topological vertex formalism \cite{AKMV} --- will not be further explored in this paper, but will rather be the topic of future work \cite{BDH}. \\ \\

We now proceed, first by reviewing the details of analytically continued Chern-Simons theory, holomorphic blocks, and A-polynomials in Section 2; and then by breaking down and reinterpreting the meaning of gluing in TQFT in Section 3. In Sections 4 and 5 we consider the classical and quantum aspects, respectively, of ideal triangulations, and show how such triangulations ultimately lead to quantized A-polynomials. Finally, in Section 6 we focus attention back on the actual wavefunctions (holomorphic blocks) of Chern-Simons theory, and use ideal triangulations and gluing to construct a state integral model.

\section{Analytically continued Chern-Simons theory}
\label{sec:CS}

In quantum field theory, one generally expects that a partition function $Z$ can be expressed as a sum of contributions from all possible classical solutions,
\be Z\;\;\sim \sum_{{\rm classical\;sol\text{'}s}\;\alpha} Z^\alpha\,. \label{ZbQFT} \ee
Each $Z^\alpha$ could be thought of as obtained by quantum perturbation theory in a fixed classical background. In general, however, an expansion such as \eqref{ZbQFT} would only strictly hold in a perturbative regime.

As first proposed in \cite{gukov-2003}, and further developed in \cite{Wit-anal}, the notion of ``summing contributions from classical solutions'' can be made much more precise in the case of Chern-Simons theory. The basic result is that for any three-manifold $M$ and gauge group $G$, there are a set of well-defined, nonperturbative pieces $Z^\alpha(M;\hbar)$ that can be used to construct the Chern-Simons partition function. We will call them holomorphic blocks. Locally, they have a holomorphic dependence on the Chern-Simons coupling (or inverse level) $\hbar$. When $M$ has a boundary, they also depend holomorphically on boundary conditions.

The holomorphic blocks $Z^\alpha(M;\hbar)$ are in one-to-one correspondence with the set of flat \emph{complexified} gauge connections $\{\CA^\alpha\}$ on $M$ \cite{gukov-2003, DGLZ, Wit-anal}. They only depend on the complexified gauge group $G_\C$. The physical partition functions for Chern-Simons theory with compact gauge group $G$, or noncompact real gauge group $G_\R$, or even complex gauge group $G_\C$, are all constructed from the same blocks. Schematically,
\be Z_G(M;\hbar) = \sum_\alpha n_\alpha^{\rm cpt} Z^\alpha(M;\hbar)\,,\qquad Z_{G_\R}(M;\hbar) = \sum_\alpha n_\alpha^{\rm split} Z^\alpha(M;\hbar)\,, \label{blocks1} \ee
\be Z_{G_\C}(M;\hbar) = \sum_{\alpha,\bar\alpha}n_{\alpha,\bar\alpha}^{\rm cx} Z^\alpha(M;\hbar)\,\ol{Z}^{\bar \alpha}(M;\bar\hbar)\,. \label{blocks2}\ee
The coefficients $n_\alpha^{\rm cpt}$, $n_\alpha^{\rm split}$, or $n_\alpha^{\rm complex}$, discussed in \cite{Wit-anal}, are the only things that depend on the precise form of the Chern-Simons theory being considered. For many three-manifolds, the set of flat $G_\C$ connections is finite, and so the sums here are finite as well. Unlike the general QFT case \eqref{ZbQFT}, the left and right hand sides in these expressions, properly interpreted, are meant to be exactly equal.

It is the blocks $Z^\alpha(M;\hbar)$ that are actually annihilated, individually, by the ``quantum Riemann surface'' $\hat{A}(\hat{\ell},\hat{m};q)$ that forms the central focus of this paper. (On a perturbative level, this statement was one of the main observations of \cite{gukov-2003, DGLZ}.) In this section, we take some time to review the structure of \eqref{blocks1}-\eqref{blocks2}, and to properly understand the relation between the classical A-polynomial $A(\ell,m)$, the quantum A-polynomial $\hat{A}(\hat{\ell},\hat{m};q)$, flat connections, and partition functions. Although the actual Riemann surface $A(\ell,m)$ is intrinsically associated to the complexified rank-one gauge group $G_\C=SL(2,\C)$ (or to $G=SU(2)$, or $G_\R=SL(2,\R)$) and a knot complement $M=\ol{M}\bs K$, there exist corresponding classical varieties $A$ and quantum operators $\hat A$ for any gauge group and any oriented three-manifold with boundary \cite{DGLZ, VCReview}, so we will try to make general statements whenever possible.

\subsection{The structure of Chern-Simons theory}
\label{sec:anal}

For compact real group $G$, such as $G=SU(2)$, the standard Chern-Simons action on an oriented three-manifold $M$ is
\be I_{CS}(\CA) = \frac{k}{4\pi}\int_{M} \Tr\Big( \CA\wedge d\CA+\frac23 \CA\wedge \CA\wedge \CA\Big)\,, \label{ICSR} \ee
where $\CA$ is a connection one-form valued in the real Lie algebra $\fg$. The partition function of quantum Chern-Simons theory is calculated by the path integral
\be Z_{CS}(M) = \int \CD \CA\, e^{iI_{CS}(\CA)}\,. \label{ZCSR} \ee
This acquires a more standard quantum-mechanical form if we identify Planck's constant $\hbar$ as the inverse of the ``level'' $k$ and rescale the action,%
\footnote{From a physical perspective, it might be more natural to set $\hbar = \pi/k$ rather than $i\pi/k$, so as to keep $\hbar$ real. For us, it does not make much difference, since we will analytically continue in $\hbar$~anyway.~The conventions for $\hbar$ here differ from those of \cite{DGLZ} by a factor of two:
\be \hbar_{\rm here} = 2\hbar_{\text{ref.\,\cite{DGLZ}}}\,. \ee}
\be \hbar = \frac{2\pi i}{k}\,,\qquad S_{CS}(\CA):=\hbar\, {\textstyle I_{CS}(\CA) =  \frac{i}{2}\int_{M} \Tr\Big( \CA\wedge d\CA+\frac23 \CA\wedge \CA\wedge \CA\Big)} \ee
\be
 \hspace{0in}\imp\qquad Z_{CS}(M;\hbar) = \int\CD \CA\,
e^{\frac{i}{\hbar}S_{CS}(\CA)}\,. \label{ZCSRh} \ee
Note that, for $G$ compact, the action \eqref{ICSR} is invariant under large gauge transformations up to shifts by $2\pi k$ times an integer; so if $k\in \Z$ the path integral \eqref{ZCSR} is well-defined. If $G$ is not compact, this quantization of the level is not always necessary \cite{Witten-cx, barnatan-1991w}.

As proposed in \cite{gukov-2003}, and further discussed and developed in \cite{DGLZ, Wit-anal, Witten-path}, the level $k$ or its inverse $\hbar$ can be analytically continued to arbitrary nonzero complex  numbers, so long as large gauge transformations are removed from the gauge group.%
\footnote{Mathematically, a somewhat different analytic continuation for Jones polynomials was considered in \cite{Gar-AC}, though its precise relation to physics is unclear.} %
This keeps the actual value of $iI_{CS}$ well defined. Simultaneous with the continuation of $\hbar$, it is useful to allow the gauge connection $\CA$ to take values in $\fg_\C$. The initial path integral \eqref{ZCSR} can be viewed as integration along a real middle-dimensional contour, or integration cycle, in the space of complexified gauge connections \cite{Wit-anal}. However, one can also consider many other integration cycles. As long as the real part of the exponent $iI_{CS}$ tends to $-\infty$ at the endpoints of a cycle, the corresponding path integral remains well defined.

At fixed $\arg(k)$ (or $\arg(\hbar)$), the set of well defined integration cycles --- \ie\ the cycles leading to a finite path integral --- forms a vector space over $\Z$ (\ie\ a lattice). A basis $\{C^\alpha\}$ for this space is simply obtained by starting at any critical point $\alpha$ of the Chern-Simons functional and flowing ``downward'' from it such that $\Re(iI_{CS})$ decreases. In other words, one forms stationary phase contours by downward flow from  saddle points.  The critical points of the complexified Chern-Simons functional are just flat $G_\C$ connections $\CA^\alpha$, and it is well known that the set of flat connections on many three-manifolds is finite. Such three-manifolds include the complements of any knot $K$ in a closed, oriented three-manifold $\ol{M}$, so long as $\ol{M}\bs K$ has no closed incompressible surfaces and appropriate boundary conditions are imposed at the excised knot or link \cite{cooper-1994}. In such cases, we immediately find that the basis $C_\alpha$ is finite.%
\footnote{To be completely precise, large gauge transformations act nontrivially on the critical points. If we remove large gauge transformations from the gauge group, then the actual critical points of the action come in a finite set of infinite \emph{families}, each family being the large-gauge-transformation orbit of a single flat connection $\alpha$. The classical Chern-Simons action only differs by $2\pi i k \Z$ when evaluated on different elements of the same family. The net effect of this is to introduce extra factors of $e^{2\pi i k}$ in the $n_\alpha$ coefficients of sums like \eqref{cptsum} below, which plays a critical role in (\eg) understanding the Volume Conjecture, but will be of minimal importance here.\label{foot:gauge}}

The outcome of the analysis of \cite{gukov-2003} and later \cite{DGLZ, Wit-anal} briefly summarized here, is that any analytically continued Chern-Simons path integral can be written as a finite sum of contributions from different critical points,
\be Z_{CS}(M;\hbar) = \sum_\alpha n_\alpha \,Z_{CS}^{{\alpha}}(M;\hbar)\,. \label{cptsum} \ee
The asymptotic expansion of each $Z_{CS}^{{\alpha}}(M;\hbar)$,
\be Z_{CS}^{{\alpha}}(M;\hbar) \overset{\hbar\to 0}{\sim} \exp\left(\frac{i}{\hbar}S_0 + \ldots\right) \ee
corresponds to a perturbative expansion of $SL(2,\C)$ Chern-Simons theory in the background of a fixed flat connection $\CA^\alpha$ \cite{gukov-2003, DGLZ}. Nonperturbatively, each $Z_{CS}^{{\alpha}}(M;\hbar)$ can in principle be obtained by evaluating a Chern-Simons path integral on the downward-flow cycle $C^\alpha$ originating from the complex $G_\C$ critical point $\CA^\alpha$ in the space of complexified gauge connections \cite{Wit-anal}. The blocks $Z_{CS}^{{\alpha}}(M;\hbar)$ are universal, in the sense that they depend on $G_\C$ but not on any particular real form $G$. Moreover, they locally have a holomorphic dependence on both $\hbar$ and on potential boundary conditions.

The coefficients $n_\alpha$ in \eqref{cptsum} do depend on and indeed encode the actual integration cycle used for a particular path integral \cite{Wit-anal}. For example, if $G_\C=SL(2,\C)$, the natural real integration cycle $C$ corresponding to non-analytically-continued $G=SU(2)$ Chern-Simons theory is written as $C^{\rm cpt}=\sum_\alpha n_\alpha^{\rm cpt} C_\alpha$ for some $n_\alpha^{\rm cpt}$ . Similarly, the natural real integration cycle $C^{\rm split}=\sum_\alpha n_\alpha^{\rm split} C_\alpha$ for non-analytically-continued $G_\R=SL(2,\R)$ theory leads to some other set of $n_{\alpha}^{\rm split}$'s. Thus, the actual $SU(2)$ and $SL(2,\R)$ partition functions are written as two different sums, as in \eqref{blocks1}.

We could also have considered honest, physical $G_\C$ Chern-Simons theory, and tried to analytically continue it. For a complex gauge group, the general Chern-Simons action takes the form (\cf\ \cite{Witten-gravCS, Witten-cx, gukov-2003})
\begin{align} I_{CS}(\CA;t,\tilde{t}) \,=\,& \frac{t}{8\pi}\int_{M}\Tr\Big( \CA\wedge d\CA+\frac23 \CA\wedge\CA\wedge \CA\Big) \\
 &\quad + \frac{\tilde{t}}{8\pi}\int_{M}\Tr\Big( \ol{\CA}\wedge d\ol{\CA}+\frac23 \ol{\CA}\wedge\ol{\CA}\wedge \ol{\CA}\Big)\,. \nno
\end{align}
In order for $I_{CS}$ to be real, ${t}$ and $\tilde{{t}}$ should be complex conjugates, but we can analytically continue them as separate, independent complex variables. Simultaneously, the $\fg_\C$-valued connection $\CA$ should be analytically continued to a $(\fg_\C)_\C$-valued connection. However, since $(\fg_\C)_\C\simeq \fg_\C\oplus \fg_\C$, a $(\fg_\C)_\C$-valued connection is really just two copies (namely $\CA$ and $\ol{\CA}$, viewed independently) of a $\fg_\C$-valued one. The analytically continued Chern-Simons partition function then takes the form \cite{gukov-2003, Wit-anal}
\be Z_{CS}^{\C}(M;\hbar,\tilde{\hbar}) = \sum_{\alpha,\tilde{\alpha}} n_{\alpha,\tilde{\alpha}}\,Z_{CS}^{\alpha}(M;\hbar)\,\ol{Z}_{CS}^{\tilde{\alpha}}(M;\tilde{\hbar})\,, \label{cpxsum} \ee
where $\alpha$ and $\tilde{\alpha}$ label flat $\CA$ and flat $\ol{\CA}$ connections, respectively, and we have set
\be \hbar = \frac{4\pi i}{{t}}\,,\qquad \tilde{\hbar}=\frac{4\pi i}{\tilde{{t}}}\,.\ee
The blocks $Z_{CS}^{\alpha}(M;\hbar)$ of \eqref{cpxsum} are identical to those of \eqref{cptsum}. The natural ``real'' integration cycle $C^{\rm cx}$ in the space of complexified connections (namely, the middle-dimensional cycle where $\ol{\CA}$ is actually the conjugate of $\CA$) leads to a coefficient matrix $n_{\alpha,\tilde{\alpha}}^{\rm cx}$ that is diagonal, although for a general integration cycle the $n_{\alpha,\tilde{\alpha}}$ can be arbitrary.

\subsection{Flat connections and the A-polynomial}
\label{sec:flatCS}

For most of this paper we specialize to a three-manifold $M$ that is the complement of a knot (or sometimes a link) in a closed, oriented manifold $\ol{M}$,
\be M = \ol{M}\bs K\,, \ee
typically with $\ol{M}=S^3$. We also take our gauge group to be nonabelian of rank one, \ie\ $SU(2)$, or $SL(2,\R)$, or even $SL(2,\C)$. It makes no difference precisely which group is chosen, since we are only interested in holomorphic blocks $Z^\alpha(M;\hbar)$. These blocks will always be labelled by flat $SL(2,\C)$ connections $\CA^\alpha$ on $M$.

What, then, are the flat $SL(2,\C)$ connections on a knot complement? Flat connections are fully characterized by their holonomies, up to gauge equivalence. Since $SL(2,\C)$ is an algebraic group, the set of flat connections forms an algebraic variety
\be \CX = \Hom\big(\pi_1(M),SL(2,\C)\big)/\raisebox{-.07cm}{conjugation}\,, \label{defX} \ee
called the $SL(2,\C)$ character variety of $M$. The complex dimension of components of $\CX$ is always $\geq 1$ for a knot complement $M=\ol{M}\bs K$ \cite{thurston-1980} (for a link complement, the dimension is at least as big as the number of link components), but, in general, it can become arbitrarily large \cite{cooper-1996}. To simplify our discussion, we can additionally assume that our knot complements $M$ have no closed incompressible surfaces, which assures that $\dim_\C\CX \equiv 1$ \cite{cooper-1994}. However, this assumption does not appear strictly necessary.

\begin{figure}[htb]
\centering
\includegraphics[width=3in]{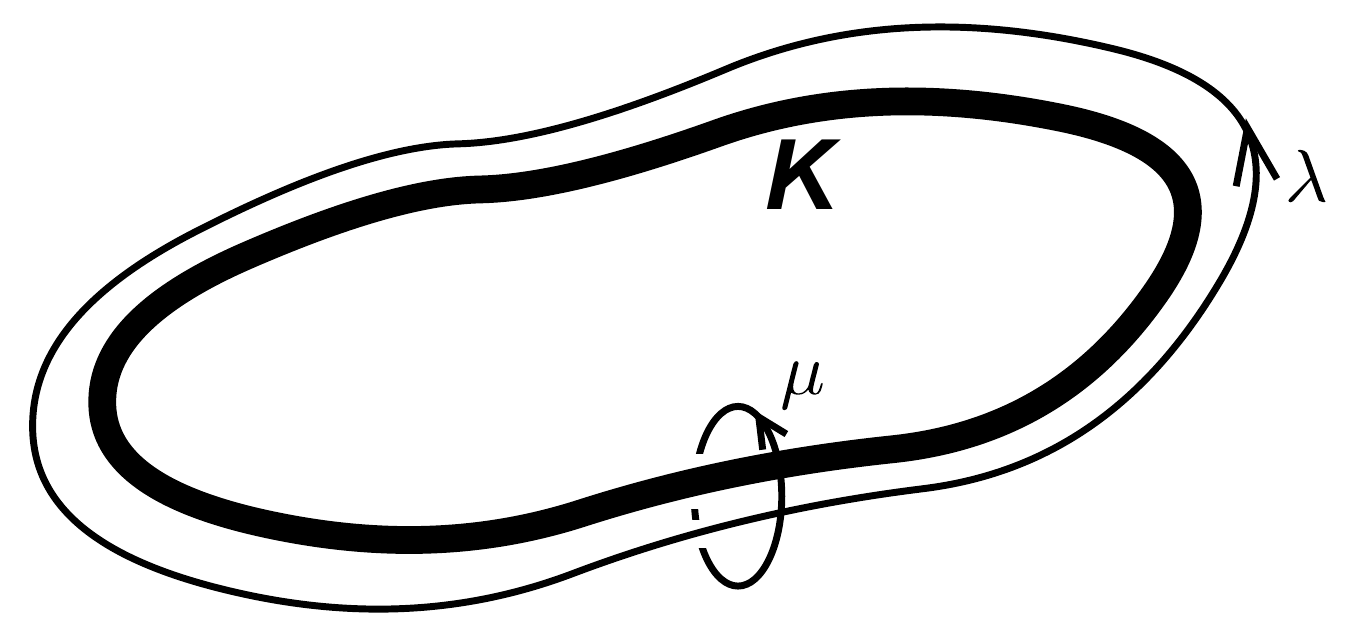}
\caption{Meridian $\mu$ and longitude $\lambda$ cycles in $M$. Here we are looking from ``inside'' $M$; the boundary torus $\pd M=T^2$ is the boundary of a neighborhood of the thickened knot $K$.}
\label{fig:LM}
\end{figure}

In Chern-Simons theory on a knot complement, one must specify gauge-invariant boundary conditions on the boundary torus $\pd M \simeq T^2$. Such boundary conditions are also given by holonomies, up to conjugation, on two independent cycles of this torus. A standard basis of cycles is given by the so-called longitude and meridian of $T^2$, which are canonically defined for a knot complement in $S^3$: the meridian $\mu$ is a small loop linking the (excised) knot once, and the longitude is a cycle $\lambda$ that intersects $\mu$ once and is null-homologous in $M$ --- essentially a projection of the knot itself to the boundary torus. These cycles are sketched in Figure \ref{fig:LM}. Since the fundamental group $\pi_1(T^2) = \Z\times\Z$ is generated by loops around the meridian and longitude cycles and is abelian, the $SL(2,\C)$ holonomies around $\mu$ and $\lambda$ can be simultaneously brought to normal form,
\be \mu \sim \mtt m * 0 {m^{-1}}\,,\qquad \lambda\sim
 \mtt \ell * 0 {\ell^{-1}}\,, \label{lambdamu} \ee
where $*$ can be $1$ if the eigenvalues coincide and otherwise $*=0$.
The Weyl group $\Z_2$ of $SL(2,\C)$, a residual gauge symmetry, acts on the matrices \eqref{lambdamu} to simultaneously exchange $(\ell,m)\leftrightarrow(\ell^{-1},m^{-1})$.

Naively, the two eigenvalues $(\ell,m)\in(\C^*)^2/\Z^2$ parametrize the classical boundary conditions in Chern-Simons theory. However, both classically and quantum mechanically, it is only possible to specify one element in this pair. Classically, this is clear when the dimension of the character variety \eqref{defX} parametrizing representations of $\pi_1(M)$ into $SL(2,\C)$ is 1: for both matrices $\mu$ and $\lambda$ of \eqref{lambdamu} to be part of the same representation some relation between $\ell$ and $m$ must be imposed. This turns out to be true even when $\CX$ has components of dimension $>1$ \cite{cooper-1994, Hodgson}. The relation between $\ell$ and $m$ is algebraic and takes the form
\be A(\ell,m) = 0\,. \ee
Aside from presently unimportant technical details, this is the definition of the classical A-polynomial \cite{cooper-1994}. It has been shown that the variety
\be  \{A(\ell,m)=0\} \quad\subset\quad (\C^*\times \C^*)/\Z^2 \label{AvarCS} \ee
is birationally equivalent to $\CX$ in many cases --- for example, there is always birational equivalence on the components of $\CX$ and $\{A(\ell,m)=0\}$ containing hyperbolic flat connections \cite{Dunfield-cyclic}.

Put a little differently, the space of flat $G_\C$ connections $\CP_{T^2}=(\C^*\times\C^*)/\Z_2$ on the boundary torus is the classical phase space of analytically continued Chern-Simons theory \cite{gukov-2003}. A ``classical state'' of Chern-Simons, \ie\ a flat connection, is described by the condition that a flat connection on the torus extends to be a flat connection on all of $M$, and this is precisely the condition $A(\ell,m)=0$. Thus, to describe a good classical boundary condition one can specify either $m$ or $\ell$, but not both independently. We will always choose to specify $m$. Then the number of flat connections on $M$ with fixed $m$ is simply equal to the degree of the A-polynomial in $\ell$. Each solution to $A(\ell,m)=0$, counted with multiplicity if necessary, corresponds to a holomorphic block in the expansion\label{Zfinite}
\be Z(M;m;\hbar)_{CS} = \sum_\alpha n_\alpha Z_{CS}^{\alpha}(M;m;\hbar)\,, \label{Zsumm} \ee
where now the $Z_{CS}^{{\alpha}}$'s are locally holomorphic functions of the boundary condition $m$.

As one varies $m$ and different branches of the surface $A(\ell,m)=0$ intersect, a block $Z_{CS}^{\alpha}(M;m;\hbar)$ may pick up contributions from other blocks $Z_{CS}^{\alpha'}(M;m;\hbar)$. Simultaneously, the coefficients $n_{\alpha'}$ will jump in order to keep the LHS continuous. This is a version of the Stokes phenomenon that was discussed at length in \cite{Wit-anal}.

Semi-classically, the holomorphic, analytically continued Chern-Simons action \eqref{ICSR} induces a holomorphic symplectic structure on the complexified classical phase space $\CP_{T^2}=(\C^*\times\C^*)/\Z_2=\{(\ell,m)\}/\Z_2$, given by (\cf\ \cite{axelrod-witten, Witten-cx, EMSS, gukov-2003})
\be \omega_{T^2} =  \frac{k}{4\pi}\,\int_{T^2}\Tr\big(\delta \CA\wedge\delta\CA\big) = \frac{2}{i\hbar} \,d\log \ell\wedge d\log m\,. \label{bdysymp} \ee
More commonly, this is written in logarithmic variables as 
\be  \boxed{\omega_{T^2} = \frac{2}{i\hbar} dv\wedge du\,,\qquad \ell=e^v\,,\quad m=e^u}\,. \label{wuv} \ee
Since the Chern-Simons action is first-order in derivatives, this symplectic structure contains no ``time derivatives'' of $u$ or $v$; rather, the conjugate momenta to coordinates $u$ and $v$ are coordinates themselves. Upon canonical quantization, the ``Hilbert space'' of analytically-continued Chern-Simons theory with torus boundary is identified with the space of functions of $u$ or of $v$, but not both. We will work in the representation where the holomorphic blocks, vectors in this ``Hilbert space,'' are functions of $u$ as in \eqref{Zsumm}. Invariance under the Weyl group action on $\CP_{T^2}$ requires the holomorphic blocks to be invariant under $m\leftrightarrow m^{-1}$, or $u\leftrightarrow -u$.

We have intentionally put ``Hilbert space'' in quotes here. In physics, a phase space is usually endowed with a \emph{real} symplectic form, not a holomorphic one. Quantization then leads to either a finite-dimensional vector space (if the phase space is compact), or to something like the space of $L^2$ functions of half the real phase space coordinates. For example, if we were quantizing honest $SU(2)$ Chern-Simons theory on the torus, the phase space would be $\CP_{\rm cpt}=(S^1\times S^1)/\Z_2$, and the Hilbert space $\CH_k$, consisting of level-$k$ representations of affine $\mathfrak{su}(2)$, would be finite-dimensional \cite{Wit-Jones, EMSS}. Similarly, if we consider $SL(2,\R)$ Chern-Simons theory, the phase space is $\CP_{\rm split}=\R^2/\Z_2\cup (S^1)^2/\Z_2$, and the Hilbert space is $\CH_{\rm split}=L^2(\R)\oplus \CH_k$.\label{SL2Rspace} In the actual case of complex $SL(2,\C)$ Chern-Simons theory, the phase space is $\CP_{\rm cx}=(\C^*)^2/\Z_2$ as above, but the real symplectic form is $\omega \sim t dv\wedge du+\tilde{t}d\bar v\wedge d\bar u$ \cite{Witten-cx, gukov-2003}. Expressing the phase space as $\CP_{\rm cx} = (\R^2\times (S^1)^2)/\Z_2$ leads to $\CH_{\rm cx}=L^2(\R)\otimes \CH_k$.

In contrast to these physical theories, the quantization that we are describing here is holomorphic. In terms of quantizing an algebra of operators and (eventually) talking about things like the quantum $\hat A$-polynomial, there is no problem with this. Indeed, it is the usual state of affairs in (\eg) deformation quantization \cite{Kont-def}. More interestingly, holomorphic quantization of the algebra of operators has a natural interpretation in terms of brane quantization \cite{gukov-2008, Gukov-quantMS}. It becomes very clear in the brane picture that the quantized algebra of operators (a space of ``($\CB_{cc},\CB_{cc}$)'' strings in \cite{gukov-2008}) depends only on the complexified form of the underlying real phase space.

In addition to the abstract algebra of operators, we find ourselves dealing here with a holomorphic version of wavefunctions themselves, namely the holomorphic blocks. These \emph{do not} live in an honest Hilbert space. They do, however, live in a vector space --- essentially a space of holomorphic functions --- that constitutes a representation of the operator algebra. In favorable circumstances, the holomorphic blocks may also be thought of as analytic continuations of wavefunctions in an actual $L^2$ Hilbert space. In our case, it is particularly tempting to consider them as analytic continuations of functions in the $L^2(\R)$ component of $\CH_{\rm split}$ above.

Coming back to the complexified phase space $\CP_{T^2}$ of the torus, the equation $A(\ell,m)=0$ that describes a classical state must be implemented as a quantum constraint on the Chern-Simons wavefunction \cite{gukov-2003}. The symplectic form \eqref{wuv} leads to a commutation relation
\be [\hat{v},\,\hat{u}]=\frac\hbar2 \ee
in the algebra of operators. For the classical coordinates $\ell$ and $m$, this implies that
\be \hat{\ell}\hat{m} = q^{1/2}\hat{m}\hat{\ell}\,,\qquad \hat{\ell}=e^{\hat v}\,,\quad \hat{m}=e^{\hat u}\,,\ee
with
\be q = e^{\hbar}\,. \ee
As described in the introduction, we then expect that the polynomial $A(\ell,m)$ is promoted to an operator $\hat{A}(\hat{\ell},\hat{m};q)$ that annihilates the Chern-Simons partition function \cite{gukov-2003, DGLZ} --- or, more precisely, the holomorphic blocks $Z_{CS}^{{\alpha}}(M;u;\hbar)$. The elementary operators $\hat{\ell}$ and $\hat{m}$ act on (locally) holomorphic functions $f(u)$ as
\bse
\begin{align}
&\hat{v}\,f(u) = \frac{\hbar}{2}f'(u) \hspace{-.3in}&& \hat u f(u) = uf(u)\,,\\
&\hat{\ell}\,f(u) = e^{\frac\hbar2\pd_u}f(u) = f(u+\hbar/2)\,,\hspace{-.3in}
&&\hat{m}\,f(u) = e^u\,f(u)\,.
\end{align}
\ese

\subsection{Recursion relations and $\hat{A}$}
\label{sec:recursion}

Up to now, almost all the known examples of operators $\hat{A}(\hat{\ell},\hat{m};q)$ have been derived by finding recursion relations for colored Jones polynomials \cite{Gar-Le, garoufalidis-2004, Gar-twist, Gar-sl3}. (A notable exception includes work using skein modules for the Kauffman bracket, \eg\ in \cite{Frohman-Gelca, Gelca-AJ, Gelca-Sain} and later \cite{Le-skein}.)
The fact that a relation of the form $\hat{A}(\hat{\ell},\hat{m};q)\,Z_{CS}(u)=0$ translates to a recursion relation for Jones polynomials has been explained in \cite{DGLZ, VCReview}. After understanding the structure of $SU(2)$, $SL(2,\R)$, and $SL(2,\C)$ partition functions as explained above, the relation simply amount to the facts that 1) the colored Jones polynomials $J_N(K,q)$ can be expressed as $SU(2)$ partition functions on knot complements, and 2) there then exists an appropriate change of variables between $(u,\hbar)$ and $(N,q)$. Let us review briefly how this works.

Physically, the colored Jones polynomial $J_N(K,q)$ is the non-analytically-continued $SU(2)$ Chern-Simons partition function on the three-manifold $\ol{M}=S^3$, with the insertion of a Wilson loop operator along a knot $K$ \cite{Wit-Jones, Resh-Tur, Kir-Mel}. The variable $q$ in $J_N(K,q)$ is the same $q$ that appears throughout this paper; it is related to the (quantized and renormalized) Chern-Simons level $k$ as
\be q = e^{\hbar} = e^{\frac{2\pi i}{k}}\,.\ee
The positive integer $N$, on the other hand, is the dimension of the $SU(2)$ representation used for the Wilson loop. By standard arguments (see \eg\ \cite{EMSS, Witten-IAS, Beasley-WL}), such a Wilson loop creates a singularity in the Chern-Simons gauge field $A$, precisely such that its holonomy on an infinitesimally small circle linking the knot is conjugate to
\be \mu \sim \mtt {e^{\frac{i\pi N}{k}}} 0 0 {e^{-\frac{i\pi N}{k}}}\,. \label{mucpt} \ee
Indeed, one can do away with the knot completely if we simply excise it from $\ol{M}=S^3$, and enforce the condition that the gauge field has a holonomy \eqref{mucpt} at the new boundary $T^2$ of the knot complement. Put differently, this is just the statement that in three dimensional Chern-Simons theory Wilson loops are interchangeable with 't Hooft loops.

From \eqref{mucpt}, we see that we should identify the standard holonomy eigenvalue $u$ with $i\pi N/k$. Therefore, the appropriate change of variables is
\be (u,\hbar) = \Big( i\pi\frac{N}{k}, \frac{2\pi i}{k}\Big)\,. \ee
The operators $\hat{\ell}$ and $\hat{m}$ then act on the set of Jones polynomials $\{J_N(K;q)\}_{N\,\in\,\mathbb{N}}$ as
\be \hat{\ell}\,J_N(K;q) = J_{N+1}(K;q)\,,\qquad \hat{m}\,J_N(K;q) = q^{N/2}J_N(K;q)\,,\ee
and the relation $\hat{A}(\hat{\ell},\hat{m};q)\,Z_{CS}(u)=\hat{A}(\hat{\ell},\hat{m};q)J_N(K;q) = 0$ is precisely a recursion relation for $J_N(K;q)$. The order of the recursion is equal to the degree of $A(\ell,m)$ in $\ell$, and hence also equal to the number of flat $SL(2,\C)$ connections on $M=\ol{M}\bs K$.

Such a recursion relation for $J_N(K;q)$ was found quite independently of analytically-continued Chern-Simons theory in \cite{Gar-Le, garoufalidis-2004}. It was argued there that the recursion operator $\hat{A}(\hat\ell,\hat m;q)$ should reproduce the classical A-polynomial $A(\ell,m)$ when $q\to 1$. From the point of view of Chern-Simons theory, it is fairly clear that there should always exist an operator $\hat{A}(\hat\ell,\hat m;q)$ with the properties that 1) it gives a recursion relation for the Jones polynomials of knots in any manifold; and 2) it reduces to the character variety in the classical limit $q\to 1$. Of course, our goal here is to actually construct $\hat A(\hat\ell,\hat m;q)$ from first principles.

\subsection{Logarithmic coordinates}
\label{sec:log}

In many places in this paper, we will find it convenient to lift complexified phase spaces like $\CP_{T^2}$, introduced in Section \ref{sec:flatCS}, to their universal covers. In other words, instead of using exponentiated coordinates $m$ and $\ell$ on $\CP_{T^2}$, we will use genuine logarithmic coordinates $u$ and $v$, with no assumption of periodicity under shifts by $2\pi i$. As far as the analysis of an operator algebra and the construction of operators like $\hat A(\hat \ell,\hat m;q)$ are concerned, the choice of logarithmic vs. exponential coordinates is unimportant. However, it ends up being highly relevant when considering analytically continued wavefunctions and holomorphic blocks. In particular, it appears that the holomorphic blocks $Z^\alpha_{\rm CS}(M;u;\hbar)$ for a knot complement $M$ naturally are non-periodic, locally holomorphic functions of $u$, rather than functions of $m=e^u$.

One way to see that holomorphic blocks should be non-periodic functions of $u$ is to extend the analysis of analytic continuation of \cite{Wit-anal} from knots in closed three manifolds to knot complements $M = \ol{M}\bs K$. For example, suppose that we consider $SU(2)$ Chern-Simons theory on knot complement $M=\ol{M}\bs K$, where the meridian holonomy has eigenvalue
\be m = e^u = e^{\frac{i\pi N}{k}}\,, \ee
as in \eqref{mucpt} above. In standard $SU(2)$ Chern-Simons theory, both $N$ and $k$ must be integers. Moreover, there exist large gauge transformations --- essentially transformations winding around the meridian loop --- that transform $N$ to $N+2k$, confirming the fact that $u$ and $u+2\pi i$ describe equivalent boundary conditions. (In the dual picture of a knot $K$ inside a closed manifold $\ol{M}$, as described in Section \ref{sec:recursion}, it is precisely these gauge transformations that assure us a representation of dimension $N$ on the knot is equivalent to one of dimension $N+2k$, \cf\ \cite{EMSS}.)

Now, both integers $N$ and $k$ of $SU(2)$ Chern-Simons theory can be analytically continued to be arbitrary nonzero complex numbers. The analytic continuation in $k$ requires one to stop quotienting out by large gauge transformations on $\ol{M}$ in the Chern-Simons path integral measure. As described briefly in Footnote \ref{foot:gauge}, this introduces multiplicative ambiguities by factors of the form $e^{2\pi i ak}=e^{-\frac{4\pi^2a}{\hbar}}$, $a\in \Z$, into the definition of an analytically continued partition function, or holomorphic block. Similarly, analytic continuation in $N$ forces one to stop quotienting out by the large gauge transformations wrapping the meridian cycle on the boundary of $M$. Fundamentally, this results in holomorphic blocks that are (locally) holomorphic but no longer periodic in $u$. Practically, the effect of not including large gauge transformations on the meridian cycle is to introduce multiplicative ambiguities of the form
\be  e^{2\pi i b N} = e^{\frac{4\pi i b u}{\hbar}}\,,\qquad b\in\Z \label{uambig} \ee
into the definition of a holomorphic block, and it is very easy to see that \eqref{uambig} is not invariant under $u\to u+2\pi i$ for arbitrary complex $\hbar$.

The setup of analytically continuing both $N$ and $k$ is the one relevant to the current paper (as it was in \cite{gukov-2003, DGLZ}), and we will eventually find that our holomorphic blocks are indeed not periodic. Thus, we will almost always use lifted logarithmic coordinates on complex phase spaces. In addition to the complexified phase space $\CP_{T^2}$ discussed in Section \ref{sec:flatCS}, we will introduce very similar, two-complex-dimensional phase spaces $\CP_{\pd\Delta}$ for tetrahedra in Sections \ref{sec:hyp}--\ref{sec:quant} (\cf\ \eqref{Pintro} in the introduction). These phase spaces are again described most naturally in logarithmic coordinates. In Section \ref{sec:duality}, we shall see very explicitly that the appropriate conformal block $\psi(z')$ for a tetrahedron is not a function of the exponentiated variable $z'$ but actually a function of $Z'=\log z'$. It breaks $Z'\to Z'+2\pi i$ periodicity by nonperturbative effects precisely of the form \eqref{uambig}.

\subsection{The structure of $A$ and $\hat{A}$}
\label{sec:Apoly}

The operator $\hat{A}(\hat{\ell},\hat{m};q)$ introduced in Section \ref{sec:flatCS} has several important but highly nontrivial properties. First, it is a polynomial in $q$ as well as in the operators $\hat{\ell}$ and $\hat{m}$. A priori, one could instead have expected an arbitrary infinite series in the coupling constant $\hbar$.%
\footnote{In the analogous case of the topological B-model, almost all the known examples of the operator $\hat{H}(\hat{x},\hat{y})$ are \emph{only} expressed as such infinite series.} %
The fact that all $\hbar$-corrections can be re-summed into a finite number of $q$'s follows from the construction of $\hat{A}(\hat{\ell},\hat{m};q)$ as a recursion relation for colored Jones polynomials. This property will also follow easily from our construction in Section \ref{sec:quant}.

Second, we implied in Section \ref{sec:flatCS} above that the operator $\hat{A}(\hat{\ell},\hat{m};q)$ annihilates not just a complete Chern-Simons partition function as in \eqref{cptsum}, but every individual holomorphic block $Z_{CS}^{{\alpha}}(M;u;\hbar)$. Perturbatively, this was already evident from the analysis of analytic continuation in \cite{gukov-2003}. Further confirmation appeared in \cite{DGLZ}, where actual solutions to $\hat{A}(\hat{\ell},\hat{m};q)\,Z(u)=0$ were constructed using a state integral model. Although the solutions of \cite{DGLZ} were described perturbatively, as saddle point expansions of finite-dimensional integrals, one could try to extend the integration contours of \cite{DGLZ} by downward flow to define nonperturbative $Z_{CS}^{{\alpha}}(M;u;\hbar)$'s as well.

More generally, we observe that in any quantization scheme the order of the difference equation $\hat{A}(\hat{\ell},\hat{m};q)\,Z(u)=0$ is  $\deg_{\ell}A(\ell,m)$, which is equal to the number of flat $SL(2,\C)$ connections on $M$. Therefore, the difference equation has a vector space of solutions of dimension $\deg_{\ell}A(\ell,m)$, and the basis elements of this vector space can be chosen to be precisely the functions $Z_{CS}^{{\alpha}}(M;\hbar;u)$. As discussed in \cite{gukov-2003, DGLZ}, the semi-classical asymptotics of the solutions $Z_{CS}^{{\alpha}}(M;\hbar;u)$ are in one-to-one correspondence with the classical solutions to $A(\ell,m)=0$ at fixed $m=e^u$. In particular
\be Z_{CS}^{{\alpha}}(M;\hbar;u) \sim \exp\left[ \frac2\hbar \int_{A(\ell,m)=0}^u v(u)du + O(\log\hbar) \right]\,, \ee
where the integral is performed over the ``$\alpha^{\rm th}$'' branch of the A-polynomial curve, and higher-order terms also have a geometric meaning corresponding in terms of flat connections $\CA^\alpha$ \cite{gukov-2003, DGLZ, gukov-2006}. (The lower limit of integration is fixed, but we do not need to specify it here. Changing it would simply multiply $Z_{CS}^{{\alpha}}(M;\hbar;u)$ by an overall constant, producing an equivalent basis element in the vector space of solutions to $\hat{A}(\hat{\ell},\hat{m};q)\,Z(u)=0$.)

In fact, a little more is true about the solutions to $\hat{A}(\hat{\ell},\hat{m};q)\,Z(u)=0$ and the structure of $\hat{A}(\hat{\ell},\hat{m};q)$. Recall that, classically, the A-polynomial of a knot in $\ol{M}=S^3$ always contains a factor $(\ell-1)$. This corresponds to an abelian component of the character variety $\CX$ --- a component where all the $SL(2,\C)$ holonomies of a flat connection are simultaneously diagonalizable, hence the representation of $\pi_1(M)$ factors through $GL(1)$. The abelianization of $\pi_1(M)=\pi_1(S^3\bs K)$ is just $H_1(S^3\bs K)\simeq\Z$, generated by the meridian loop in the knot complement. Therefore, the equation $\ell-1=0$ simply reflects the fact that for an abelian connection the holonomy along the longitude loop must be trivial. For example, the classical A-polynomial of the unknot complement is
\be \mb{U}\,:\qquad A(\ell,m)=\ell-1\,, \ee
since the longitudinal holonomy in $S^3\bs \mb{U}$ is always trivial; whereas the trefoil $(\mb{3_1})$, figure-eight knot $(\mb{4_1})$, and $\mb{5_2}$ knot complements have A-polynomials%
\footnote{It is known that any nontrivial knot in $S^3$ has a nontrivial A-polynomial; in other words, there are always components besides $(\ell-1)$ \cite{Gar-Dun-AS3}.}
\begin{align}
 \mb{3_1}\,&: \quad (\ell-1)(\ell+m^6)\,, \nno \\
 \mb{4_1}\,&: \quad (\ell-1)\big(m^4\ell^2-(1-m^2-2m^4-m^6+m^8)\ell+m^4\big)\,, \\
 \mb{5_2}\,&: \quad (\ell-1)\big(m^{14}\ell^3+m^4(1-m^2+2m^6+2m^8-m^{10})\ell^2 \nno \\
 &\qquad\qquad\qquad\qquad\qquad -(1-2m^2+2m^4+m^8-m^{10})\ell+1\big)\,. \nno
\end{align}

Quantum-mechanically, in the case of knot complements in $S^3$, it is {still} the case that $\hat{A}(\hat{\ell},\hat{m};q)$ has a factor of $(\hat{\ell}-1)$. This factor always appears on the left of the quantum operator, and factors out in a nontrivial manner. To be more precise, the recursion relations of \cite{Gar-Le, Gar-twist} for colored Jones polynomials always take the form
\be \hat{A}^{\rm na}(\hat{\ell},\hat{m};q)\, J_N(K;q) = B(m;q)\,, \label{recinhom} \ee
where the operator $\hat{A}^{\rm na}(\hat{\ell},\hat{m};q)$ is a quantization of the classical A-polynomial with the factor $(\ell-1)$ removed. This \emph{inhomogeneous} recursion implies the homogeneous recursion
\be (\hat{\ell}-1)\frac{1}{B(\hat{m};q)}\hat{A}^{\rm na}(\hat{\ell},\hat{m};q)\, J_N(K;q) = 0 \label{rechomB} \ee
\be \imp\qquad \big( B(\hat{m};q)\hat{\ell}-B(q^{1/2}\hat{m};q)\big)\hat{A}^{\rm na}(\hat{\ell},\hat{m};q)\, J_N(K;q) = 0\,. \label{rechom} \ee
The operator on the left-hand side of \eqref{rechom} is what we have been calling $\hat{A}(\hat{\ell},\hat{m};q)$.

The inhomogeneous recursion \eqref{recinhom} actually carries a little more information than the homogeneous version \eqref{rechom}. Most importantly for us, it seems to be the case that the operator $\hat{A}^{\rm na}(\hat{\ell},\hat{m};q)$ identically {annihilates} all blocks $Z_{CS}^{{\alpha}}(M;\hbar;u)$ \emph{except} for the block corresponding to the abelian flat connection. The abelian block $Z_{CS}^{(\alpha={\rm abel})}(M;\hbar;u)$, in contrast, satisfies \eqref{recinhom} with nonzero $B(m;q)$. We therefore have a situation that is very familiar from the theory of inhomogeneous differential equations: the functions $Z_{CS}^{(\alpha\neq{\rm abel})}(M;\hbar;u)$ constitute a vector space of general solutions to the homogeneous equation
\be \hat{A}^{\rm na}(\hat{\ell},\hat{m};q)\,Z_{CS}^{(\alpha\neq{\rm abel})}(M;\hbar;u) = 0\,, \label{recnonab} \ee
whereas the abelian block is a special solution (with fixed normalization!) to the inhomogeneous equation 
\be \hat{A}^{\rm na}(\hat{\ell},\hat{m};q)\,Z_{CS}^{(\alpha={\rm abel})}(M;\hbar;u) = B(m;q)\,. \label{recab} \ee
Any linear combination of nonabelian solutions plus one copy of the abelian block will then solve the inhomogeneous equation. Presumably, the colored Jones polynomial is a linear combination precisely of this type.\footnote{This fact was actually verified for the figure-eight knot in \cite{Wit-anal}.}

The structure appearing in equations \eqref{recnonab}-\eqref{recab} and the fact that $\hat{A}^{\rm na}(\hat{\ell},\hat{m};q)$ alone is sufficient to annihilate nonabelian blocks of the Chern-Simons partition function is by no means proven. Such a structure became apparent%
\footnote{We thank H. Fuji for very useful discussions on this topic and for sharing important examples related to this structure.} %
from studying examples of partition functions built with the state integral model of \cite{DGLZ, hikami-2006}. It is very important for us, since, in the remainder of the paper, \uline{it is the \emph{nonabelian} operator $\hat{A}^{\rm na}(\hat{\ell},\hat{m};q)$ that we actually construct}.

Our methods for quantizing $SL(2,\C)$ Chern-Simons theory will use the relation between flat $SL(2,\C)$ connections and hyperbolic metrics. Although only a single flat $SL(2,\C)$ connection on a three-manifold can correspond to a global hyperbolic metric \cite{Witten-cx, gukov-2003}, we will see that the tools of ideal hyperbolic triangulation can construct more general flat connections as long as they are nonabelian. Unfortunately, hyperbolic geometry can never detect an abelian flat connection, and this is why, for a knot complement in $S^3$, we at best find a quantized version 
$\hat{A}^{\rm na}(\hat{\ell},\hat{m};q)$ of the reduced A-polynomial, with the $\ell-1$ factor removed.

For knot complements in more general manifolds $\ol{M}\neq S^3$, abelian connections should again factor out as a component of the classical A-polynomial, though perhaps not in the form $(\ell-1)$.
Again, the ideal hyperbolic triangulations of Sections \ref{sec:hyp}-\ref{sec:quant} will only be able to describe and quantize reduced A-polynomials, where these factors have been removed. Something interesting can be gained from this statement. The fact that we \emph{can} always quantize a reduced, nonabelian A-polynomial by itself implies that in general, for a knot complement in any three-manifold, the full quantum A-polynomial should always have a left-factorized structure as in \eqref{rechomB}-\eqref{rechom}.

The precise relation between flat $SL(2,\C)$ connections and hyperbolic geometry will be discussed further in Section \ref{sec:classrem}. It is the hyperbolic ``gluing variety'' there that corresponds to $A^{\rm na}$ here. It is unfortunately not yet clear how to quantize the entire A-polynomial, \ie\ including abelian factors like $(\ell-1)$. The answer no doubt rests on understanding the physical basis for the inhomogeneity of \eqref{recinhom} or \eqref{recab}. With the exception of this subsection and Section \ref{sec:classrem} we remove the distinction ``na'' from $A^{\rm na}(\ell,m)$, simply referring to this reduced object as the ``A-polynomial.'' We hope that this will cause no confusion.

\subsection{Generalizations}
\label{sec:CSgen}

Although most of this paper focuses on rank-one nonabelian Chern-Simons theory on knot complements, hence on quantization of A-polynomials, much of the discussion in this section extends easily to more general situations (\cf\ \cite{DGLZ, VCReview}).

The simplest generalization would be to let a three-manifold $M$ be the complement of a link, $M=\ol{M}\bs L$. Then, instead of describing classical flat connections on $M$ by a single equation $A(\ell,m)=0$, there would be a system of equations
\be A_{1}(\ell_1,m_1,...,\ell_\nu,m_\nu) = \ldots= A_{\nu}(\ell_1,m_1,...,\ell_\nu,m_\nu) = 0\,, \label{linkA} \ee
where $\nu$ denotes the number of components of $L$. There is one pair of meridian and longitude holonomies for each of the $\nu$ torus boundaries. Algebraically, these equations generate an ideal. Geometrically, they describe a Lagrangian submanifold of the phase space
\be \CP_{\pd M} = \{(u_1,v_1,...,u_\nu,v_\nu)\} \simeq ((\C^*\times \C^*)/\Z_2)^{\nu}\,, \ee
with symplectic structure
\be \omega = \frac{2}{i\hbar}\sum_{i=1}^\nu dv_i\wedge du_i\,.\ee
Upon quantization, the $\nu$ equations \eqref{linkA} become quantum operators acting on a ``Hilbert'' space that, in analytic continuation, can be described as a space of holomorphic functions $f(u_1,...,u_\nu)$. The quantum operators $\hat A_j(\hat\ell_i,\hat m_i;q)$ generate a \emph{left} ideal in the noncommutative ring $\C(q)[\hat\ell_1{}^{\pm 1},\hat m_1{}^{\pm 1},...,\hat\ell_\nu{}^{\pm 1},\hat m_\nu{}^{\pm 1}]$, defined by the equations
\be \hat A_j(\hat\ell_1,\hat m_1,...,\hat\ell_\nu,\hat m_\nu;q)\simeq 0\,, \label{qlinkA} \ee
where ``$\simeq$'' means ``annihilates holomorphic blocks when acting on the left.'' In the classical limit $q\to 1$, this ideal reduces to the commuting ideal \eqref{linkA}.

Since ideal triangulations of link complements are no more complicated than ideal triangulations of knot complements, extending the methods of the present paper to the case of link complements is trivial. We usually ignore this generalization for simplicity of presentation.

Two further generalizations would be to three-manifolds with general Riemann surface boundaries, and to higher-rank gauge groups. From the point of view of Chern-Simons theory, still not much changes. For a boundary that is a higher-genus surface, one must carefully choose holonomies on dual cycles to build a phase space. Once that is done, there must again be a Lagrangian submanifold describing the flat connections on the boundary that extend to the bulk. In the case of higher-rank gauge groups, the increase in rank simply increases the number of independent holonomy eigenvalues that one should keep track of for any given boundary cycle. For example, on a torus, a simple Lie group of rank $r$ will lead to $r$ meridian eigenvalues, $r$ longitudinal eigenvalues, and a $2r$-dimensional phase space. We expect that the subset of flat connections that extend to the bulk always contains a Lagrangian submanifold as its highest-dimensional component; then the defining equations for the Lagrangian should be quantized as a system of $\hat A$ operators (\cf\ \cite{Gar-sl3}).

From the point of view of ideal triangulations, our practical building blocks for operator quantization in Sections \ref{sec:hyp} and Section \ref{sec:quant}, both higher-genus surfaces and higher-rank groups require some refined methods. Allowing higher-genus surfaces will necessitate modifying what we call ``vertex equations'' in Sections \ref{sec:hyp}-\ref{sec:quant}, because the standard hyperbolic structures on ideal tetrahedra cause all triangular pieces of boundary around ideal vertices to be Euclidean --- and Euclidean triangles cannot be glued together to form anything but a torus. In the case of higher-rank gauge groups, the triangulations themselves will require a refinement and further decoration, essentially a three-dimensional version of the two-dimensional refinement suggested by Fock and Goncharov in \cite{FG-Teich}. Another perspective on this necessary refinement appears in \cite{Zickert-sl3}. We hope to implement such generalizations in the future.

\section{Gluing with operators in TQFT}
\label{sec:opglue}

\begin{figure}[htb]
\centering
\includegraphics[width=3.6in]{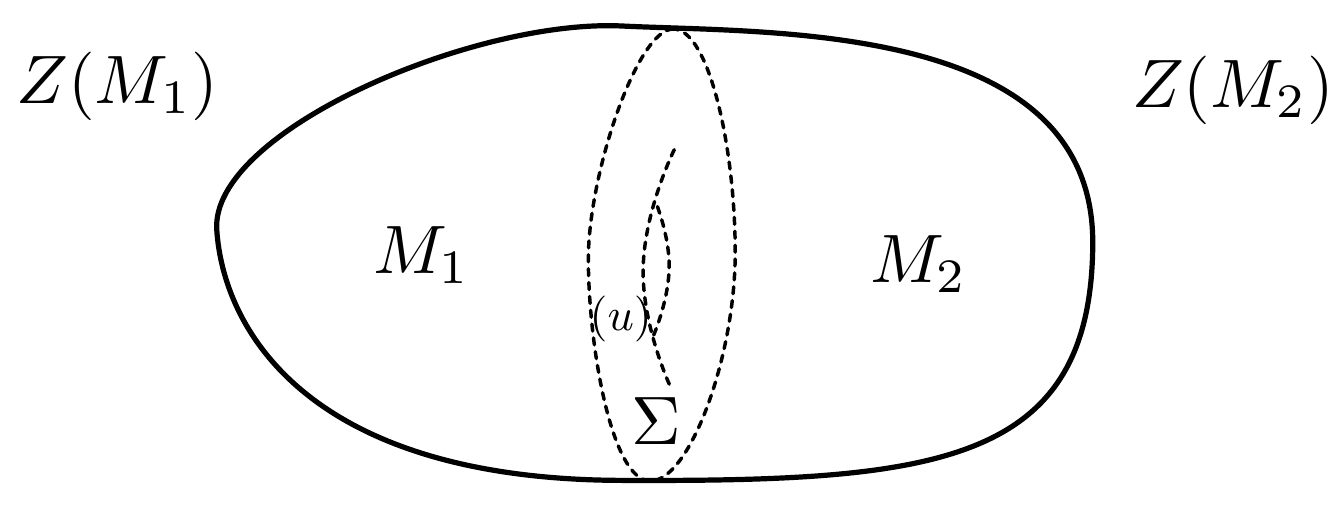}
\caption{Gluing wavefunctions in QFT}
\label{fig:cut}
\end{figure}

The partition function of any quantum field theory on a spacetime manifold $M$ can be constructed by cutting $M$ into pieces, calculating a partition function as a function of boundary conditions on each piece, and integrating out over boundary conditions to glue the pieces back together. Quantum mechanically, ``integrating out boundary conditions'' is precisely expressed as taking an inner product of wavefunctions in the Hilbert space associated to a boundary. For example, if an $n$-dimensional manifold $M$ is cut into pieces $M_1$ and $M_2$ along an $(n-1)$-dimensional $\Sigma$ as in Figure \ref{fig:cut}, then
\be Z(M) = \langle Z(M_1)\,|\,Z(M_2) \rangle_{\CH(\Sigma)}\,. \ee
Alternatively, since basis elements in $\CH(\Sigma)$ are just choices of quantum mechanical boundary conditions, labelled (say) by some symbol ``$u$,'' we can consider both $Z(M_1;u)$ and $Z(M_2;u)$ to be functions of $u$. Then
\be Z(M) = \int du\,Z(M_1;u)\,Z(M_2;u)\,, \ee
possibly with some complex conjugation of $Z(M_1;u)$ if appropriate.

When a quantum field theory is topological, the process of cutting and gluing becomes especially simple. In particular, since nothing in the theory depends on a metric, a Hilbert space $\CH(\Sigma)$ can be \emph{canonically} associated to the topological class of a boundary $\Sigma$. Similarly, wavefunctions such as $Z(M)$ and $Z(M_1),Z(M_2)\in\CH(\Sigma)$ only depend on the topologies of $M,\,M_1,\,M_2$. These ideas led to the mathematical axiomatization of TQFT by Atiyah and Segal \cite{atiyah-1990}.

In the case of Chern-Simons theory, the boundary Hilbert spaces $\CH(\Sigma)$ can be obtained systematically by geometric quantization. The classical phase space of $\Sigma$ is, by definition, the space of flat gauge connections on $\Sigma$ modulo gauge equivalence,
\be \CP(\Sigma) = \{\text{flat connections on $\Sigma$}\}\big/\text{gauge}\,, \ee
and Chern-Simons theory induces a symplectic form $\omega\sim\int_\Sigma \Tr(\delta A\wedge\delta A)$ on this space, \cf\ \eqref{bdysymp}. Geometric quantization then turns $\CP(\Sigma)$ into a Hilbert space $\CL(\Sigma)$, roughly thought of as the space of $L^2$ functions that depend on half the coordinates of $\CP(\Sigma)$. We described this in Section \ref{sec:CS} for the case $\Sigma=T^2$.

Now, in many quantum field theories, one can work not only with wavefunctions but with operators (``Schr\"odinger equations'') that annihilate the wavefunctions. Indeed, wavefunctions could be implicitly \emph{defined} as the solutions to Schr\"odinger equations, up to some normalization. In Section \ref{sec:CS}, we saw how this worked for Chern-Simons theory. The set of flat connections on a boundary $\Sigma$ that can extend to be flat connections throughout the bulk manifold $M$ forms a Lagrangian submanifold
\be \CL(M) = \{\text{flat connections on $M$}\}\big/\text{gauge}\quad\subset \quad \CP(\Sigma)\,. \ee
The equations that cut out this submanifold are (somehow) promoted to quantum operators, which in turn should all annihilate the partition function, or physical wavefunction.

Unfortunately, although cutting and gluing in terms of wavefunctions is a very familiar process in TQFT, cutting and gluing in terms of {operators} is not. This is what we mean to investigate in the present section. In particular, we want to know what happens to operators when two manifolds $M_1$ and $M_2$ are glued together along a common boundary $\Sigma$. If either of $M_1$ or $M_2$ has additional (unglued) boundaries, then the glued manifold $M_1\cup M_2$ still has a boundary, and there should therefore be a new operator that annihilates $Z(M_1\cup M_2)$ as a wavefunction. We want to explain how this new operator is obtained in terms of the original ones for $M_1$ and $M_2$.

\subsection{A toy model}

To begin, let us consider an example where the gluing of wavefunctions is already fairly well understood. Since we have just reviewed analytically-continued rank-one (\eg\ $SU(2)$ or $SL(2,\R)$) Chern-Simons theory on three-manifolds with torus boundary, we can take this as our TQFT.

(As discussed in Section \ref{sec:flatCS}, it does not quite make sense to talk about Hilbert spaces in an analytically continued theory. Rather, one should consider the analytic continuation of functions in a real Hilbert space. This really makes no difference to the illustrative construction here. For the reader's complete peace of mind, we can assume to be discussing an honest $SL(2,\R)$ Chern-Simons theory, and focus only on the $L^2(\R)$ part of the Hilbert space $\CH_{\rm split}$ defined on page \pageref{SL2Rspace}. That is, we assume that all phase space coordinates $u$ and $v$ are real and take all Hilbert spaces to be $L^2(\R)$ or $L^2(\R^N)$, as appropriate. Some more serious and practical implications of analytic continuation to a gluing construction for holomorphic blocks will be taken up in Section \ref{sec:wf}.)

\begin{figure}[h]
\centering
\includegraphics[width=5in]{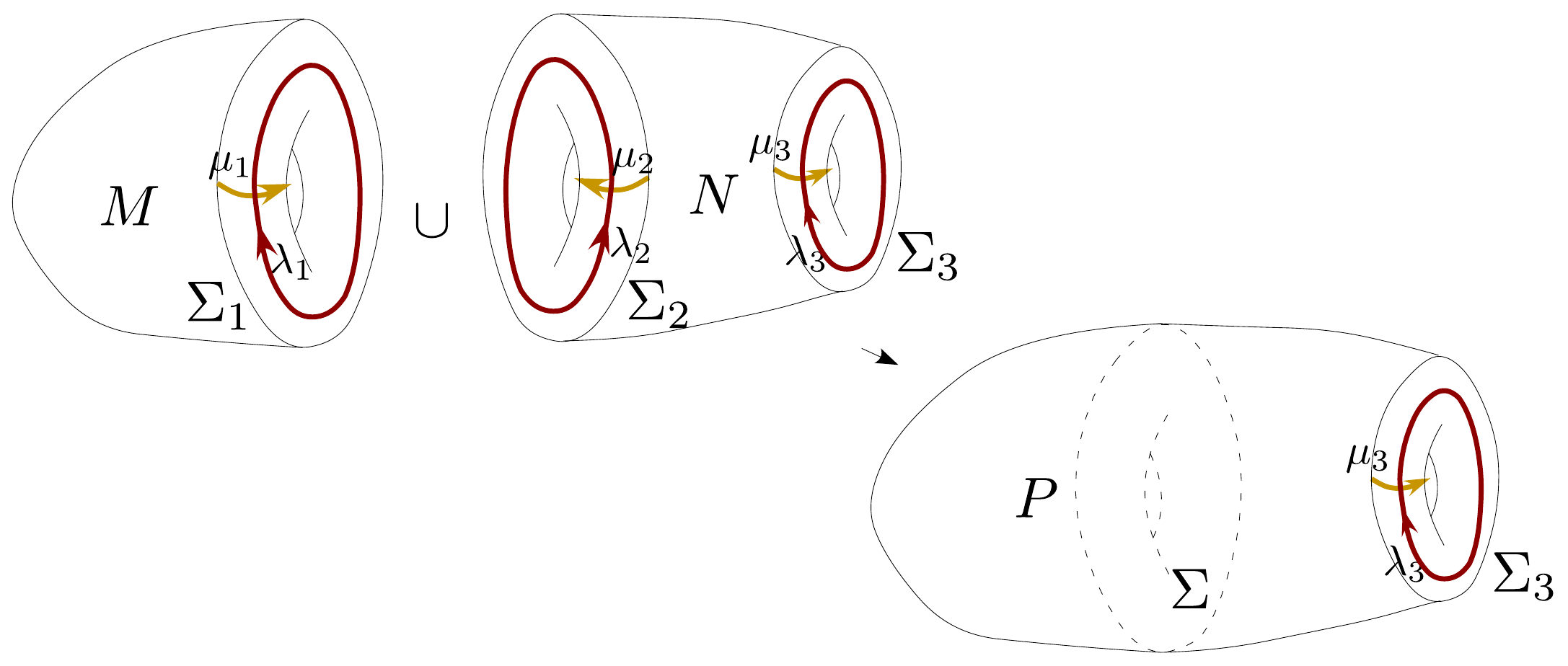}
\caption{The TQFT gluing setup. Holonomies around the cycles $\lambda_i$ and $\mu_i$ (not necessarily longitudes and meridians as defined in Section \protect\ref{sec:CS}) have eigenvalues $\ell_i$ and $m_i$ respectively.}
\label{fig:cutCS}
\end{figure}

The three-manifold to be considered appears in Figure \ref{fig:cutCS}.
We begin with two oriented manifolds $M$ and $N$, which have torus boundaries $\Sigma_1$ and $(-\Sigma_2)\cup \Sigma_3$, respectively. The minus sign in front of $\Sigma_2$ indicates a reversal of orientation, which will be quite important. These two manifolds are glued together by identifying $\Sigma_1=-\Sigma_2=\Sigma$, producing a manifold $P$ whose boundary is the torus $\Sigma_3$. We expect that the wavefunction $Z(P)$ can be expressed as an inner product, or an integral over boundary conditions at $\Sigma$,
\be Z_P = \langle Z_M\,|\,Z_N\rangle_{\CH_\Sigma}\,, \label{ZPMN} \ee 
and we want to recast this statement in terms of operators. Namely, given an operator $\hat{A}_M$ and operators $\hat{A}_{N,1}$, $\hat{A}_{N,2}$ (there are two of them, since $N$ is a link complement) that annihilate the partition functions $Z_M$ and $Z_N$, respectively, we want to construct the operator $\hat{A}_P$ that annihilates $Z_P$.

Semi-classically, we know that for each boundary $\Sigma_i$ there is a phase space $\CP_{\Sigma_i}$ consisting of flat connections at that boundary. We can choose a basis of ``longitude'' and ``meridian'' cycles for each torus. These are not necessarily the actual longitude and meridian as defined in Section \ref{sec:flatCS} (our manifolds are not necessarily knot or link complements in $S^3$). We can, however chose the cycles such that $(\lambda_1,\mu_1)$ on $\Sigma_1$ are identified with $(\lambda_2,\mu_2)$ on $\Sigma_2$ during the gluing. Each $\CP_{\Sigma_i}$ can then be described by holonomy eigenvalues as $\{(\ell_i,m_i)\}\in\C^*\times \C^*$, or, in lifted logarithmic coordinates $\ell_i=e^{v_i}$ and $m_i=e^{u_i}$ (\cf\ Section \ref{sec:log}), as
\be \CP_{\Sigma_i} = \{(v_i,u_i)\} \simeq \C\times \C\,, \ee
modulo a $\Z_2$ Weyl group action. The Weyl group quotient simply requires that wavefunctions (as in \eqref{MNPH} below) ultimately be invariant under $u_i\leftrightarrow -u_i$.
The phase spaces associated to the boundaries of $M$ and $N$ become
\be \CP_{\pd M} = \CP_{\Sigma_1} = \{(v_1,u_1)\}\,,\qquad
 \CP_{\pd N} = \CP_{-\Sigma_2}\times\CP_{\Sigma_3} = \{(v_2,u_2,v_3,u_3)\}\,. \ee
The symplectic forms on these spaces are
\be \frac{i\hbar}{2}\omega_{\pd M} = dv_1\wedge du_1\,,\qquad \frac{i\hbar}{2}\omega_{\pd N} = -dv_2\wedge du_2+dv_3\wedge du_3\,, \label{sympMN}\ee
where $dv_2\wedge du_2$ acquires an extra minus sign due to the orientations.

After gluing together $M$ and $N$, we construct a manifold $P$ whose boundary phase space is
\be \CP_{\pd P} = \CP_{\Sigma_3} = \{(v_3,u_3)\}\,, \ee
with symplectic structure
\be \frac{i\hbar}{2}\omega_{\pd P} = dv_3\wedge du_3\,.\ee
Although we know this must be the end result of the gluing, it is useful to understand how $\CP_{\pd P}$ can be systematically obtained from $\CP_{\pd M}$ and $\CP_{\pd N}$.

To this end, observe that the classical identification of boundary conditions $\ell_1=\ell_2$ and $m_1=m_2$ during the gluing can be expressed as the vanishing of two gluing constraints
\be C_1 := u_1-u_2 = 0\qquad\text{and}\qquad C_2:= v_1-v_2 = 0\,. \ee
As functions on the product phase space $\CP_{(M,N)}:=\CP_{\pd M}\times \CP_{\pd N}$, they have trivial Poisson bracket
\be [C_1\,,C_2]_{\rm P.B.} = 0\,. \ee
This suggests that we could use $C_1$ and $C_2$ simultaneously as moment maps to perform a (holomorphic version of) symplectic reduction on $\CP_{(M,N)}$. A basic counting of coordinates shows that the (complex) dimension of the quotient will be $6-2\times 2=2$, exactly right for the phase space $\CP_{\pd P}$. Moreover, the coordinate functions $u_3$ and $v_3$ have trivial Poisson brackets with $C_1$ and $C_2$, so they are invariant under the flow of these moment maps and descend to be good coordinates on the quotient. Thus,
\be \CP_{\pd P} = \big( \CP_{\pd M}\times \CP_{\pd N}\big)\big/\!\!\big/(\C_{C_1}\times \C_{C_2})\,. \ee

To be a little more explicit, it is convenient to introduce canonical conjugates
\be \Gamma_1 := v_1 \qquad\text{and}\qquad \Gamma_2:=-u_2 \ee
to $C_1$ and $C_2$, respectively. These satisfy
\be [\Gamma_1,C_1]_{\rm P.B.}=[\Gamma_2,C_2]_{\rm P.B.}=\frac{i\hbar}{2}\,,\qquad [\Gamma_i,\Gamma_j]_{\rm P.B.}=[\Gamma_i,u_3]_{\rm P.B.}=[\Gamma_i,v_3]_{\rm P.B.}=0\,.\ee
Therefore, the $\Gamma_i$ are interpreted as coordinates along the flows generated by the respective moment maps $C_i$, and
\be \CP_{\pd P} = \left(\big(\CP_{\pd M}\times \CP_{\pd N}\big)\big|_{C_1=C_2=0}\right)\big/\raisebox{-.1cm}{\small$(\Gamma_1\sim\Gamma_1+t_1,\Gamma_2\sim \Gamma_2+t_2)$}\,. \ee

Still staying semi-classical, let us next consider the Lagrangian manifolds in the phase spaces $\CP_{\pd M}$ and $\CP_{\pd N}$ that describe semi-classical states. For $M$, which has a single torus boundary, the set of flat connections that extend from the boundary to the bulk is given by the standard A-polynomial $A_M(\ell_1,m_1)=0$. For $N$, which has two boundaries, the set of flat connections in the bulk is described by two equations $A_{N,1}(\ell_2,m_2,\ell_3,m_3)=A_{N,2}(\ell_2,m_2,\ell_3,m_3)=0$. These equations cut out a Lagrangian submanifold of $\CP_{\pd N}$, just as $A_M=0$ cuts out a Lagrangian submanifold of $\CP_M$.
 It is then easy to see that upon setting $m_1=m_2=m$ and $\ell_1=\ell_2=\ell$ and eliminating $m$ and $\ell$ from all three equations $A_M=A_{N,1}=A_{N,2}=0$ we should find the classical A-polynomial for $P$, $A_P(\ell_3,m_3)=0$. However, we could also describe this elimination a little differently and more suggestive of the symplectic reduction on phase spaces.

In order to pull Lagrangian submanifolds through symplectic reduction, let us start with a symplectic basis of coordinates $(v_1,u_1,-v_2,-u_2,v_3,u_3)$ on the product phase space $\CP_{(M,N)}$, and change coordinates to a new symplectic basis $(\Gamma_1,C_1,\Gamma_2,C_2,v_3,u_3)$, with
\bse \label{MNPoldnew}
\begin{align}
v_1 &= \Gamma_1\,,\\
u_1 &= C_1-\Gamma_2\,,\\
-v_2 &= C_2-\Gamma_1\,, \\
u_2 &= -\Gamma_2\,,\\
v_3 &= v_3\,. \\
u_3 &= u_3\,.
\end{align}
\ese
The A-polynomials for $M$ and $N$ cut out a product Lagrangian submanifold $\CL_{(M,N)}$ in $\CP_{(M,N)}$, described by
\bse \label{classidealMNP}
\begin{align}
 A_M(\ell_1,m_2) &= 0\,,\\
 A_{N,1}(\ell_2,m_2,\ell_3,m_3) &= 0\,, \\
 A_{N,2}(\ell_2,m_2,\ell_3,m_3) &= 0\,.
\end{align}
\ese
Then, the process of finding $A_P(\ell_3,m_3)$ consists of 1) 
using \eqref{MNPoldnew} to rewrite equations \eqref{classidealMNP} in terms of new variables $\ell_3$, $m_3$ and
\be \gamma_1:=e^{\Gamma_1}\,,\qquad c_1:=e^{C_1}\,,\qquad \gamma_2:=e^{\Gamma_2}\,,\qquad c_2:=e^{C_2}\,; \ee
2) eliminating all the $\gamma_i$ from the equations, so that one equation in $(c_1,c_2,\ell_3,m_3)$ remains; and 3) setting $c_1=c_2=1$ in this last equation.
 After eliminating $\gamma_1$ and $\gamma_2$, the one remaining equation in $(c_1,c_2,\ell_3,m_3)$ has trivial Poisson bracket with the $\Gamma_i$, so it descends to a well-defined function on the slice $C_1=C_2=0$, and that function is the A-polynomial for $P$.

Geometrically, we have projected $\CL_{(M,N)}$ perpendicular to flow lines and intersected it with the zero-locus of the moment maps.
We can also say this somewhat more algebraically. The equations \eqref{MNPoldnew} define an ideal in $\C[v_1^{\pm 1},u_1^{\pm 1},v_2^{\mp 1},u_2^{\pm 1},v_3^{\pm 1},u_3^{\pm 1}]$, which after changing to new symplectic variables is an ideal in the isomorphic ring $\C[\gamma_1^{\pm 1},c_1^{\pm 1},\gamma_2^{\pm 1},c_2^{\pm 1},\ell_3^{\pm 1},m_3^{\pm 1}]$. Eliminating $\gamma_1$ and $\gamma_2$ produces the intersection of this ideal with the subring $\C[c_1^{\pm 1},c_2^{\pm 1},\ell_3^{\pm 1},m_3^{\pm 1}]$, a so-called elimination ideal. In this case, the elimination ideal is generated by a single equation, and setting $c_1=c_2=1$ in this equation recovers $A_P(\ell_3,m_3)=0$.

Now, let us quantize. The phase spaces $\CP_{\pd M}$, $\CP_{\pd N}$, and $\CP_{\pd P}$ give rise to respective ``Hilbert'' spaces
\be \CH_{\pd M} \sim \{f(u_1)\}\,,\qquad \CH_{\pd N} \sim \{f(u_2,u_3)\}\,,\qquad \CH_{\pd P} \sim \{f(u_3)\}\,. \label{MNPH} \ee
For concreteness, we can suppose that $\CH_{\pd M}$, $\CH_{\pd N}$, and $\CH_{\pd P}$ consist of meromorphic functions that are square integrable on the real line.
We can also form a product space
\be \CH_{(M,N)} = \ol{\CH_{\pd M}\otimes\CH_{\pd N}} \sim \{f(u_1,u_2,u_3)\}\,.\ee
On any of these these spaces, operators $\hat u_i$ and $\hat v_i$ act as
\be \hat u_i\,f(...)=u_i f(...)\,,\qquad \hat v_i\, f(...) = \frac{\hbar}{2}\,\pd_{u_i}\,f(...)\,.\ee

We expect that the Chern-Simons partition functions $Z_M(u_1)$ and $Z_N(u_2,u_3)$ are annihilated by some quantized operators $\hat A_M(\hat\ell_1,\hat m_1;q)$ and a pair $\hat A_{N,i}(\hat \ell_2,\hat m_3,\hat \ell_3,\hat m_3;q)$, $i=1,2$, respectively:
\be \hat A_M\cdot Z_M = 0\,, \qquad \hat A_{N,1}\cdot Z_N=\hat A_{N,2}\cdot Z_N = 0\,. \ee
In the semi-classical symplectic reduction above, we began by creating a product phase space $\CP_{(M,N)}$ with a product Lagrangian submanifold. Here, it similarly makes sense to define a product wavefunction
\be \CZ_{(M,N)}(u_1,u_2,u_3) = Z_M(u_1)Z_N(u_2,u_3) \in \CH_{(M,N)}\,, \ee
which is annihilated by all three operators $\hat A_M$, $\hat A_{N,1}$, and $\hat A_{N,2}$. We will write this suggestively as
\bse \label{qidealMNP}
\begin{align}
 \hat A_M(\ell_1,m_2;q) &\simeq 0\,,\\
 \hat A_{N,1}(\hat \ell_2,\hat m_2,\hat \ell_3,\hat m_3;q) &\simeq 0\,, \\
 \hat A_{N,2}(\hat \ell_2,\hat m_2,\hat \ell_3,\hat m_3;q) &\simeq 0\,,
\end{align}
\ese
where "$\simeq0$'' means ``annihilates the wavefunction when acting on the left.'' Indeed, these three equations are the generators of an entire \emph{left} ideal of operators that annihilate $\CZ(u_1,u_2,u_3)$: we can add, subtract, and multiply by other operators on the \emph{left} while staying within the ideal. Being precise, this is a left ideal in the $q$-commutative ring $\C(q)[\hat \ell_1{}^{\pm 1},\hat m_1{}^{\pm 1},\hat \ell_2{}^{\mp 1},\hat m_2{}^{\pm 1},\hat \ell_3{}^{\pm 1},\hat m_3{}^{\pm 1}]$.

In order to perform the quantum gluing, we cannot simply set $\hat C_1 = \hat u_1-\hat u_2$ or $\hat C_2 = \hat v_1-\hat v_2$ to be zero in the full algebra of operators on $\CH_{(M,N)}$, because these elements are clearly not central. However, just as in the semi-classical case, we \emph{could} set $\hat C_1=\hat C_2=0$ in an operator equation that only involved generators that commute with $\hat C_1$ and $\hat C_2$. So, let us do this. In the algebra of linear, ``logarithmic'' operators on $\CH_{(M,N)}$, the only generators that do not commute with $\hat C_1$ and $\hat C_2$ are
\be \hat \Gamma_1 := \hat v_1\qquad\text{and}\qquad \hat \Gamma_2:=-\hat u_2\,. \ee
We can perform a canonical change of basis in the operator algebra by inverting these relations, \ie\ setting
\bse \label{MNPqoldnew}
\begin{align}
\hat v_1 &= \hat \Gamma_1\,,\\
\hat u_1 &= \hat C_1-\hat \Gamma_2\,,\\
-\hat v_2 &= \hat C_2-\hat \Gamma_1\,, \\
\hat u_2 &= -\hat \Gamma_2\,,\\
\hat v_3 &= \hat v_3\,,\\
\hat u_3 &= \hat u_3\,.
\end{align}
\ese
Exponentiating, we have
\be \hat \ell_1=\hat \gamma_1\,,\qquad \hat m_1= \hat c_1\hat \gamma_2{}^{-1}\,,\qquad \hat \ell_2=\hat c_2{}^{-1}\hat \gamma_1\,,\qquad \hat m_2=\hat \gamma_2{}^{-1}\,.\ee
Then, replacing $\hat\ell_i$ and $\hat m_i$ with the new exponentiated operators, equations \eqref{qidealMNP} define a left ideal in the isomorphic $q$-commutative ring $\C(q)[\hat \gamma_1{}^{\pm 1},\hat c_1{}^{\pm 1},\hat \gamma_2{}^{\pm 1},\hat c_2{}^{\pm 1},\hat \ell_3{}^{\pm 1},\hat m_3{}^{\pm 1}]$. By adding, subtracting, and multiplying on the left, we can eliminate $\hat\gamma_1$ and $\hat\gamma_2$ from the new equations \eqref{qidealMNP}, leaving (ideally) a single equation
\footnote{We should note that polynomial algebra and elimination of variables in a $q$-commutative ring work much the same way as their classical fully commutative cousins. We will say more about this in Sections \ref{sec:qglue} and \ref{sec:ex}.}
\be \hat A(\hat c_1,\hat c_2,\hat\ell_3,\hat m_3;q) \simeq 0\,. \label{defAMNP} \ee
More formally, \eqref{defAMNP} is the generator of the intersection of our left ideal with the subring $\C(q)[\hat c_1{}^{\pm 1},\hat c_2{}^{\pm 1},\hat \ell_3{}^{\pm 1},\hat m_3{}^{\pm 1}]$.

By construction, the product wavefunction $\CZ(u_1,u_2,u_3)$ is annihilated by \eqref{defAMNP},
\be \hat A(\hat c_1,\hat c_2,\hat\ell_3,\hat m_3;q)\,\CZ(C_1,C_2,u_3)=0 \label{MNPfinalA} \ee
To understand this equation a little better, though, we should perform the symplectic transformation \eqref{MNPqoldnew} on the ``Hilbert'' space $\CH_{(M,N)}$ as well as on the algebra of operators. This requires some version of a Fourier transform on $\CH_{(M,N)}$ to be defined. Our previous stipulation that wavefunctions be $L^2$ on the real line should be sufficient for this.
Switching to a representation of the operator algebra that consists of functions $\{f(C_1,C_2,u_3)\}$, with
\bse
\begin{align} \hat\Gamma_i f(C_1,C_2,u_3) &= \frac\hbar2 \pd_{C_i} f(C_1,C_2,u_3)\,,\quad &\hat C_i f(C_1,C_2,u_3) = C_i f(C_1,C_2,u_3)\,, \\
 \hat v_3 f(C_1,C_2,u_3) &= \frac\hbar2 \pd_{u_3} f(C_1,C_2,u_3)\,,\quad &\hat u_i f(C_1,C_2,u_3) = u_3 f(C_1,C_2,u_3)\,,
\end{align}
\ese
the product wavefunction $\CZ(u_1,u_2,u_3)$ formally becomes
\be \CZ(u_1,u_2,u_3) \;\mapsto\; \CZ(C_1,C_2,u_3)
 = \frac{1}{\sqrt{2\pi i\hbar}} \int du\, \CZ(u+C_1,u,u_3)\,e^{\frac{1}{\hbar}C_2 u}\,. \label{MNPwf}
\ee
This expression follows systematically from the Weil representation of the symplectic group \cite{Shale-rep, Weil-rep}, discussed further in Section \ref{sec:Weil}.

The actual wavefunction that we know we should obtain for the glued manifold $P$ is just the integral \eqref{MNPwf} with $C_1=C_2=0$,
\begin{align} Z_P(u_3) &= \CZ(C_1,C_2,u_3)\big|_{C_1,C_2=0} = \frac{1}{\sqrt{2\pi i\hbar}} \int du\,\CZ(u,u,u_3) \nno \\ &= \frac{1}{\sqrt{2\pi i\hbar}}\int du\, Z_M(u)Z_N(u,u_3)\,. \end{align}
However, because the operator $\hat A(\hat c_1,\hat c_2,\hat\ell_3,\hat m_3;q)$ in \eqref{MNPfinalA} is a function of generators that all \emph{commute} with $\hat c_1$ and $\hat c_2$, we can also consistently set $C_1=C_2=0$ in \eqref{MNPfinalA} to find that
\be \hat A(1,1,\hat\ell_3,\hat m_3;q)\, Z_P(u_3) = 0\,. \ee
This leads us to the conclusion that the ``glued'' operator $\hat A_P$ must be
\be \boxed{\hat A_P(\hat \ell_3,\hat m_3;q) = \hat A(\hat c_1,\hat c_2,\hat\ell_3,\hat m_3;q)\,\big|_{\hat C_1,\hat C_2=0}}\,. \ee
By construction, the classical $q\to 1$ limit of this final operator is simply the classical A-polynomial $A_P(\ell_3,m_3)=0$.

\subsection{The toy is real}
\label{sec:realtoy}

The above example contains all the features of a generic gluing in any TQFT --- particularly in any physical TQFT with honest Hilbert spaces. It also contains all the ingredients that we will need to glue tetrahedra in our analytically continued context. Let us therefore summarize schematically but generally what should happen when two oriented manifolds $M$ and $N$ are to be glued together along a common boundary component $\Sigma$,
\be \Sigma \subset \pd M,\,\qquad \Sigma \subset -\pd N\,,\ee
to form an oriented manifold
\be P = M\cup_{\Sigma} N\,,\ee
possibly with $\pd P\neq 0$.

A TQFT typically assigns phase spaces $\CP_{\pd M}$ and $\CP_{\pd N}$ to the full boundaries of $M$ and $N$, respectively. These are symplectic manifolds. Semiclassical states for the TQFT on $M$ and $N$ are described by Lagrangian submanifolds
\be \CL_M\subset \CP_{\pd M}\,,\qquad \CL_N\subset \CP_{\pd N}\,. \ee
The equations that cut out $\CL_M$ and $\CL_N$ can be thought of as generating ideals $\CI_M$ and $\CI_N$ in the algebras of functions on $\CP_{\pd M}$ and $\CP_{\pd N}$, respectively.

Upon quantization, the boundary phase spaces become Hilbert spaces $\CH_{\pd M}$ and $\CH_{\pd N}$, and the complete quantum wavefunctions or partition functions of $M$ and $N$ are elements of these Hilbert spaces,
\be Z_M \in \CH_{\pd M}\,,\qquad Z_N \in \CH_{\pd N}\,.\ee
Each wavefunction is annihilated by the quantization of the functions that define the semi-classical Lagrangians $\CL_M$ and $\CL_N$. Thus, corresponding to $\CI_M$ and $\CI_M$, there are \emph{left} ideals $\hat\CI_M$ and $\hat\CI_N$ in the algebras of operators on $\CH_{\pd M}$ and $\CH_{\pd N}$ such that
\be \hat\CI_M\cdot Z_M = 0\,,\qquad \hat \CI_N\cdot Z_N = 0\,.\ee

In order to glue together $M$ and $N$ to form $P$ semi-classically and quantum mechanically, one should:

\begin{enumerate}

\item Semiclassically, form the product phase space $\CP_{\pd M\sqcup \pd N} = \CP_{\pd M}\times \CP_{\pd N}$.

\item Select $g$ functions $\{C_j\}_{j=1}^g$ on $\CP_{\pd M\sqcup \pd N}$ to be gluing constraints, so that setting $C_j=0$ for all $j$ classically identifies the boundary conditions on $\Sigma\subset \pd M$ with the corresponding boundary $\Sigma\subset -\pd N$. The number of constraints is
\be g = \dim\CP_{\Sigma}\,,\ee
where $\CP_{\Sigma}$ is the semi-classical phase space of $\Sigma$, a subfactor of both $\CP_{\pd M}$ and $\CP_{\pd N}$. The gluing functions should have trivial Poisson brackets among themselves.

\item Construct the phase space $\CP_{\pd P}$ as a symplectic quotient, using the $g$ $C_j$'s as moment maps. Schematically, if $G_j$ is the group action generated by the vector field $\omega^{-1}dC_j$, then
\be \CP_{\pd P} = \CP_{\pd M\sqcup \pd N}\big/\!\!\big/({\textstyle\prod_j G_j}) = \CP_{\pd M\sqcup \pd N}\big/({\textstyle \prod_j G_j})\,\big|_{C_j=0}\,. \ee
Note that $\dim \CP_{\pd P} = \dim \CP_{\pd M}+\dim \CP_{\pd N}-2\dim\CP_{\Sigma}$.

\item \label{item:classL} Form the product Lagrangian $\CL_{(M,N)} = \CL_M\times \CL_N \subset \CP_{\pd M\sqcup \pd N}$. Then construct the Lagrangian $\CL_P\in \CP_{\pd P}$ by first projecting $\CL_{(M,N)}$ onto the quotient $\CP_{\pd M\sqcup \pd N}\big/({\textstyle \prod_j G_j})$, then intersecting with $C_j=0$\; $\forall\,j$.

\item[\ref{item:classL}a.] Algebraically, the ideal $\CI_P$ corresponding to $\CL_P$ in the algebra of functions on $\CP_{\pd P}$ is formed by starting with $\CI_M\cup \CI_N$ (as an ideal in the algebra of functions on $\CP_{\pd M\sqcup \pd N}$), removing all elements that have nontrivial Poisson bracket with the $C_j$ (\ie\ forming an elimination ideal), and setting $C_j=0$.

\item Form the product Hilbert space $\CH_{\pd M\sqcup \pd N} = \ol{\CH_{\pd M}\otimes \CH_{\pd N}}$. This is a quantization of $\CP_{\pd M\sqcup \pd N}$, with some polarization induced from the constructions of $\CH_{\pd M}$ and $\CH_{\pd N}$. Recall that in geometric quantization a polarization consists of $\dim \CP_{\pd M\sqcup \pd N}/2$ commuting vector fields. To form the glued Hilbert space $\CH_{\pd P}$, first \emph{change} the polarization on $\CP_{\pd M\sqcup \pd N}$ so that $g$ of the commuting vector fields are the moment map vector fields $\omega^{-1}dC_j$, leading to an isomorphic Hilbert space $\widetilde\CH_{\pd M\sqcup \pd N}\simeq \CH_{\pd M\sqcup \pd N}$. In $\widetilde\CH_{\pd M\sqcup \pd N}$, wavefunctions depend explicitly on $C_j$ as ``coordinates,'' so it makes sense to set
\be \CH_{\pd P} = \widetilde\CH_{\pd M\sqcup \pd N}\big|_{C_j=0}\,.\ee

The change of polarization here is typically implemented via some version the Weil representation of the symplectic group.

\item \label{item:wf} It follows that the wavefunction of $P$ is
\be Z_P = \widetilde{Z_M\times Z_N}\,|_{C_j=0}\,, \ee
where $f\mapsto \tilde{f}$ is the preceding isomorphism of Hilbert spaces.

\item \label{item:qelim} Finally, construct the operator(s) that annihilate $Z_P$ by using the quantum version of Step \ref{item:classL} above.

\item[\ref{item:qelim}a.] Algebraically, form the union left ideal $\hat \CI_{(M,N)}=\hat \CI_M\times \hat \CI_N$ in the algebra of operators on $\CH_{\pd M\sqcup \pd N}$, or (equivalently) on $\widetilde\CH_{\pd M\sqcup \pd N}$. Remove all elements of $\hat\CI_{(M,N)}$ that do not commute with the quantized gluing constraints $\hat C_j$ to obtain an elimination ideal $\hat \CJ_P$. Set $\hat C_j=0$ in $\hat \CJ_P$ to find the ideal of operators $\hat\CI_P$ that annihilate $Z_P$,
\be \hat \CI_P = \left(\hat \CI_{(M,N)}\cap \{\text{operators commuting with $\hat C_j$'s}\}\right)\cap (\hat C_j=0)\, \label{qIideal} \ee
\be \imp \qquad \hat \CI_P\cdot Z_P=0\,. \nno \ee
While the order of operations done in finding the semi-classical Lagrangian $\CL_P$ was not important, the order of operations in \eqref{qIideal} is critical.

\end{enumerate}

Note that the construction here makes sense even when $P$ has no boundary, and $Z_P$ is just a number. Then the ideal $\hat{\CI}_P$ is empty, and Step \ref{item:wf} simply reproduces the usual TQFT inner product $Z_P=\langle Z_M\,|\,Z_N\rangle_{\CH_\Sigma}$. \\

Our goal in the remaining sections is to apply the above gluing scheme to a three-manifold $P$ that is a knot complement with an ideal triangulation. After properly understanding the phase space, Hilbert space, and wavefunction of individual ideal tetrahedra, we will find that following the above steps yields both the quantum $\hat{A}$-polynomial of $P$ and the wavefunction --- a holomorphic block --- that it annihilates in an extremely straightforward manner. (In the rest of the paper, the glued knot complement is usually called `$M$' rather than `$P$.')

\section{Classical triangulations}
\label{sec:hyp}

As described in the introduction, our approach to finding both the Chern-Simons partition function on a knot complement $M$ and the operator $\hat A$ that annihilates it relies on cutting $M$ into ideal tetrahedra. In Section \ref{sec:opglue}, we learned how to systematically obtain the annihilating operator and wavefunction on a glued manifold in terms of the operators and wavefunctions of pieces. In order to apply this machinery to tetrahedra, however, we must first understand how Chern-Simons theory on tetrahedra behaves. In the current section, we therefore begin by studying ideal triangulations (semi)classically.

In the beginning, we will simply review a well-known mathematical procedure of constructing a flat $SL(2,\C)$ connection on a three-manifold in terms of flat $SL(2,\C)$ structures on tetrahedra. Since $SL(2,\C)$ is the double cover of the isometry group $PSL(2,\C)$ of hyperbolic three-space, one can describe $SL(2,\C)$ structures much more easily and intuitively by using hyperbolic geometry. We follow standard references, such as the classic notes of W. Thurston \cite{thurston-1980} and the work of Neumann and Zagier \cite{NZ}, and well as (\eg) the more recent \cite{Neumann-combinatorics, neumann-2004, Zickert-rep}. However, we will try to recast the classic constructions in just the right language to make the eventual quantization of Chern-Simons theory on triangulations (Section \ref{sec:quant}) both easy and natural.

\subsection{Ideal triangulation}
\label{sec:triang}

\begin{figure}[htb]
\centering
\hspace{.2in}\includegraphics[width=5in]{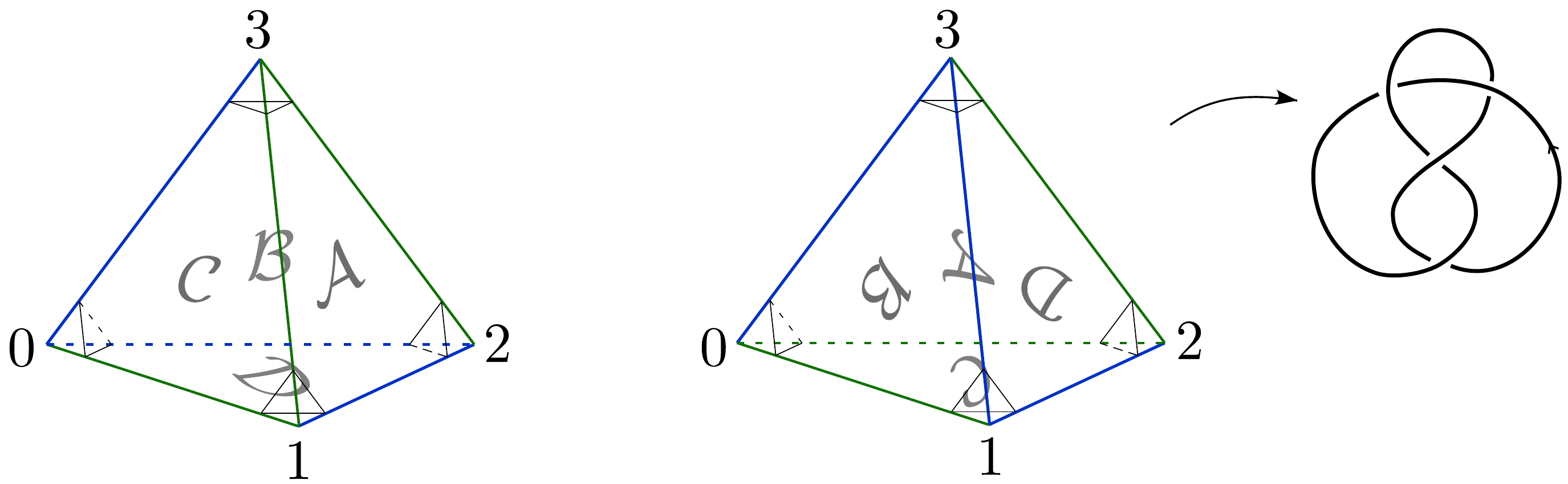}
\caption{Triangulation of the $\mb{4_1}$ knot complement. The gluing of faces is indicated by calligraphic letters.}
\label{fig:top41tet}
\end{figure}

An ideal topological tetrahedron is an ordinary tetrahedron with neighborhoods of its vertices removed. Two such tetrahedra are shown in Figure \ref{fig:top41tet}. It is possible to glue ideal tetrahedra together to form any knot (or link) complement $M=\ol{M}\bs K$, for $\ol{M}$ oriented and compact, in such a way that the small triangular pieces of boundary around the vertices join together to form the torus boundary of $M$ (\cf\ \cite{thurston-1980}). This is called an ideal triangulation of $M$.
The edges and faces of tetrahedra in this triangulation are part of $M$, so the gluing must be continuous there. The vertices, however, do not belong to $M$, and can be thought of as lying instead on the excised knot $K\subset\ol{M}$. Therefore, the gluing need not be (and generally is not) continuous at the vertices themselves.

\begin{figure}[htb]
\centering
\includegraphics[width=4in]{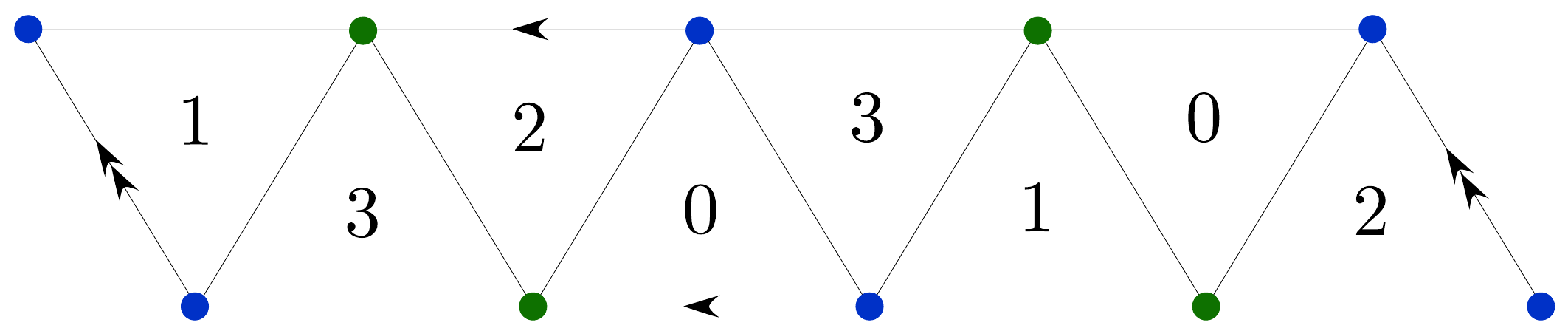}
\caption{Developing map for the boundary of the $\mb{4_1}$ knot complement. The four triangles $(\Delta)$ on the bottom come from the vertices of the tetrahedron on the left of Figure \protect\ref{fig:top41tet}, and the triangles $(\nabla)$ on top come from the tetrahedron on the right. The torus is being viewed from outside of $M$ (from inside of the thickened knot)}
\label{fig:top41dev}
\end{figure}

As an example, consider the complement of the figure-eight knot $\mb{4_1}$ in the three-sphere, $M=S^3\bs \mb{4_1}$. This knot complement can be built from just two tetrahedra, as shown in Figure \ref{fig:top41tet} \cite{thurston-1980}. If we number the vertices of the two tetrahedra as in this Figure, then the small triangular boundaries around the eight tetrahedron vertices glue together to form the torus of Figure \ref{fig:top41dev}. Such a drawing of the triangulated boundary torus is called a developing map. Notice that the final, glued triangulation of $M$ has only two distinct edges (blue and green in Figure \ref{fig:top41tet}), which each intersect the boundary torus twice.

\begin{figure}[htb]
\centering
\includegraphics[width=4.5in]{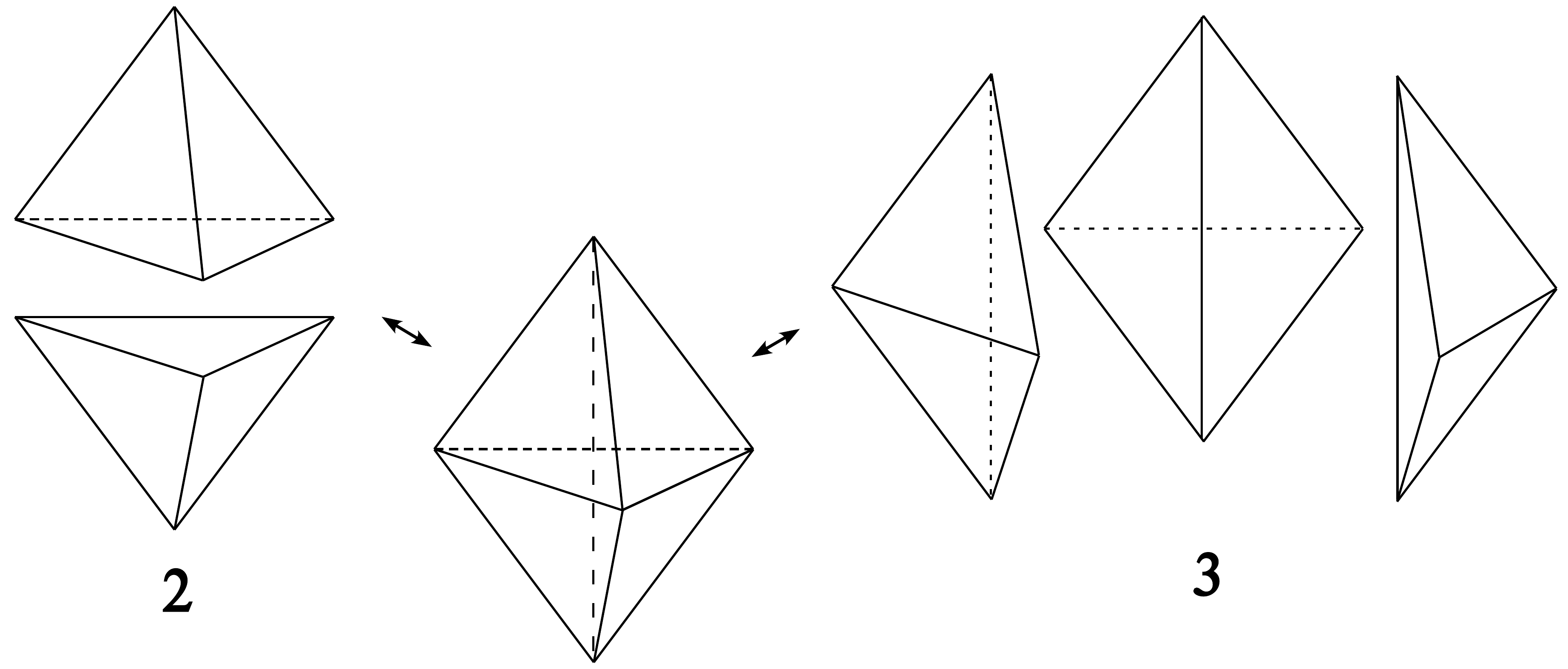}
\caption{The 2-3 Pachner move.}
\label{fig:Pachner}
\end{figure}

Any two ideal triangulations of a knot complement are related by a sequence of so-called 2-3 Pachner moves, illustrated in Figure \ref{fig:Pachner}. A non-ideal simplicial triangulation (which includes its vertices) would admit a ``1-4'' move as well, which places a vertex at the center of a single tetrahedron to subdivide it into four new ones. However, a 1-4 move in an \emph{ideal} triangulation would create or destroy spherical boundary components (around newly created vertices), thereby changing the glued manifold, so it cannot be allowed. For ideal triangulations, the 2-3 moves are sufficient.

\begin{figure}[h]
\centering
\includegraphics[width=3.7in]{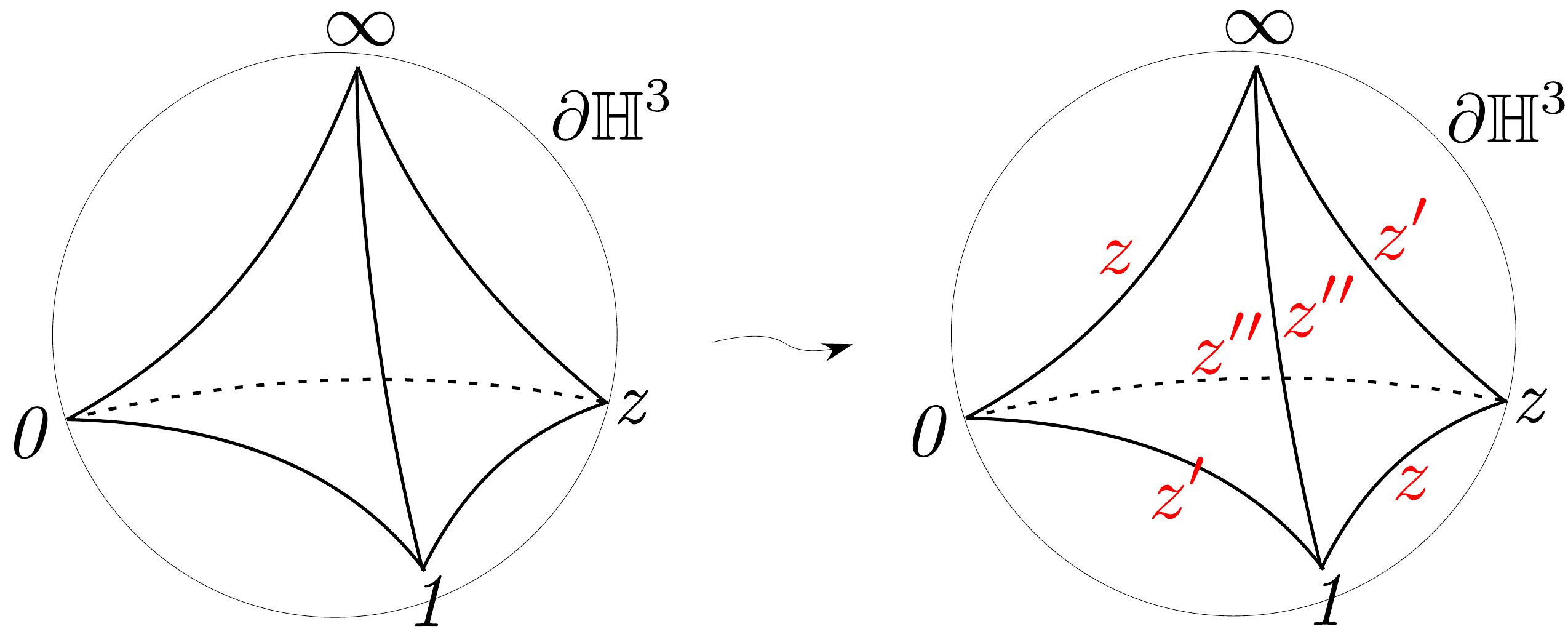}
\caption{An ideal hyperbolic tetrahedron.}
\label{fig:hyptet}
\end{figure}

Now, let us put hyperbolic structures on ideal tetrahedra. Recall that hyperbolic three-space $\H^3$ can be visualized as the upper half-three-space, or, conformally, as the interior of a three-ball. The boundary of $\H^3$ is a two-sphere, thought of as the Riemann sphere, or $\C\cup\{\infty\}$. By definition, an \emph{ideal hyperbolic tetrahedron} is a tetrahedron in $\H^3$ all of whose vertices lie on $\pd\H^3$ and all of whose faces are geodesic surfaces. An ideal hyperbolic tetrahedron is illustrated in Figure \ref{fig:hyptet}.

The positions of the vertices of this tetrahedron on $\pd\H^3=\C\cup\{\infty\}$ fully determine its geometric structure. In fact they overdetermine it: the isometry group $PSL(2,\C)$ acts as the M\"obius group (\ie\ by fractional linear transformations) on the boundary, and allows any three points to be fixed. Thus, the only independent parameter of the hyperbolic structure on an ideal tetrahedron is a single complex cross ratio, the so called \emph{shape parameter} of the tetrahedron.

If we place three vertices of the tetrahedron at $0$, $1$ and $\infty$ as in Figure \ref{fig:Pachner}, the fourth vertex lies at the shape parameter $z$. It then turns out that the dihedral angles on two opposing edges of the tetrahedron are actually equal to $\arg(z)$, and we label these edges by `$z$' as indicated. However, there exist two other pairs of edges, and it is only natural to associate to them their own parameters $z'$ and $z''$, so that the dihedral angle around any edge is equal to the argument of its shape parameter. It is easy to see that $z'$ and $z''$ are just conjugate cross ratios, given by
\be z' = \frac{1}{1-z}\,,\qquad z''=1-\frac{1}{z}\,. \label{zpzpp} \ee

The three shape parameters or edge parameters $z,\,z'$, and $z''$ should really be treated symmetrically. From \eqref{zpzpp}, we see that they must satisfy two relations (giving one independent parameter in the end). The two relations, however, are \emph{not} on equal footing. First, the product of shape parameters around any vertex of the tetrahedron is
\be zz'z''=-1\,. \label{triple} \ee
This ensures, in particular, that the sum of angles%
\footnote{In a hyperbolic triangulation, the vertices are truncated by geodesic horospheres, so the boundary triangles are Euclidean. Note that only Euclidean triangles could line up as in Figure \ref{fig:top41dev} to form a torus.} %
in the little boundary triangle that is formed by truncating any vertex is $\pi$; hence we call \eqref{triple} the \emph{vertex equation}. The second relation between shape parameters can be written in any one of the three equivalent forms
\bse \label{interior}
\begin{align} z+(z')^{-1}-1&=0\,, \label{interiora} \\
z'+(z'')^{-1}-1&=0\,, \\
z''+z^{-1}-1 &=0\,.
\end{align}
\ese
Roughly, these equations contain the requirement that a hyperbolic structure is consistent through the interior of a tetrahedron.

In terms of $SL(2,\C)$ or $PSL(2,\C)$ structures, the interpretation of equations \eqref{triple} and \eqref{interior} (and the distinction between them) becomes much clearer. The shape parameters $z$, $z'$, and $z''$ can actually be thought of as squared partial holonomy eigenvalues along small bits path running from one face to another (through a dihedral angle) around an edge. Their product on any closed path is an honest gauge-invariant holonomy eigenvalue. The vertex equation, the fact that the product of shape parameters at any vertex is $-1$, is simply the condition that a flat $PSL(2,\C)$ connection exists on the \emph{boundary} of an ideal tetrahedron. On the other hand, equations \eqref{interior} are precisely the conditions that a flat connection from the boundary extends through the interior.

One way to justify this interpretation of equations \eqref{triple} and \eqref{interior} is to think of the boundary of a tetrahedron $\Delta$ as a triangulated, four-punctured sphere $S^2_4$. The moduli space of flat $PSL(2,\C)$ structures on $S^2_4$ is a natural complexification of its Teichm\"uller space. Moreover, our edge parameters $z,\,z',\,z''$ are nothing but complexifications of Checkov-Fock coordinates \cite{Fock-Teich, FockChekhov} (\emph{a.k.a.} Thurston's shear coordinates) on this triangulated surface --- with the restriction that holonomy eigenvalues at each puncture equal $-1$.%
\footnote{We would also like to thank R. Kashaev for extremely enlightening discussions regarding the connection between 3d hyperbolic geometry and 2d moduli spaces.}${}^{,}$%
\footnote{For interesting and possibly related recent applications of shear coordinates in other areas of physics, see \cite{Teschner-TeichMod} and \cite{GMN-II}.} %
A standard counting argument immediately shows that the dimension of Teichm\"uller space, equal to the expected complex dimension of our phase space, is (\# edges $-$ \# punctures) $= 6-4=2$. A little further thought leads to the conclusion that this phase space is indeed $\CP_{\pd \Delta}=\{(z,z',z'')\in (\C\bs \{0,1,\infty\})^3\;|\;zz'z''=-1\}$.

Equations \eqref{interior} also have an interpretation in terms of Teichm\"uller theory. Namely, they are related to a ``diagonal flip'' transformation that pushes a $PSL(2,\R)$ structure from one hemisphere of $\pd\Delta=S^2_4$ through to the other. Hence our claim that equations \eqref{interior} are precisely the requirements that a complexified flat $PSL(2,\C)$ connection extends through the bulk of $\Delta$.

One great advantage of using two-dimensional shear coordinates is that they automatically come with a representation of the Weil-Petersson symplectic form, which is precisely the symplectic structure induced by Chern-Simons theory. We find that the classical phase space for Chern-Simons theory on a tetrahedron, $\CP_{\pd \Delta}=\{(z,z',z'')\in (\C\bs \{0,1,\infty\})^3\;|\;zz'z''=-1\}$, has the symplectic form
\be \omega = (i\hbar)^{-1}\frac{dz}{z}\wedge \frac{dz'}{z'}\,. \ee
(This is a complexification of the Weil-Petersson form on Teichm\"uller space, written in shear coordinates \cite{Fock-Teich}.)
Better still, we can lift to linear, logarithmic coordinates $Z,\,Z',\,Z''$ such that
\be z=e^Z\,,\qquad z'=e^{Z'}\,,\qquad z''=e^{Z''}\,. \label{cllog} \ee
As discussed in Section \ref{sec:log}, holomorphic blocks will explicitly depend on these lifted coordinates.
Then
\be \boxed{\CP_{\pd\Delta} := \{(Z,Z',Z'')\in (\C\bs\,2\pi i\Z)^3\;|\; Z+Z'+Z''=i\pi \}}\,, \label{defPD} \ee
with
\be (i\hbar)\,\omega_\Delta = dZ\wedge dZ' = dZ'\wedge dZ''=dZ''\wedge dZ\,. \ee
Equations \eqref{interior} define a Lagrangian submanifold
\be \boxed{\CL_\Delta := \{z+z'{}^{-1}-1=0\} \quad\subset\quad \CP_{\pd\Delta}}\,\ee
that parametrizes the set of ``classical solutions in the bulk'' of $\Delta$.

In defining $\CP_{\pd \Delta}$ in \eqref{defPD}, we have purposely excluded the points $z,z',z''=1$, as well as $z,z',z''\in\{0,\infty\}$. It is clear from Figure \ref{fig:hyptet} that these values lead to degenerate tetrahedra, whose hyperbolic volumes are ill-defined. In terms of flat connections, the values of the classical Chern-Simons action would become ill-defined.

As we have defined them, both the phase space $\CP_{\pd\Delta}$ and Lagrangian $\CL_{\Delta}$ are completely invariant under cyclic permutations of the shape parameters,
\be z\mapsto z' \mapsto z'' \mapsto z\,. \label{zcyclic} \ee
The cyclic order \eqref{zcyclic} is determined by the orientation of a hyperbolic tetrahedron, and it will be important for us to give all tetrahedra in the triangulation of an oriented manifold $M$ the common orientation induced from that of $M$. Then $z$, $z'$, and $z''$ are always assigned to edges in the order appearing in Figure \ref{fig:hyptet}.

\subsection{Gluing, holonomies, and character varieties}
\label{sec:glue}

When gluing ideal hyperbolic tetrahedra together to form a three-manifold $M$, extra conditions must be imposed to ensure that the hyperbolic structures of different tetrahedra match up globally. These are the equivalent of the TQFT gluing conditions discussed in Section \ref{sec:opglue}. They require that the total angle circling around each distinct edge in the triangulation of $M$ is $2\pi$ and that the hyperbolic ``torsion'' around the edge vanishes. Equivalently, in terms of flat connections, they simply require that the $PSL(2,\C)$ holonomy circling around any edge in $M$ be the identity, which must be the case since this holonomy loop is contractible.

\begin{figure}[htb]
\centering
\includegraphics[width=3.5in]{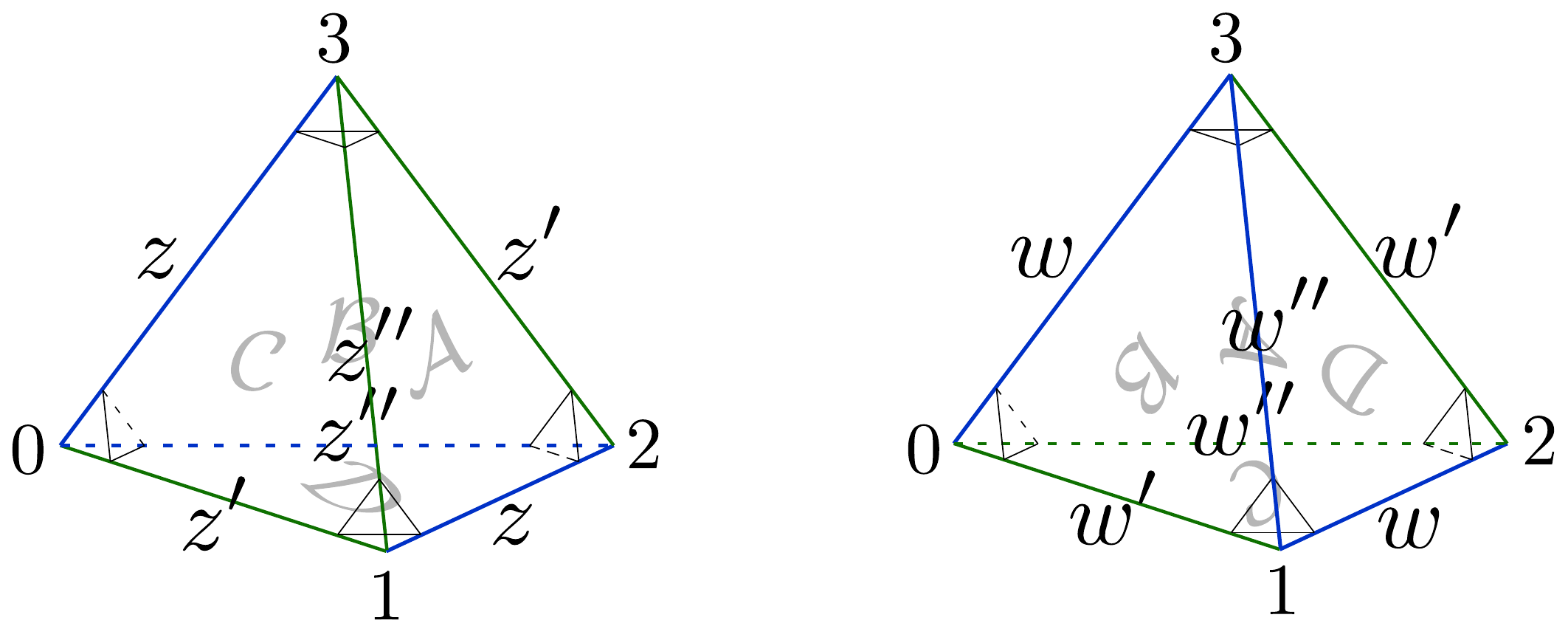}
\caption{Hyperbolic triangulation of the $\mb{4_1}$ knot complement.}
\label{fig:hyp41tet}
\end{figure}

To translate this to equations, suppose that an oriented knot complement $M$ is composed from $N$ tetrahedra $\Delta_i$,\, $i=1,...,N$. Each tetrahedron initially has its own independent set of shape parameters $(z_i,z_i',z_i'')$ with $z_iz_i'z_i''=-1$.
 Computing the Euler character of the triangulation quickly shows that there must be exactly $N$ distinct edges in the triangulation. (In the $\mb{4_1}$ knot example of Figure \ref{fig:hyp41tet}, the two distinct edges were colored green and blue.) Then, at the $j^{\rm th}$ edge in $M$, the square of the $PSL(2,\C)$ holonomy eigenvalue --- or the exponential of the complexified metric quantity [torsion + $i$ angle] --- is given by the product of all shape parameters that meet this edge.
 
To be precise, we can define $\epsilon(i,j)$ to be the number of times (0, 1, or 2) that an edge with edge parameter $z_i$ in tetrahedron $\Delta_i$ is identified with edge $j$ in $M$. Similarly, define $\epsilon(i,j)'$ and $\epsilon(i,j)''$ to be the number of times $z_i'$ and $z_i''$ meet $j$. Then the gluing constraint at edge $j$ is that
\be c_j := \prod_{i=1}^N z_i^{\epsilon(i,j)}(z_i')^{\epsilon(i,j)'}(z_i'')^{\epsilon(i,j)''}\ee
must equal $1$.

\begin{figure}[htb]
\centering
\includegraphics[width=5in]{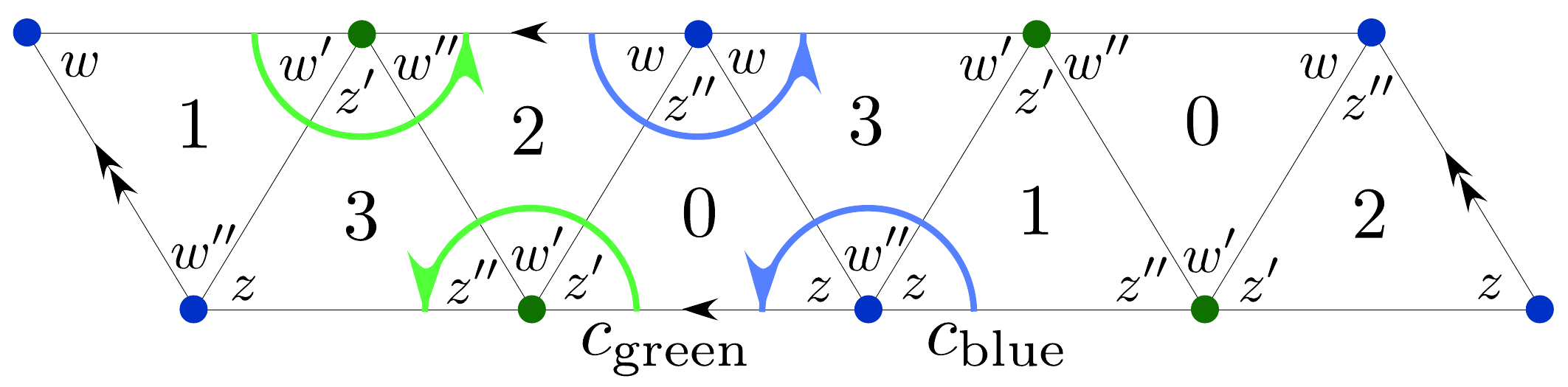}
\caption{Hyperbolic developing map of the $\mb{4_1}$ knot complement.}
\label{fig:hyp41dev}
\end{figure}

As an example, consider the ideal triangulation of the figure-eight knot complement. Let us assign shape parameters $(z,z',z'')$ and $(w,w',w'')$ to the two tetrahedra in its triangulation, as in Figure \ref{fig:hyp41tet}. Every dihedral angle of these two tetrahedra appears twice in the developing map, since each edge intersects the boundary torus twice, so we can also label angles in the developing map with $z$'s and $w$'s as in Figure \ref{fig:hyp41dev}. From the developing map, it is clear that the products at the two edges are
\bse \label{cbgcl}
\begin{align} c_{\rm blue} &= {z}^2z''{w}^2w''\,, \label{cblue} \\
c_{\rm green} &= {z'}^2z''{w'}^2w''\,. \label{cgreen}
\end{align}
\ese

Upon using vertex equations $zz'z''=-1$ and $ww'w''=-1$, the two constraints $c_{\rm blue}=1$ and $c_{\rm green}=1$ become equivalent. 
In general, there is a single constraint among the $N$ gluing functions, coming from the fact that every edge parameter in a tetrahedron meets two edges, \ie\ $\sum_{j=1}^N \epsilon(i,j)=2$. Then the constraint is
\be \prod_{j=1}^N c_j = \prod_{i=1}^N (z_iz_i'z_i'')^2 \;\overset{vx\,eqs}{=}\; 1\,.\ee
In the case of the figure-eight knot, we see that $c_{\rm blue}\,c_{\rm green} = (zz'z''ww'w'')^2\;\overset{vx\,eqs}{=} \;1$.

We can also lift the gluing constraints to logarithmic coordinates \eqref{cllog}. They take the form of $N$ \emph{linear} functions 
\be C_j := \sum_{i=1}^N \left( \epsilon(i,j)Z_i+\epsilon(i,j)'Z_i'+\epsilon(i,j)''Z_i''\right)\,,\qquad j=1,...,N\,, \label{Ccllog} \ee
which must classically satisfy
\be C_j = 2\pi i\,,\qquad j=1,...,N\,. \label{clCcons} \ee
This certainly implies that
\be c_j := e^{C_j} = 1\,. \ee
Notice that if we view the $C_j$ as functions on the product phase space
\be \CP_{(M,\Delta)} := \CP_{\Delta_1}\times\cdots\times \CP_{\Delta_N}\,, \label{PMDcl} \ee
which includes the vertex equations, imposing conditions \eqref{clCcons} is compatible with the constraint
\be \sum_{j=1}^N C_j = 2\sum_{i=1}^N (Z_i+Z_i'+Z_i'') = 2N\pi i\,. \ee

\begin{figure}[htb]
\centering
\includegraphics[width=5in]{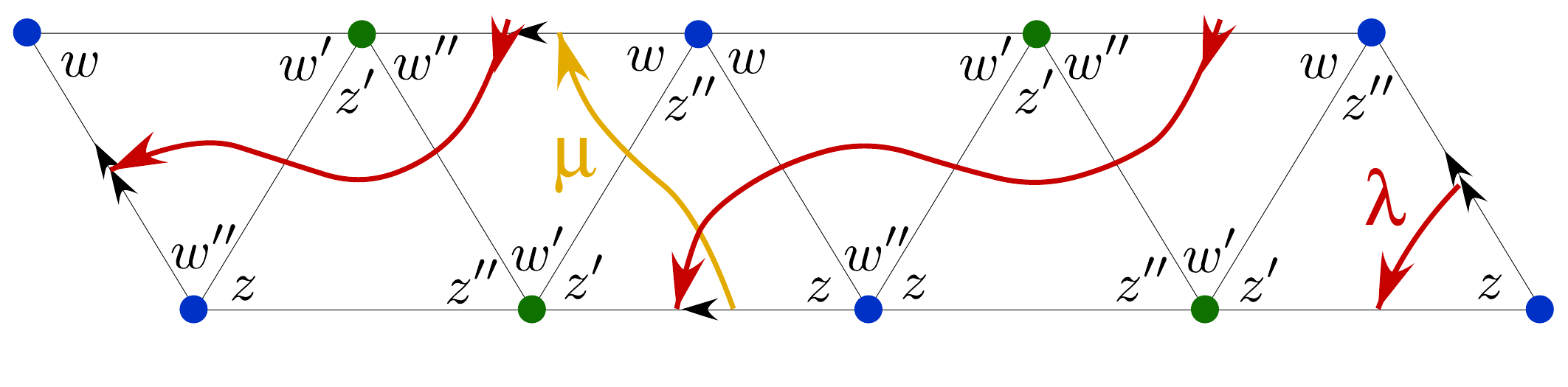}
\caption{Meridian and longitude paths for the $\mb{4_1}$ knot complement.}
\label{fig:hyp41devLM}
\end{figure}

Of course, there are two more important holonomy eigenvalues that we would like to compute in terms of the shape parameters: the longitude and meridian of the boundary torus. The longitude and meridian paths can be drawn on the developing map, as in Figure \ref{fig:hyp41devLM}. The rule for computing the squared eigenvalues $\ell^2$ or $m^2$ is to multiply by shape parameters that are encircled clockwise and to divide by shape parameters that are encircled counterclockwise \cite{NZ, Neumann-combinatorics}.%
\footnote{Of course ``multiplication'' and ``division'' are relative, since holonomy eigenvalues are only well defined up to the final Weyl group action $(\ell,\,m)\mapsto (\ell^{-1},m^{-1})$.} %
Alternatively, in lifted logarithmic variables $U$ and $V$ such that
\be \boxed{m^2 = e^U\,,\qquad \ell^2=e^V}\,, \ee
one just adds or subtracts. Thus, for the figure-eight knot we find
\bse \label{UV41}
\begin{align}
U&= Z'-W\,,\\
V&= Z-W''-Z'+W''+Z-W''-Z'+W'=2Z-2Z'\,,
\end{align}
\ese
which implies
\bse \label{ML41}
\begin{align} m^2 = e^U= z'w^{-1} \label{m41}  \\
 \ell^2 =e^V = z{w''}^{-1}{z'}^{-1}w''z{w''}^{-1}{z'}^{-1}w'' = z^2{z'}^{-2}\,.
\end{align}
\ese

Now, for any oriented, triangulated knot complement $M$ the $N-1$ gluing constraints $c_j=1$, the $N$ vertex equations $z_iz_i'z_i''=-1$, and the $N$ Lagrangian equations $z_i+z_i'{}^{-1}-1=0$ together describe a one-complex-dimensional subvariety of $(\C^*\bs\{1\})^{3N} = \{(z_i,z_i',z_i'')\}$. Up to some small caveats that we discuss in Section \ref{sec:classrem}, this is, abstractly, the $PSL(2,\C)$ character variety of $M$. In the case of the figure-eight knot, we can easily eliminate variables to obtain a single equation
\be z(1-z)w(1-w)-1=0 \qquad\subset\quad\C^*\times \C^*\,. \label{XP41}\ee
In order to lift from $PSL(2,\C)$ to an $SL(2,\C)$ character variety, one should in general take appropriate square roots of all the shape parameters in these equations. For the figure-eight knot, the $SL(2,\C)$ character variety $\CX$ is then described abstractly by a component of
\be z^2(1-z^2)w^2(1-w^2)-1=0  \,. \label{X41}\ee

Since we also have at our disposal the concrete relation between the meridian and longitude holonomies \eqref{ML41} and shape parameters, we can combine them with \eqref{XP41} or its lifted version to obtain A-polynomials (\cf\ \eqref{defX} vs. \eqref{AvarCS} in Section \ref{sec:CS}). In general, to obtain the lifted $SL(2,\C)$ A-polynomial of a knot complement in $S^3$, the only square root one ever needs to take is that of the $\ell^2$ equation. In the case of the figure-eight knot, the proper square root is
\be \ell = -zz'{}^{-1}\,, \label{Lsq41} \ee
which, combined with \eqref{m41} and \eqref{XP41} immediately leads to (\cf\ \cite{Champ-hypA, Culler-Apolys})
\be A_{\mb 4_1}(\ell,m) = m^4\ell^2-(1-m^2-2m^4-m^6+m^8)\ell+m^4 =0\qquad\subset\quad\C^*\times \C^*\,. \label{A41cl} \ee
Of course, this also descends to a variety in the Weyl-group quotient $(\C^*\times \C^*)/\Z_2$.

For knot complements in $S^3$, or in any homology sphere $\ol{M}$, the A-polynomial is always a polynomial in $m^2$ rather than just $m$ \cite{cooper-1994}. (A physical explanation of this fact appears in Section 4.2.7 of \cite{Wit-anal}.) This is precisely why taking a square root of $m^2$ is unnecessary. We will often emphasize the dependence on $m^2$ and write $A(\ell,m^2)$ rather than $A(\ell,m)$.

\subsubsection{The A-polynomial, symplectically}
\label{sec:hypsemicl}

Above, we reviewed the classical construction of the A-polynomial from hyperbolic triangulation data. In order to truly understand its quantization, however, we must view the A-polynomial not just as a variety in $(\C^*\times \C^*)/\Z_2$ but as a Lagrangian submanifold in the phase space of the torus, obtained by symplectic reduction. In other words, we should recast the classical construction of the A-polynomial in the semi-classical spirit of Section \ref{sec:opglue}.

To begin, for any given triangulation $(M,\{\Delta_i\}_{i=1}^N)$, we form a product phase space $\CP_{(M,\Delta)}$ as in \eqref{PMDcl}. This is an affine linear symplectic space, with product symplectic structure
\be \omega_{(M,\Delta)} = (i\hbar)^{-1}\sum_{i=1}^N dZ_i\wedge dZ_i'\,. \label{Omd} \ee
The coordinate functions $Z_i$ and $Z_i'$ then have Poisson brackets
\be [Z_i,Z_{i'}]_{\rm P.B.} = i\hbar\,\delta_{ii'}\,. \ee
The individual Lagrangian equations $z_i+z_i'{}^{-1}-1=0$ define a product Lagrangian submanifold
\be \CL_{(M,\Delta)} = \{z_i+z_i'{}^{-1}-1=0\quad\forall\;i\} \subset \CP_{(M,\Delta)}\,.  \ee

The phase space of the final $T^2$ boundary of $M$ can be lifted to logarithmic coordinates as
\be \CP_{T^2} = \{(V,U)\in \C^2 \}\,,\qquad \omega_{T^2} = (2i\hbar)^{-1}dV\wedge dU\,.\ee
We ignore the Weyl quotient for the moment.
Recalling the standard convention $\ell=e^v$ and $m=e^u$, it is convenient to define $v$ to be either
\be v= \frac{V}{2} \qquad\text{or}\qquad v=\frac{V}{2}+i\pi\,, \label{vVclass} \ee
depending on which square root of $\ell^2$ we chose.%
\footnote{We deeply regret breaking the general rule that capital letters are used for logarithmic variables and lowercase for exponentiated ones, but unfortunately there is already a convention in place for the logarithms $v=\log\ell$ and $u=\log m$ in much of the literature.} %
Then the symplectic form becomes
\be \omega_{T^2}=(i\hbar)^{-1}dv\wedge dU\,, \ee
with a canonical normalization.

The space $\CP_{T^2}$ should be a symplectic reduction of $\CP_{(M,\Delta)}$ with the $N$ linear gluing functions $C_j$ used as moment maps. In fact, due to the relation \eqref{Ccllog}, it suffices to take any set of $N-1$ $C_j$'s as linearly independent functions. However, in order for such a set of the $C_j$ to be used as simultaneous moment maps, they must all have trivial Poisson brackets with each other. It is an important theorem of Neumann and Zagier \cite{NZ} that under the symplectic structure \eqref{Omd} this is precisely the case:
\be [C_j,C_{j'}]_{\rm P.B.} = 0\qquad \forall\;j,j'\in\{1,...,N\}\,.\ee
Moreover, it was shown in \cite{NZ} that the affine linear functions $U$ and $v$ have trivial Poisson brackets with all the $C_j$'s, and precisely the desired canonical bracket
\be [v,U]_{\rm P.B.} = i\hbar\,, \label{vUpb} \ee
with each other.

In the spirit of Section \ref{sec:opglue}, we therefore choose $N-1$ independent $C_j$'s, say the first $N-1$, and complete the set $\{v,U,C_1,...,C_{N-1}\}$ to a new symplectic basis of $\CP_{(M,\Delta)}$ by defining $N-1$ new linear functions $\Gamma_j$ such that
\be [\Gamma_j,C_{j'}]_{\rm P.B.} = i\hbar\,\delta_{jj'}\,,\qquad
 [\Gamma_j,\Gamma_{j'}]=[\Gamma_j,U]=[\Gamma_j,v] = 0\,. \ee
In the new affine linear symplectic coordinates $(v,U,C_j,\Gamma_j)$ on $\CP_{(M,\Delta)}$, each moment map $C_j$ generates a $\C$ translation $\Gamma_j\mapsto \Gamma_j+t_j$. Therefore, precisely as desired, we find%
\footnote{Note that the phase space $\CP_{T^2}$ here is described in lifted, logarithmic coordinates --- as anticipated in Section \ref{sec:log}.}
\be \CP_{(M,\Delta)}\big/\!\!\big/(\Gamma_j\mapsto\Gamma_j+t_j) = \CP_{(M,\Delta)}\big/(\Gamma_j\sim\Gamma_j+t_j)\,\big|_{C_j=2\pi i} =\CP_{T^2}\,. \label{Predcl} \ee

In order to reduce the product Lagrangian $\CL_{(M,\Delta)}\subset \CP_{(M,\Delta)}$ to a Lagrangian in $\CP_{T^2}$, we must first project $\CL_{(M,\Delta)}$ to a codimension-one submanifold of the quotient space $\CP_{(M,\Delta)}\big/(\Gamma_j\sim\Gamma_j+t_j)$, then take the intersection with the locus $C_j=2\pi i$\;\;$\forall\;j$.

Algebraically, we can first generate an ideal with the defining equations of $\CL_{(M,\Delta)}$,
\be \CI_{(M,\Delta)} := (z_i+z_i'{}^{-1}-1)_{i=1}^N\,. \label{IMDcl} \ee
This is an ideal in the ring of functions on $\CP_{(M,\Delta)}$,
\be \CR_{(M,\Delta)} =  \C[z_i^{\pm1},z_i'{}^{\pm1},(1-z_i)^{-1},(1-z_i')^{-1},(z_iz_i'+1)^{-1}]_{i=1}^N\,, \label{RDcl}
\ee
where the extra inverted elements $z_i^{-1}$, $(1-z_i)^{-1}$, etc. are present to reflect the fact that none of the $z_i,z_i',z_i''$ ever take the values $\{0,1,\infty\}$. The (inverse) transformation of affine symplectic basis $\varphi:(Z_1,Z_1',...,Z_N,Z_N') \mapsto (v,U,\Gamma_1,C_1,...,\Gamma_{N-1},C_{N-1})$ for $\CP_{(M,\Delta)}$ then allows us to pull back the ideal $\CI_{(M,\Delta)}$ to an isomorphic ring $\C[\ell^{\pm 1},m^{\pm 2},c_j^{\pm 1},\gamma_j^{\pm 1},...]$, where now
\be \gamma_j := e^{\Gamma_j}\,. \ee
This is just a monomial transformation: each of the $z_i$ and $z_i'$ become monomials in $\ell^{\pm 1},m^{\pm 2},c_j^{\pm 1},\gamma_j^{\pm 1}$. Working in these new variables, we eliminate the $N-1$ $\gamma_j$'s from $\CI_{(M,\Delta)}$, producing an elimination ideal $\CJ_M$. Finally, we intersect $\CJ_M$ with $c_j=1$\; $\forall\,j$ to get $\CI_M$. From what we know about ideal hyperbolic triangulations and character varieties, the ideal $\CI_M$ should be generated by a single equation, the A-polynomial of $M$,
\be \CI_M = \CJ_M\big|_{c_j=1} =(A_M(\ell,m^2))\,. \label{IMcl} \ee \\

We could now put the $\Z_2$ Weyl quotient back in the definition of $\CP_{T^2}$. The fact that $A_M(\ell,m^2)=0$ descends to a well-defined equation on the quotient space $(\C\times\C)/\Z_2=\{(U,V)\}\big/$\raisebox{-.1cm}{\small $(U,V)\sim(-U,-V)$} results from the observation that the overall (artificial) orientation of tetrahedra cannot affect the final construction of the A-polynomial. Flipping the orientation of all tetrahedra is equivalent to sending all $(z_i,z_i',z_i'')\mapsto (z_i^{-1},z_i'{}^{-1},z_i''{}^{-1})$ and reversing the cyclic ordering $z_i \leftarrow z_i' \leftarrow z_i'' \leftarrow z_i$ everywhere. This produces an A-polynomial $A_M(\ell^{-1},m^{-2})$ that must be equivalent to $A_M(\ell,m^2)$.

\subsubsection{Example: $\mb{4_1}$ knot}

Since this description of symplectic reduction may have been very abstract, let us go through it explicitly for the figure-eight knot. We can choose the ``green'' edge as the one independent gluing function. We already know
\bse
\begin{align} C := C_{\rm green} &= 2Z'+Z''+2W'+W''=2\pi i-Z+Z'-W+W'\,,\\
U &= Z'-W\,, \label{41U} \\
v &= Z-Z'+i\pi \label{41v} \,
\end{align}
\ese
(note the extra $i\pi$ in $v$ that results from taking the negative square root of $\ell^2$),
and it suffices to chose the dual variable
\be \Gamma = W\,. \ee
Then the inverse transformation is
\be Z = -i\pi+v+U+\Gamma\,,\quad Z'=U+\Gamma\,,\quad W=\Gamma\,,\quad W'=-3\pi i+v+C+\Gamma \,,\ee
or
\be
z=-\gamma m^2\ell,\qquad z'=\gamma m^2\,,\qquad w=\gamma\,,\qquad w'=-c\gamma\ell\,.
\ee
The ideal $\CI_{(M,\Delta)} = (z+z'{}^{-1}-1,\,w+w'{}^{-1}-1)$ can be rewritten as
\be \CI_{(M,\Delta)} = (-\gamma m^2\ell+m^{-2}\gamma^{-1}-1,\; 
 \gamma-c^{-1}\gamma^{-1}\ell^{-1} -1)\,. \ee
Eliminating $\gamma$ results in
\be \CJ_M = \big(c^2m^4\ell^2-(c^2-c^2m^2-2cm^4-cm^6-m^8)\ell+cm^4\big)\,, \ee
and finally setting $c\to 1$, we retrieve the A-polynomial \eqref{A41cl}\,.

\subsection{Hyperbolic versus $SL(2,\C)$}
\label{sec:classrem}

There are several technical but important differences between hyperbolic structures and $SL(2,\C)$ structures, or flat connections, on a three-manifold. In order to tell a complete story, we must now draw some attention to them.

We begin by recalling the famous fact that on a so-called hyperbolic three-manifold --- defined as a three-manifold that admits \emph{any} hyperbolic metric of finite volume --- the hyperbolic metric is unique. This is the statement of Mostow rigidity \cite{mostow-1973}. If the three-manifold in question happens to have a boundary and one chooses appropriate boundary conditions, then the uniqueness property continues to hold. For example, on a knot complement, one should fix a conjugacy class for the meridian holonomy of the metric; then, if a finite-volume hyperbolic metric with such a meridian holonomy exists it will be unique.

One can also consider flat $SL(2,\C)$ connections on a knot complement $M = \ol{M}\bs K$. After fixing the conjugacy class of the meridian holonomy, one often finds (\eg\ when $M$ has no closed incompressible surfaces \cite{cooper-1994}, \cf\ Section \ref{sec:flatCS}) that the flat connections on $M$ form a discrete, finite set $\{\CA^{{\alpha}}\}$. If $M$ admits a hyperbolic metric of finite volume, then precisely \emph{one} element of this set, say $\CA^{(\rm geom)}$, corresponds to the hyperbolic metric. (By ``corresponds to'' we mean that the flat $SL(2,\C)$ connection can be transformed into the hyperbolic metric and vice versa, with all holonomies matching, as detailed in \cite{Witten-gravCS, gukov-2003}.) Other flat connections correspond to ``metrics'' that have curvature -1 but may not be everywhere positive definite; from an $SL(2,\C)$ standpoint, positive definiteness is simply not a natural constraint.

When asking for the classical A-polynomial of a knot complement $M$, we want to know about \emph{all} the flat $SL(2,\C)$ connections, not just the (potential) geometric one. Global hyperbolic geometry on all of $M$ is unlikely to help us in this regard. However, the construction of hyperbolic structures in terms of ideal tetrahedra, as described in Sections \ref{sec:triang}-\ref{sec:glue}, is almost sufficient. To understand this, note that an ideal hyperbolic tetrahedron with shape parameter $z$ (or $z'$, or $z''$) has a hyperbolic volume \cite{milnor-1982, thurston-1980}
\begin{align} \Vol(\Delta_z) &= D(z):= \Im\,\Li_2(z)+\arg(1-z)\log|z|\,\label{BW} \\ &= D(z')=D(z'')\,. \nno \end{align}
When $\Im(z)>0$, the volume is positive; but if $\Im(z)=0$ or $\Im(z)<0$, the volume is correspondingly zero or negative. In the construction of Sections \ref{sec:triang}-\ref{sec:glue}, we put no restriction whatsoever on $z$ aside from requiring $z\neq 0,1,\infty$ (in order to avoid completely \emph{degenerate} tetrahedra). This allows us the freedom to access many of the $SL(2,\C)$ connections on $M$ that do not correspond to positive-definite metrics. Moreover, it becomes possible to analyze $SL(2,\C)$ structures on manifolds such as the complements of torus knots in $S^3$, for which no complete positive-definite hyperbolic structure exists at all. As an example, in Section \ref{sec:ex}, we will explicitly use ideal triangulations to quantize the A-polynomial for the trefoil.

Unfortunately, the shape parameters of ideal tetrahedra cannot be used to parametrize quite all $SL(2,\C)$ structures. Even more regrettably, it is not completely known yet precisely \emph{which} flat connections can be obtained, or even how the set of attainable connections depends on the choice of an ideal triangulation. We can list several facts that are known, and formulate a guess at what the general picture could be.

First, let us note that ideal hyperbolic tetrahedra by themselves can never be used to construct abelian (or reducible) flat connections. The basic reason behind this is that the $SL(2,\C)$ holonomies corresponding to an ideal hyperbolic triangulation of a three-manifold are constructed from matrices that are either upper or lower triangular --- hence possibly parabolic --- but never diagonal. This can be seen explicitly either from constructions involving three-dimensional developing maps \cite{Champ-hypA, Dunfield-Mahler}, or from more combinatorial viewpoints as in \cite{Zickert-rep}. In the case of a knot complement in $S^3$, \ie\ $M = S^3\bs K$, this means that an ideal hyperbolic triangulation never detects the single abelian $(\ell-1)$ factor of the $SL(2,\C)$ A-polynomial.

In general, given a knot complement $M$ with an ideal triangulation $\{\Delta_i\}$, one can use the methods of Section \ref{sec:glue} to construct a ``gluing variety'' that is cut out by a single polynomial equation as in \eqref{A41cl} or \eqref{IMcl}. For clarity, let us denote this polynomial as $A_{M,\Delta}^{\rm glue}(\ell,m^2)$. It divides the actual $SL(2,\C)$ A-polynomial $A_M(\ell,m^2)$, but it is not equal to $A_M(\ell,m^2)$. At the very least, by what was said above,
\be A_M(\ell,m^2) = (\ell-1)\times A_{M,\Delta}^{\rm glue}(\ell,m^2)\times(\text{possibly other factors})\,. \label{AAtrue} \ee
In other words, $A_{M,\Delta}^{\rm glue}(\ell,m^2)$ contains \emph{some combination} of the irreducible factors of $A_M(\ell,m^2)$, with the exception of $(\ell-1)$. It was argued in \cite{Champ-hypA} (see also \cite{Dunfield-Mahler}) that if $M$ admits a finite-volume hyperbolic metric \emph{and} the triangulation $\Delta$ is sufficiently generic, then $A_{M,\Delta}^{\rm glue}(\ell,m^2)$ must contain at least the irreducible component of $A_M(\ell,m^2)$ containing the geometric flat connection $\CA^{(\rm geom)}$.

The condition that the triangulation $\Delta$ be ``sufficiently generic'' is motivated by the fact that certain ``bad'' triangulations give an empty gluing variety, \ie\ $A_{M,\Delta}^{\rm glue}(\ell,m^2)=1$. This happens when an edge $e$ in the triangulation belongs to only one tetrahedron. Suppose that in this one tetrahedron the edge has a parameter $z_e$. Then the condition that the total angle around edge $e$ is $2\pi$ results in the classical gluing equation $z_e = 1$, which contradicts the general non-degeneracy condition on shape parameters, $z_e\neq 0,1,\infty$.

In Section \ref{sec:ind}, we will argue (but not quite prove) that if two triangulations $\Delta,\Delta'$ are related by a 2-3 Pachner move and both triangulations are sufficiently generic --- in particular, neither triangulation is ``bad'' in the above sense of containing a degenerate edge --- then the two gluing varieties  $A_{M,\Delta}^{\rm glue}$ and $A_{M,\Delta'}^{\rm glue}$ must be equal. If this were indeed proven, \emph{and} if it were possible to go from one good triangulation of a knot complement to any other by using a chain of 2-3 Pachner moves that in turn only involved good triangulations, then the gluing variety $A_{M,\Delta}^{\rm glue}$ would be a unique invariant of $M$. It could simply be defined by choosing any good triangulation. This is the scenario that we hope is actually realized. In addition, we could imagine that at least for knot complements in $S^3$, we have exactly
\be A_{M}(\ell,m^2) = (\ell-1)\,A_{M,\Delta}^{\rm glue}(\ell,m^2)\qquad (\text{for ``good'' $\{\Delta_i\}$})\,, \label{AAhope} \ee
with no other factors involved.

So far, to the best of our knowledge, no data has invalidated \eqref{AAhope}. However, not much data is available. In general, we only know that \eqref{AAtrue} holds. One can sort the infinite triangulations of $M$ into a \emph{finite} number of classes according to which irreducible components of $A_M(\ell,m^2)$ are contained in the corresponding $A_{M,[\Delta]}^{\rm glue}(\ell,m^2)$. Then the gluing variety $A_{M,[\Delta]}^{\rm glue}(\ell,m^2)$ becomes an invariant of $M$ and its triangulation class.\label{triclass} \\

For the remainder of this paper, we will largely ignore issues of triangulation class and whether or not various factors of $A_M(\ell,m^2)$ appear in gluing varieties $A_{M,\Delta}^{\rm glue}(\ell,m^2)$. On one hand, for many simple knots (including all the examples in this paper) the A-polynomial $A_M(\ell,m^2)$ has exactly two factors, one being $(\ell-1)$ and the other containing the geometric flat connection; then for any ``good'' triangulation, \eqref{AAhope} must hold. On the other hand, it is not entirely clear whether the symplectic construction of Section \ref{sec:hypsemicl} (and in particular its quantum version that will appear in Section \ref{sec:qglue}) might not be able to bypass the problems of bad gluing conditions such as $z_e=1$ conflicting with constraints $z_e\neq 0,q,\infty$. This requires some further study. Physically, \emph{any} triangulation of a knot or link complement $M$ should lead to a triangulation-independent wavefunction for $M$ and a system of operators that annihilates it. This is yet another reason to suspect that gluing varieties, at least when defined as in Sections \ref{sec:hypsemicl} or \ref{sec:qglue}, are not nearly as triangulation-dependent as they seem.

Henceforth, just as in preceding sections, we hide the potential dependence of the gluing variety $A_{M,\Delta}^{\rm glue}(\ell,m^2)$ on the triangulation $\{\Delta_i\}$, and denote it simply by $A_M(\ell,m^2)$. It is the object we quantize. It is to be understood that it differs from the actual $SL(2,\C)$ A-polynomial in at least (and probably at most) an abelian factor like $(\ell-1)$.

\subsection{Independence of triangulation and path}
\label{sec:ind}

The topological nature of the A-polynomial is inherent in its definition as the projection to $\CP_{T^2}$ of the (nonabelian) character variety of a knot complement $M$. In the preceding subsections, however, we presented an alternative combinatorial definition of the A-polynomial, as a ``gluing variety,'' using shape parameters of ideal triangulations. From the combinatorial point of view, topological invariance is not quite obvious, and several things should be checked. In particular, within a given triangulation
\begin{itemize}

\item one chooses labels $z_i,\,z_i',\,z_i''$ in some order for each tetrahedron;

\item one chooses certain paths for the meridian and longitude on the triangulated boundary torus (the homology classes of these paths are fixed, but their particular representatives are not); and

\item in the symplectic, semi-classical approach of Section \ref{sec:hypsemicl}, one chooses $N-1$ of the $N$ $C_j$ with which to do the reduction, and then chooses $N-1$ $\Gamma_j$'s as conjugate coordinates.

\end{itemize}
Moreover, at the beginning of the whole construction
\begin{itemize}

\item one also chooses a particular triangulation of $M$.

\end{itemize}

In this subsection, we sketch proofs that the combinatorial construction of the classical A-polynomial is independent of the first three choices, and mostly independent of the last. Many of these arguments are already familiar from the literature on hyperbolic triangulation, but we find it useful to review them here as models for the discussion of topological invariance of $\hat A$ in Sections \ref{sec:qpath}--\ref{sec:qPach}.

Let us begin then with the standard topological setup of an oriented knot complement $M$ that is triangulated into $N$ ideal tetrahedra $\{\Delta_i\}_{i=1}^N$. Within this triangulation, the cyclically symmetric constructions of the phase spaces $\CP_{\pd\Delta_i}$ and Lagrangians $\CL_{\Delta_i}$ in Section \ref{sec:triang} guarantee invariance under the precise labeling of $z_i,$ $z_i',$ and $z_i''$ for each tetrahedron. As noted below \eqref{zcyclic}, this works as long as edges of tetrahedra are labeled in the cyclic order determined by the orientation of $M$.

Within the triangulation $\{\Delta_i\}$, it is also easy to see that any choice of $N-1$ moment maps $C_j$ will work for the reduction of the phase space \eqref{Predcl} due to the classical constraint $\sum_{j=1}^N C_j = 2\pi i N$. The subsequent choice of $\Gamma_j$'s is completely irrelevant in the symplectic reduction. Similarly, in the classical construction of the reduced Lagrangian ideal $\CI_M$, we could first have intersected with $c_j=1$ and then eliminated the $\gamma_j$'s. The intersection does not depend on which $N-1$ $c_j$'s we choose, and the choice of $\gamma_j$'s is irrelevant precisely because they are eliminated.

\begin{figure}[htb]
\centering
\includegraphics[width=4.5in]{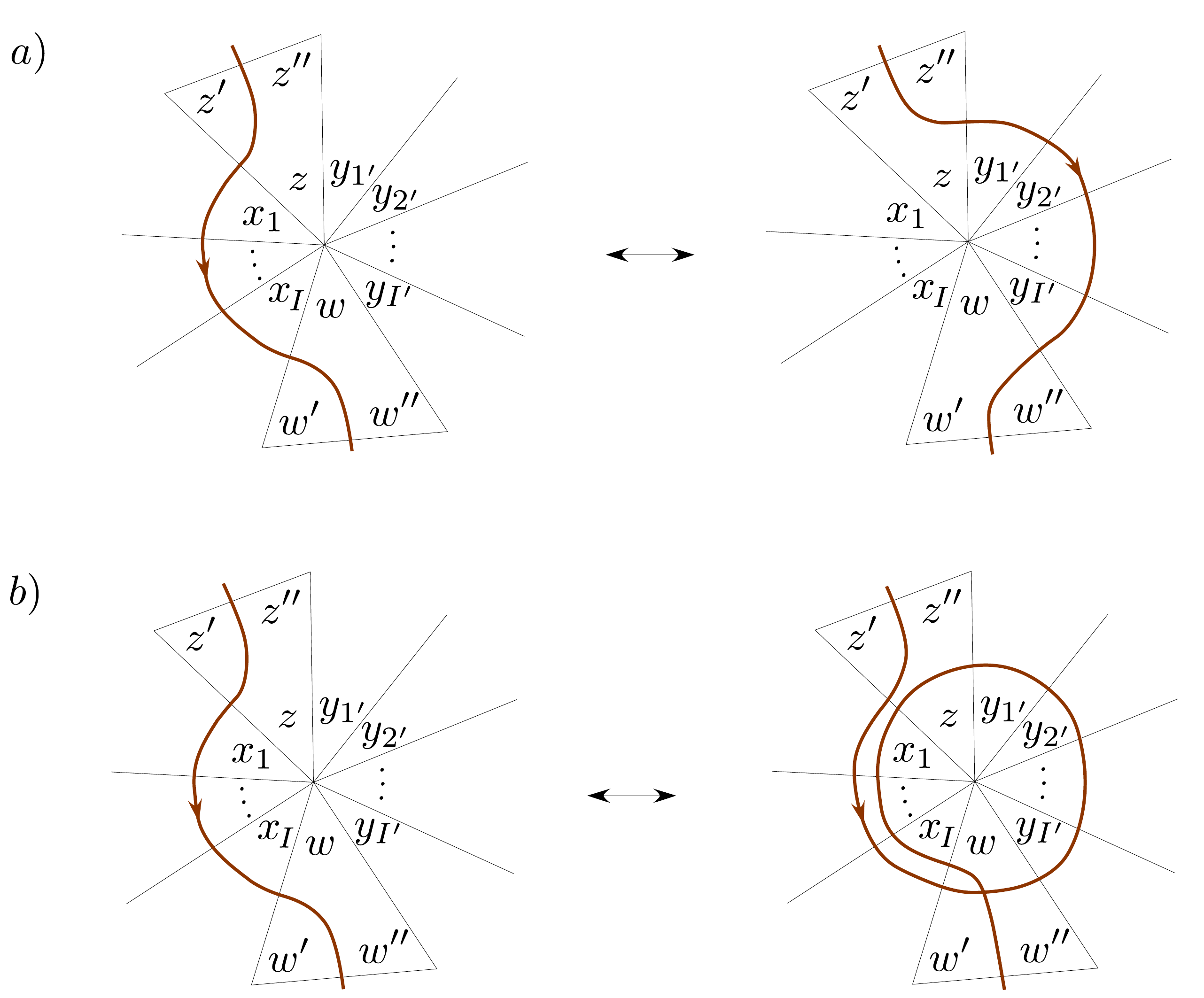}
\caption{Elementary moves for deforming a path}
\label{fig:pathelem}
\end{figure}

The only nontrivial thing to verify within a single triangulation is the independence under a change of meridian or longitude paths on the boundary torus. Suppose then that we have two paths on the boundary torus which both represent the same homotopy class, and hence should have the same holonomy eigenvalues. Any two such paths on a developing map can be related by repeated applications of the two elementary moves depicted in Figure \ref{fig:pathelem}.

The move in Figure \ref{fig:pathelem}(a) deforms a path through a point where an edge in $M$ meets the boundary $T^2$. The partial logarithmic holonomy on this section of path corresponding to the left side of the move is
\be -Z'+\sum_i X_i-W''\,, \label{pathdef1} \ee
while on the right side of the move it is
\be Z''- \sum_j Y_j + W'\,. \label{pathdef2} \ee
We know that $Z+Z'+Z''=i\pi$ and $W+W'+W''=i\pi$, and that $C:=Z+W+\sum_i X_i+\sum_j Y_j =2\pi i$ due to the gluing condition on the central edge. Therefore, the difference between \eqref{pathdef1} and \eqref{pathdef2} is
\be \textstyle -Z'-Z''+\sum_i X_i+\sum_j Y_j-W''-W \overset{vx\, eqs}{=} -2\pi i + Z+W+\sum_i X_i+\sum_j Y_j \overset{C\to 2\pi i}{=} 0\,. \label{pathdiff} \ee
Thus, upon imposing vertex and gluing equations, move (a) does nothing to the logarithmic holonomy along the path.

The second move, in Figure \ref{fig:pathelem}(b), adds a full $C=Z+W+\sum_i X_i+\sum_j Y_j \to 2\pi i$ to the logarithmic holonomy along a path. Obviously, this does not leave the logarithmic holonomy invariant, although it cannot change the final equations for the Lagrangian ideal $\CI_\Delta$ simply because these depend on exponentiated variables, where shifts of $2\pi i$ are invisible. Nevertheless, we do find it very useful to have well-defined logarithmic holonomies as coordinates on the affine linear (logarithmic) phase space $\CP_{(M,\Delta)}$, as discussed in Section \ref{sec:log}.
We observe then that move (b) is fundamentally different from move (a) in that it introduces a self-intersection of the path on the developing map. If we simply require that meridian and longitude paths be drawn in such a way that they have no self-intersections (which is certainly always possible), then move (a) alone is sufficient to translate between all possible paths, and invariance of the logarithmic holonomy is ensured. \\

Having examined the choices to be made within a single triangulation, let us now consider a change in the triangulation itself. Although our combinatorial A-polynomial is not (in our present understanding) quite independent of triangulation, as discussed in Section \ref{sec:classrem}, it comes very close to being so.

\begin{figure}[htb]
\centering
\includegraphics[width=5in]{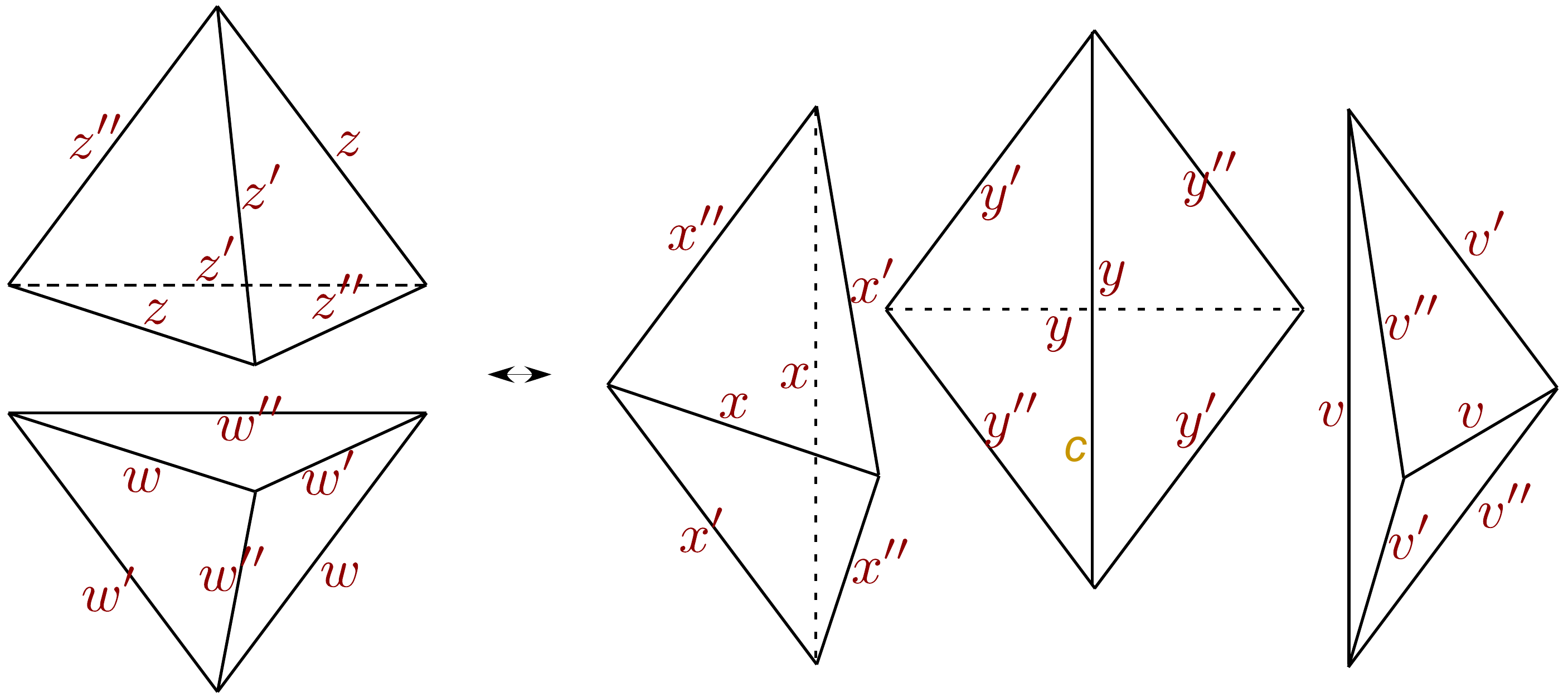}
\caption{The 2-3 Pachner move, with shape parameters.}
\label{fig:Pachnerhyp}
\end{figure}

We know that changes in triangulation are generated by 2-3 Pachner moves (Figure \ref{fig:Pachner}), so let's suppose that after a $2\to 3$ move there is a new triangulation $\{\tilde\Delta_i\}_{i=1}^{N+1}$ of $M$, such that the first $N-2$ tetrahedra are identical to those of the $\{\Delta_i\}$ triangulation,
\be \tilde\Delta_i = \Delta_i\,,\qquad i=1,...,N-2\,,\ee
while $\Delta_{N-1}$ and $\Delta_{N}$ participate in the 2-3 move to become $\tilde \Delta_{N-1}$, $\tilde \Delta_{N}$, and $\tilde \Delta_{N+1}$. These tetrahedra that are involved in the move are given shape parameters $z,w,x,y,v$, respectively, as in Figure \ref{fig:Pachnerhyp}. 

In the $\{\tilde\Delta_i\}$ triangulation, the fact that there are $N+1$ tetrahedra means that there are $N+1$ edges in $M$. The new edge is obviously the central one on the right side of Figure \ref{fig:Pachnerhyp}, the internal vertical edge of the 2-3 ``hexahedron.'' The holonomy around it is
\be C:=\tilde C_{N+1} = X+Y+V\,. \label{23glue} \ee
In addition the nine external edges of the hexahedron, six upright diagonal and three equatorial, are involved in the move. The holonomies around them depend on whether we are in the $\{\Delta_i\}$ or $\{\tilde\Delta_i\}$ triangulations. In order for these holonomies to be calculated consistently, and to even have a hope for obtaining the same A-polynomial before and after the move, we must require six relations
\bse \label{eqs6}
\begin{align} Z = V'+Y''\,,\quad Z'=X'+V''\,,\quad Z''=Y'+X''\,,\;\;\\ W=Y'+V''\,,\quad W'=X'+Y''\,,\quad W''=V'+X''\,.\end{align}
\ese
These come from the diagonal edges.
Together with the vertex equations for the $\{\tilde\Delta_i\}$ triangulation, equations \eqref{eqs6} imply
\be X = C-2\pi i+Z+W\,,\quad Y=C-2\pi i+Z'+W''\,,\quad V=C-2\pi i+Z''+W'\,, \label{eqs3}\ee
which at $C=2\pi i$ are the conditions for consistent calculation of holonomies around the three equatorial edges.

In order to argue for invariance under the 2-3 move, we will show that a \emph{partial} symplectic reduction of the $2(N+1)$-dimensional phase space $\CP_{(M,\tilde \Delta)}$, using only $C$ as a moment map, produces the $2N$-dimensional phase space $\CP_{(M,\Delta)}$ of the $\{\Delta_i\}$ triangulation. More importantly, we show that the product Lagrangian submanifold $\CL_{(M,\tilde{\Delta})}\, \subset\, \CP_{(M,\tilde \Delta)}$ reduces to the product Lagrangian $\CL_{(M,\Delta)}\,\subset\, \CP_{(M,\Delta)}$.

For the statement about phase spaces, it suffices to observe that due to the Poisson brackets
\be [X,X']_{\rm P.B.}=[Y,Y']_{\rm P.B.}=[V,V']_{\rm P.B.}=i\hbar\qquad\text{(+ cyclic)} \ee
on $\CP_{(M,\tilde\Delta)}$, equations \eqref{eqs6} ensure that
\be [Z,Z']_{\rm P.B.}=[W,W']_{\rm P.B.}=i\hbar\qquad \text{(+ cyclic)}, \ee
and that $Z,Z',Z''$ have trivial brackets with $W,W',W''$. Here, we are viewing the $Z$'s and $W$'s as functions on $\CP_{(M,\tilde\Delta)}$.
Moreover, since
\be Z+Z'+Z''=W+W'+W''\;\overset{\eqref{eqs6},\eqref{23glue}}{=}\;
3\pi i-C\,,\ee
and
\be [C,Z]_{\rm P.B.}=[C,Z']_{\rm P.B.}=[C,Z'']_{\rm P.B.}=[C,W]_{\rm P.B.}=[C,W']_{\rm P.B.}=[C,W'']_{\rm P.B.}=0\,, \ee
any two of $(Z,Z',Z'')$ and any two of $(W,W',W'')$ form good coordinates on the symplectic quotient of $\CP_{(M,\tilde\Delta)}$ generated by $C$, and the vertex equations for $Z$ and $W$ will be obeyed on the quotient space after sending $C\to 2\pi i$.

It is especially convenient to pick five coordinates $(Z'',Z,W'',W,C)$ on the part of $\CP_{(M,\tilde\Delta)}$ corresponding to the tetrahedra involved in the 2-3 move. This set can be completed to a symplectic basis for the full 6-dimensional subspace of $\CP_{(M,\tilde\Delta)}$ involved in the 2-3 move by adding
\be \Gamma := X''\,. \label{23gamma} \ee
Noting that
\be [\Gamma,C]_{\rm P.B.}=i\hbar\,,\qquad
[\Gamma,Z'']_{\rm P.B.}=[\Gamma,Z]_{\rm P.B.}=[\Gamma,W'']_{\rm P.B.}=[\Gamma,W]_{\rm P.B.}=0\,,\ee
it is clear that
\be \CP_{(M,\Delta)} = \CP_{(M,\tilde\Delta)}\big/\raisebox{-.1cm}{\small $(\Gamma\sim\Gamma+t)$}\big|_{C=2\pi i}\,.\ee

To analyze the Lagrangians, let us write the part of $\CL_{(M,\tilde\Delta)}$ involved in the 2-3 move as
\be x'+x''{}^{-1}-1 =0\,,\qquad  y'+y''{}^{-1}-1=0\,,\qquad v'+v''{}^{-1}-1=0\,. \label{Pachideal} \ee
This defines an ideal in the ring
\begin{align} \CR_{xyv} &= \C[x'{}^({}'{}^){}^{\pm 1},y'{}^({}'{}^){}^{\pm 1},v'{}^({}'{}^){}^{\pm 1},(1-x'{}^({}'{}^))^{- 1},(1-y'{}^({}'{}^))^{- 1},(1-v'{}^({}'{}^))^{- 1}, \label{Pachring} \\
&\hspace{2in} (x'x''+1)^{-1},\,(y'y''+1)^{-1},\,(v'v''+1)^{-1}]\,. \nno
\end{align}
By inverting equations \eqref{eqs6} as well as the gluing equation \eqref{23glue} and the definition of $\Gamma$ \eqref{23gamma}, we can change coordinates to $(z'',z,w'',w,\gamma=e^\Gamma,c=e^C)$ using
\be x'=\frac{-1}{czw\gamma}\,,\quad x''=\gamma\,,\quad y'=\frac{z''}{\gamma}\,,\quad y''=\frac{z\gamma}{w''}\,,\quad v'=\frac{w''}{\gamma}\,,\quad v''=\frac{w \gamma}{z''}\,. \label{xyvsols}\ee
Rewriting the ideal \eqref{Pachideal} in the new coordinates, eliminating $\gamma$, and setting $c=1$, we are left with two independent equations
\bse \label{prezweqs}
\begin{align} (1-zw)(z''+z^{-1}-1)&=0\,, \\
(1-zw)(w''+w^{-1}-1)&=0\,.
\end{align}
\ese
However, due to the invertible elements in \eqref{Pachring} and relations \eqref{eqs3}, these equations define an ideal in a ring where the element $(1-zw)$ is invertible, so equations \eqref{prezweqs} are actually equivalent to
\be z''+z^{-1}-1 =0\,,\qquad  w''+w^{-1}-1=0\,, \ee
the desired Lagrangian equations for $\CL_{(M,\Delta)}$. As functions on $\CP_{(M,\Delta)}$ we know that these are completely equivalent to
\be z+z'{}^{-1}-1=0\,,\qquad w+w'{}^{-1}-1=0\,. \ee

\subsubsection{What's missing}
\label{sec:missing}

The desired invariance of the A-polynomial under changes of triangulation could be summarized in the commutative diagram
\be \label{cldiag}
\begin{diagram}
(\CP_{(M,\Delta)},\CL_{(M,\Delta)}) & & \\
  & \rdTo^{{\rm quotient},\; C_j\to 2\pi i} & \\
\uTo^{{2-3}\atop{\rm Pach.}} \dTo & & \qquad\qquad (\CP_{T^2},A=0) \\
  & \ruTo_{{\rm quotient},\;\tilde C_j\to 2\pi i} & \\
 (\CP_{(M,\tilde\Delta)},\CL_{(M,\tilde{\Delta})}) & &
\end{diagram}
\ee
Unfortunately, the commutativity here is not quite true. Although Lagrangian submanifolds, or ideals in the algebra of functions, behave exactly as expected under 2-3 moves, the algebras of functions themselves do not!

To be more precise, in \eqref{Pachring} we introduced a ring $\CR_{xyz}$ that is a subring of the full ring $\CR_{(M,\tilde\Delta)}$ of ``algebraic'' functions on $\CP_{(M,\tilde\Delta)}$. It is the part of $\CR_{(M,\tilde\Delta)}$ that is expected to change under the $3\to2$ move. After the $3\to2$ move, we would expect $\CR_{xyz}$ to reduce to the ring
\be \CR_{zw}=\C[z{}^({}''{}^){}^{\pm 1},\,w{}^({}''{}^){}^{\pm 1},\,(1-z{}^({}''{}^))^{\pm 1},\,(1-w{}^({}''{}^))^{\pm 1},\,(zz''+1)^{-1},\,(ww''+1)^{-1}]\,, \ee
a subring of the full ring $\CR_{(M,\Delta)}$ for the $\{\Delta_i\}$ triangulation. Instead, however, the local $3\to2$ symplectic reduction leads to
\be \CR_{zw}^{3\to2}=\C[z{}^({}''{}^){}^{\pm 1},\,w{}^({}''{}^){}^{\pm 1},\,
(1-zw)^{-1},\,(ww''+z'')^{-1},\,(zz''+w'')^{-1}]\,.
\ee
The rings $\CR_{zw}$ and $\CR_{zw}^{3\to2}$ are generated by the same monomials, but they have different invertible elements. Indeed, the ability to invert $(1-zw)^{-1}$ in $\CR_{zw}^{3\to2}$ was what allowed us to remove the leading factors in Equations \eqref{prezweqs}.

The distinction between $\CR_{zw}$ and $\CR_{zw}^{3\to2}$ can be crucial when triangulations are sufficiently non-generic. For example, suppose that $\{\Delta_i\}$ is one of the ``bad'' triangulations of $M$ discussed in Section \ref{sec:classrem}. In particular, suppose that one of the ``diagonal edges'' of the 2-3 hexahedron, (say) the one with parameter $z$, doesn't become identified with any other edge but itself during the gluing. Then the classical gluing equation $z=1$ cannot actually be imposed in the phase space acted on by $\CR_{zw}$ (which contains $(z-1)$ as an invertible element), though it \emph{can} be imposed in $\CR_{zw}^{3\to2}$. If we are to take invertible elements seriously, then in this case it looks like the $\{\tilde\Delta_i\}$ triangulation could lead to a reasonable A-polynomial, or ``gluing variety,'' while the $\{\Delta_i\}$ triangulation would lead to a trivial, empty one. This example illustrates how the caveats of Section \ref{sec:classrem} can show up in symplectic gluing.

In the final step of our symplectic construction of $A(\ell,m^2)$, we typically forget the information about the precise invertible elements in the rings that we started with.%
\footnote{In some cases, however, invertible elements may be relevant for removing prefactors from the A-polynomial in a penultimate step, much as prefactors were removed from \eqref{prezweqs}. This will occur with our triangulation of the trefoil knot complement in Section \ref{sec:ex}.} %
We just think of $A(\ell,m^2)$ as an element in $\CR_{T^2}=\C[\ell^{\pm1},\,m^{\pm 2}]$. 
We might then expect that for sufficiently general triangulations it should be possible to ignore the details of ``precursor'' rings such as $\CR_{(M,\Delta)}$. The extent to which this is true is still unclear. It is an interesting but fundamentally classical issue, and beyond the scope of the present paper. In the following sections, we will show how to \emph{quantize} any A-polynomial (\emph{a.k.a} gluing variety) obtained from an ideal triangulation, and show that the construction is at least as independent of combinatorial choices as it would be classically.

\section{Quantization}
\label{sec:quant}

The semi-classical description of ideal tetrahedra and A-polynomials presented in Section \ref{sec:hyp} is ripe for quantization. In parallel to Section \ref{sec:hyp}, we begin here by discussing the quantization of Chern-Simons theory on a single tetrahedron, then show how several tetrahedra can be glued together to build an operator $\hat A_M(\hat\ell,\hat m;q)$ (also denoted $\hat A_M(\hat\ell,\hat m^2;q)$) that annihilates the Chern-Simons partition function on an entire knot complement $M$. In this section, our focus is almost entirely on the operator side of the story, which can be analyzed abstractly in its own right. Later, in Section \ref{sec:wf}, we will return to the technicalities of wavefunctions and holomorphic blocks, for individual tetrahedra and for all of $M$.

\subsection{A single tetrahedron}
\label{sec:qtet}

Recall from Section \ref{sec:triang} that the complexified Chern-Simons phase space associated to the boundary of a tetrahedron is
\be \CP_{\pd\Delta} = \{(Z,Z',Z'')\in (\C^3\bs\,2\pi i\Z)^3\;|\;Z+Z'+Z''=i\pi\}\,,\qquad (i\hbar)\,\omega_{\pd\Delta} =dZ\wedge dZ'\,,\label{defPDq} \ee
and that a flat connection, or semi-classical state, in the bulk of the tetrahedron is described by a Lagrangian submanifold
\be \CL_\Delta := \{z+z'{}^{-1}-1=0\}\quad\subset\quad\CP_{\pd \Delta}\,, \label{defLq} \ee
with $z=e^Z\,,z'=e^{Z'}$, and $z''=e^{Z''}$.

Upon quantization, the coordinate functions $Z,\,Z',\,Z''$ should be promoted to generators $\hat Z,\,\hat Z',\,\hat Z''$ of an operator algebra, which satisfy commutation relations 
\be [\hat Z,\hat Z']=[\hat Z',\hat Z'']=[\hat Z'',\hat Z]=\hbar\,. \label{Zcomm} \ee
Since the space $\CP_{\pd\Delta}$ is linear, we expect no further quantum corrections to \eqref{Zcomm} at subleading order in $\hbar$.
The classical coordinates obey the vertex equation $Z+Z'+Z''=i\pi$, which should imply that the corresponding quantized operators also obey
\be \hat Z+\hat Z'+\hat Z''=i\pi + a\hbar, \label{qvertex} \ee
where $a$ is a potential quantum correction. It turns out that that the nontrivial correction 
\be \boxed{a=\frac12}\, \label{a12} \ee
is necessary; it will be uniquely determined either by asking for topological invariance of our combinatorial construction of $\hat A$ (Sections \ref{sec:qpath}--\ref{sec:qPach}) or by requiring a certain S-duality in the algebra of operators (Section \ref{sec:Heis}).
For now, however, let us just keep $a$ as an undetermined constant.

The logarithmic commutation relations \eqref{Zcomm} imply $q$--commutation relations for exponentiated operators,
\be \hat z\hat z'=q \hat z'\hat z\,,\qquad
 \hat z'\hat z''=q \hat z''\hat z'\,,\qquad \hat z''\hat z=q\hat z\hat z''\,, \label{zcomm} \ee
where 
\be \hat z := e^{\hat Z}\,,\qquad \hat z':=e^{\hat Z'}\,,\qquad \hat z'':=e^{\hat z''}\,,\ee
and
\be q = e^{\hbar}\,.\ee
Furthermore, the quantum vertex equation \eqref{qvertex} implies
\be \hat z\hat z'\hat z'' = -q^{a+\frac12}\,. \label{qzvertex} \ee
Note that in the abstract algebra of operators with commutation relations \eqref{Zcomm} or \eqref{zcomm}, the elements $\hat Z+\hat Z'+\hat Z''$ and $\hat z\hat z'\hat z''$ are \emph{central}. This is necessary in order to consistently set them equal to constants as in  \eqref{qvertex} and \eqref{qzvertex}.

The Lagrangian equation \eqref{defLq} should be promoted to a quantum operator $\hat\CL_\Delta$ that annihilates the Chern-Simons wavefunction of a tetrahedron. Our basic ansatz, which we will see justified in many ways, is that the quantization of this operator is \emph{simple}. In particular, we assume that
\be \hat \CL_\Delta = q^b \hat z+q^{b'}\hat z'{}^{-1}-q^{b''} \simeq 0\,, \label{qLb} \ee
for some mild $q$-corrections parametrized by $b,b',b''$. As in Section \ref{sec:opglue}, the symbol ``$\simeq$'' means ``annihilates wavefunctions when acting on the left.''
Just as in the semi-classical case, we would like equation \eqref{qLb} to be equivalent to the two other cyclic permutations
\be q^b \hat z'+q^{b'}\hat z''{}^{-1}-q^{b''} \simeq 0\,,\qquad
q^b \hat z''+q^{b'}\hat z{}^{-1}-q^{b''} \simeq 0\,,\ee
and this condition fixes the constants $b,b',b''$. One can check (by multiplying on the \emph{left} of \eqref{qLb} with $\hat z'$ and $\hat z''\hat z'$) that the unique equation consistent with \eqref{qzvertex} and cyclic permutations is
\be \boxed{\hat \CL_\Delta:=\alpha\hat z+\alpha^{-1}\hat z'{}^{-1}-1\simeq 0\,,\qquad \alpha := q^{\frac{1-2a}{6}}}\,, \label{qLatet} \ee
up to an irrelevant overall constant prefactor. At the proper value $a=1/2$, we simply have $\boxed{\alpha=1}$.

The algebra of operators described here can be represented on any physical Hilbert space $\CH_\Delta^{\rm phys.}$ obtained by quantizing a real slice of $\CP_{\pd \Delta}$ where $\omega_{\pd\Delta}$ is nondegenerate. Different physical theories (for example, $SU(2)$ vs. $SL(2,\R)$ Chern-Simons theory) correspond to quantizing different real slices. In any such representation, it is the \emph{same} operator $\hat \CL_\Delta$ that annihilates the wavefunction of a tetrahedron.

If we analytically continue wavefunctions, then the algebra of operators can be taken to act on a functional space $\CH_\Delta'$ that contains locally holomorphic functions of $Z'$,%
\footnote{Alternatively, one could choose equivalent holomorphic representations $\CH_\Delta\sim\{f(Z)\}$ or $\CH_\Delta''\sim\{f(Z'')\}$; the relations between these choices will be discussed in Section \ref{sec:Weil}.} %
\be \CH_\Delta' \sim \{f(Z')\}\,,\ee
in the representation
\begin{align} &\hat Z\,f(Z') = \hbar\pd_{Z'} f(Z')\,,\qquad
\hat Z'\,f(Z') = Z'f(Z')\,, \\ &\quad\hat Z''\,f(Z') = (i\pi+a\hbar-\hbar\pd_{Z'}-Z')f(Z')\,. \end{align}
The wavefunction or ``holomorphic block'' of a tetrahedron then obeys
\be  \hat \CL_\Delta\,\cdot \psi(Z') = \big( \hat z+\hat z'{}^{-1}-1\big)\,\psi(Z') = 0\,, \label{qLangeq} \ee
where we have set $a=1/2$. One solution to this equation is easily seen to be
\be \psi(Z') = \prod_{r=1}^\infty \big(1-q^rz'{}^{-1}\big)\,. \label{quant-qdl} \ee
This is a quantum dilogarithm function%
\footnote{This function is periodic in $Z'\to Z'+2\pi i$. As previewed in Section \ref{sec:log}, it is actually not quite the right conformal block for the tetrahedron. We will see in Section \ref{sec:duality} that nonperturbative corrections to \eqref{quant-qdl} ultimately break its periodicity.} %
\cite{Fad-Kash, Fad-Volkov-preqdl},
with leading asymptotic behavior as $\hbar\to 0$ given by
\be \psi(Z') \sim \exp\left(\frac1\hbar\,\Li_2(z'{}^{-1})\right). \ee

The function $-\Li_2(z'{}^{-1}) = \Li_2(z)+\log(z)\log(1-z)-\frac{\pi^2}{6}$ (using the classical relation $z'{}^{-1}=1-z$) is a holomorphic version of the Bloch-Wigner function \eqref{BW}, which gives the classical volume of an ideal tetrahedron. More precisely, $-\Li_2(z'{}^{-1})$ differs from an analytic continuation of the Bloch-Wigner function by a rational multiple of $\pi^2$.\label{Volrat} This is the correct leading asymptotic that one would expect in analytically continued Chern-Simons theory: the classical Chern-Simons action evaluated at a flat, complexified $SL(2,\C)$ connection on an ideal tetrahedron should reproduce its analytically continued volume (\cf\ \cite{gukov-2003} or \cite{DGLZ}, Section 2). This agreement is the first justification that the simple quantization ansatz \eqref{qLb} is on the right track. Much stronger justification will come from the internal consistency and topological invariance of our scheme for gluing tetrahedra together, and from the fact that this scheme actually reproduces known $\hat A$-polynomials on the nose.

\subsection{Quantum gluing}
\label{sec:qglue}

Now, let's try to glue $N$ tetrahedra $\{\Delta_i\}_{i=1}^N$ together to form a knot complement $M=\ol{M}\bs K$. Specifically let us try to glue together $N$ operator equations of the form \eqref{qLatet} to find the quantum $\hat A$-polynomial for $M$. The general methods of Section \ref{sec:opglue} tell us how to proceed.

We begin with a product phase space $\CP_{(M,\Delta)}=\prod_{i=1}^N\CP_{\pd\Delta_i}$ and a product Lagrangian submanifold $\CL_{(M,\Delta)} = \prod_{i=1}^N\CL_{\Delta_i}$.
The algebra of functions on $\CP_{(M,\Delta)}$ is quantized to a product algebra of operators, generated by $\{\hat Z_i,\hat Z_i',\hat Z_i''\}_{i=1}^N$ with
\be [\hat Z_i,\hat Z_{i'}']=[\hat Z_i',\hat Z_{i'}'']=[\hat Z_i'',\hat Z_{i'}]=\hbar\,\delta_{ii'}\,, \label{qZicomm} \ee
\be \hat Z_i+\hat Z_i'+\hat Z_i'' = i\pi +a\hbar\qquad\forall\;i\,. \label{qvxai} \ee
Since there is nothing to distinguish the quantization of one tetrahedron from another, the constant $a$ appearing in \eqref{qvxai} must be the same for all $i$. (Recall that we will eventually set $a=1/2$.) The product Lagrangian submanifold $\CL_{(M,\Delta)}$ is promoted to a left ideal in the operator algebra, generated by the equations
\be \hat \CL_{\Delta_i} = \alpha \hat z+\alpha^{-1}\hat z'{}^{-1}-1 \simeq 0\,,\qquad i=1,...,N\,, \label{quantideal} \ee
with $\hat z_i=e^{\hat Z_i}$, $\hat z_i'=e^{\hat Z_i'}$, $\hat z_i''=e^{\hat Z_i''}$.

The algebra \eqref{qZicomm}-\eqref{qvxai} can be represented on a space of holomorphic functions $\CH_{(M,\Delta)}'\sim \{f(Z_1',...,Z_N')\}$, which is essentially the tensor product of the spaces $\CH_{\pd\Delta_i}$ for $i=1,...,N$. Then, the left ideal \eqref{quantideal} contains precisely the elements of the operator algebra that annihilate the analytically continued product wavefunction
\be \Psi(Z_1',...,Z_N') :=\psi(Z_1')\cdots \psi(Z_N') \in \CH_{(M,\Delta)}'\,.\ee

Algebraically, we could also write the left ideal \eqref{quantideal} as
\be \hat \CI_{(M,\Delta)} := \big(\alpha \hat z+\alpha^{-1}\hat z'{}^{-1}-1\big)_{i=1}^N\,.\ee
It is then useful to consider it as an ideal in the $q$-commutative ring of exponentiated variables
\be \hat \CR_{(M,\Delta)} = \C(q)[\hat z_i{}^{\pm 1},\hat z_i'{}^{\pm 1},(1-\hat z_i)^{-1},(1-\hat z_i')^{-1},(1+\hat z_i\hat z_i')^{-1}]\,. \label{qringz}\ee
Note that if a ring generator like $(1-\hat z_i)$ is invertible, then $(1-q^{\#}\hat z_i')$ is also invertible for any power of $q$.

In the semi-classical gluing of Section \ref{sec:glue}, an identification among phase spaces of different tetrahedra was imposed through $N$ gluing conditions $C_j=2\pi i$. The $N$ functions $\{C_j\}_{j=1}^N$ served as moment maps for a symplectic quotient of $\CP_{(M,\Delta)}$, and a corresponding reduction of the product Lagrangian submanifold $\CL_{(M,\Delta)}$. Quantum mechanically, we should promote $C_j$ to elements $\hat C_j$ in the operator algebra. Then the ``reduction'' of the ideal $\hat \CI_{(M,\Delta)}$ consists of removing all elements that do not commute with the $\hat C_j$ and then setting $\hat C_j\to 2\pi i$.

Recall that the classical, logarithmic gluing functions $C_j$ are always linear in $Z_i$, $Z_i'$, and $Z_i''$. They can therefore be quantized in a very straightforward manner, simply by modifying \eqref{Ccllog} to
\be \hat C_j := \sum_{i=1}^N \left( \epsilon(i,j)\hat Z_i+\epsilon(i,j)'\hat Z_i'+\epsilon(i,j)''\hat Z_i''\right)\,,\qquad j=1,...,N\,. \label{qCgen}
\ee
For example, for the figure-eight knot complement, we define (\cf\ \eqref{cbgcl})
\bse \label{qCbg}
\begin{align}
 \hat C_{\rm blue} &:= 2\hat Z+\hat Z''+2\hat W+\hat W'' \\
 \hat C_{\rm green} &:= 2\hat Z'+\hat Z''+2\hat W'+\hat W''\,. \label{qCbgg}
\end{align}
\ese

In principle, there could be mild quantum corrections to equations \eqref{qCgen} in the operator algebra. Rather than correcting definitions like \eqref{qCbg} directly, we will instead modify the substitutions
\be \hat C_j \to 2\pi i+\kappa_j\hbar\, \label{qCsubs} \ee
that occur as the final step in the reduction of the ideal $\hat \CI_{(M,\Delta)}$. The $N$ edges in a triangulation of $M$ could have many factors distinguishing them from one another, such as the number of dihedral angles touching each edge. Therefore, the corrections $\kappa_j$ might depend on the edge $j$ rather than being universal. However, in order for the substitutions \eqref{qCsubs} to be consistent, they do have to be compatible with the operator algebra constraint
\be \sum_{j=1}^N\hat C_j = 2\sum_{i=1}^N(\hat Z+\hat Z'+\hat Z'') = 2N\pi i + 2Na\hbar\,. \ee
This implies that
\be \boxed{\textstyle \sum_{j=1}^N\kappa_j = 2Na}\,. \label{qKcons} \ee
After checking invariance under changes of path and Pachner moves (Sections \ref{sec:qpath}--\ref{sec:qPach}), we will find that the only possibility for the $\kappa_j$'s is ultimately to assume an edge-independent value
\be \boxed{\kappa_j \equiv 1\quad\forall\;j}\,,\ee
compatible with $a=1/2$.

The logarithmic longitude and meridian holonomies $v$ and $U$ can also be quantized in a straightforward manner, by simply adding hats to all the variables that appear in their affine-linear semi-classical definitions. Thus, for the figure-eight knot complement (\cf\ \eqref{ML41} or rather \eqref{41U}-\eqref{41v}), we find that
\bse \label{qUv41}
\begin{align}
 \hat v &:= \hat Z-\hat Z'+i\pi\,,\\
 \hat U &:= \hat Z'-\hat W\,.
\end{align}
\ese
Unlike in the case of quantum gluing operators $\hat C_j$, the expressions \eqref{qUv41} for $\hat v$ and $\hat U$ really are \emph{definitions}. The operators $\hat v$ and $\hat U$ generate the final operator algebra for the glued knot complement $M$, and any modification (such as $\hbar$ corrections) to expressions like \eqref{qUv41} would simply result in a new action of $\hat v$ and $\hat U$ on the final glued ``Hilbert'' space $\CH_{\pd M}=\CH_{T^2}$.

Due to the Poisson brackets of their semi-classical counterparts, the operators $\hat C_j$, $\hat U$, and $\hat v$ will satisfy commutation relations
\be [\hat v,\hat U]=\hbar\,,\qquad [\hat C_j,\hat C_{j'}]=[\hat C_j,\hat v]=[\hat C_j,\hat U]=0\qquad\forall\;j,j'\,. \label{qCcomm}\ee
As in the semi-classical case, we can choose any set of $N-1$ $\hat C_j$'s (say the first $N-1$) to be linearly independent in the operator algebra. Then the set $\{\hat v,\hat U,\hat C_1,...,\hat C_{N-1}\}$ can be completed to an (affine) linear basis for (affine) linear operators by adding $N-1$ more operators $\hat \Gamma_j$, such that
\be [\hat \Gamma_j,\hat C_{j'}]=\delta_{jj'}\hbar\,,\qquad [\hat \Gamma_j,\hat \Gamma_{j'}]=[\hat \Gamma_j,\hat v]=[\hat\Gamma_j, \hat U]=0\qquad \forall\;j,j'=1,...,N-1\,. \label{qGcomm} \ee
This guarantees that the change of generators
\be (\hat Z_1,\hat Z_1',...,\hat Z_N,\hat Z_N')\overset{\varphi_*}{\longmapsto} (\hat v,\hat U,\hat\Gamma_1,\hat C_1,...,\hat\Gamma_{N-1},\hat C_{N-1}) \label{qbasischange} \ee
in the operator algebra preserves the commutator.

The inverse of the map \eqref{qbasischange} allows us to write the equations generating the ideal $\hat \CI_{(M,\Delta)}$ in new exponentiated variables
\be \hat \ell:=e^{\hat v}\,,\qquad \hat m^2=e^{\hat U}\,,\qquad
 \hat \gamma_j := e^{\hat \Gamma_j}\,,\qquad \hat c_j:= e^{\hat C_j}\,. \label{hatexpcg} \ee
This transformation simply replaces every $\hat z$ or $\hat z'{}^{-1}$ in $(\hat\ell^{\pm1},\hat m^{\pm 2},\hat\gamma_j^{\pm 1},\hat c_j^{\pm 1})$, possibly multiplied by a power of $q$. The ordering of the operators in these monomials is unambiguous because inverting the map \eqref{qbasischange} is an affine linear operation (with no ordering ambiguities), and all exponentiated operators are well-defined in terms of linear ones. In terms of the new operators \eqref{hatexpcg}, the ideal $\hat I_{(M,\Delta)}$ becomes an ideal in the isomorphic $q$-commutative ring
\be \hat {\tilde \CR} = \C(q)[\hat\ell{}^{\pm1},\hat m{}^{\pm 2},\hat\gamma_j{}^{\pm 1},\hat c_j{}^{\pm 1},...], \label{qringc} \ee
where ``$...$'' are additional elements that must be inverted due to the $(1-\hat z_i{}^({}'{}^))^{-1}$'s and $(1+\hat z_i\hat z_i')^{-1}$'s in \eqref{qringz}.

For example, for the figure-eight knot we can choose one gluing operator
\be \hat C:=\hat C_{\rm green} = 2\pi i+2a\hbar - \hat Z+\hat Z'-\hat W+\hat W'\,, \label{qC41} \ee
where we have used the quantum vertex equations in the operator algebra to eliminate $\hat Z''$ and $\hat W''$ from \eqref{qCbgg}. The new operator $\hat\Gamma$ can be chosen as
\be \hat \Gamma = \hat W\,. \label{qG41} \ee
Inverting the equations \eqref{qUv41}, \eqref{qC41}, and \eqref{qG41}, we find
\bse
\begin{align}
 \hat Z &= -i\pi +\hat v+\hat U+\hat \Gamma\,,\\
 \hat Z' &= \hat U+\hat \Gamma\,,\\
 \hat W &= \hat \Gamma\,,\\
 \hat W' &= -3\pi i-2a\hbar+\hat v+\hat C+\hat \Gamma\,,
\end{align}
\ese
or, after exponentiating,\footnote{Any time operators are exponentiated, one should keep in mind the Baker-Campbell-Hausdorff formula \be e^{\hat A+\hat B} = e^{-\frac12[\hat A,\hat B]}e^{\hat A}e^{\hat B}\,,\qquad\text{if $[\hat A,\hat B]$ is central}. \nno \ee}
\be \hat z=-q^{\frac12}\hat\gamma \hat m^2\hat\ell\,,\qquad
 \hat z'=\hat\gamma\hat m^2\,,\qquad
 \hat w = \hat\gamma\,,\qquad
 \hat w' = -q^{\frac12-2a}\hat c\hat\gamma\hat\ell\,.
\ee
The equations generating $\hat \CI_{(M,\Delta)}$ then become
\bse \label{ideal41}
\begin{align}
\hat\CL_{z} = -\alpha q^{\frac12}\hat\gamma\hat m^2\hat\ell+\alpha^{-1}\hat m{}^{-2}\hat\gamma{}^{-1} - 1\simeq 0\,,\\
\hat\CL_{w} = \alpha\hat\gamma-\alpha^{-1}q^{\frac12+2a}\hat c{}^{-1}\hat\gamma{}^{-1}\hat\ell{}^{-1}-1 \simeq 0\,.
\end{align}
\ese

From here, one must eliminate variables $\hat \gamma_j$ from $\hat \CI_{(M,\Delta)}$ to obtain an ideal $\hat \CJ_M$ all of whose elements commute with the $\hat c_j$. Then, setting $\hat c_j=\exp(2\pi i+\kappa_j\hbar)=q^{\kappa_j}$ in $\hat \CJ_M$ provides the final ideal $\hat \CI_M$ that annihilates the holomorphic blocks of Chern-Simons theory on $M$. Eliminating variables in the left ideal $\hat \CI_{(M,\Delta)}$ is somewhat trickier than in the classical case of Section \ref{sec:hyp}, because one is only allowed to multiply on the left, and new factors of $q$ arise when operators are commuted past each other. However, since the ring \eqref{qringc} is just \emph{barely} noncommutative, many standard techniques for eliminating variables in polynomial equations are easily adapted to $\hat \CI_{(M,\Delta)}$. In particular, the ring \eqref{qringc} (or the isomorphic ring \eqref{qringz}) is Noetherian, and ideals in it have Gr\"obner bases, \cf\ \cite{Ore-NC, Takayama-qWeyl, KRW-Grob}.%
\footnote{Near the completion of this project, we were introduced to an extraordinarily efficient Mathematica package \cite{Koutschan-thesis, Koutschan-guide} that, among other operations, can compute Gr\"obner bases and eliminate variables in $q$--commuting algebras.}

We expect that the final ideal $\hat \CI_M$ obtained after setting $\hat c_j=q^{\kappa_j}$ is generated by a single element, the quantum $\hat A$-polynomial of $M$,
\be \hat \CI_M = \big(\,\hat A_M(\hat \ell,\hat m^2;q)\,\big)\,. \label{finalqI} \ee
Our claim is that at the special values of vertex and gluing constants $a=1/2$ and $\kappa_j\equiv 1$ \;$\forall\,j$, the ideal $\hat \CI_M$ is a well-defined topological invariant of $M$ itself, independent of any triangulation or any choices made in a given triangulation. We now proceed to check this, finishing the computation of $\hat\CI_M$ for the figure-eight knot complement and exhibiting a few other examples in Section \ref{sec:ex}.

\subsection{Independence of path and more}
\label{sec:qpath}

We begin by examining the choices made within a single triangulation $\{\Delta_i\}_{i=1}^N$ of $M$. As in the semi-classical case of Section \ref{sec:ind}, these are:
\begin{itemize}

\item a choice of labelings $z_i,z_i',z_i''$ of tetrahedron edges;

\item a choice of $N-1$ of the $N$ gluing operators $\hat C_j$ to use as new generators of the operator algebra;

\item a choice of $N-1$ linearly independent elements $\hat \Gamma_j$ satisfying commutation relations \eqref{qGcomm}; and

\item a choice of paths for the longitude and meridian holonomies (within a fixed homology class) on the developing map.

\end{itemize}

Due to our cyclically symmetric constructions of both the operator algebra and the annihilating operators $\hat \CL_{\Delta_i}$ for single tetrahedra $\Delta_i$, it is clear (just like in the semi-classical case) that any cyclic relabeling
\be \hat z_i\mapsto \hat z_i'\mapsto \hat z_i''\mapsto \hat z\qquad\text{(for any $i$)} \ee
leaves the definition of $\hat \CI_{M,\Delta}$, and ultimately $\hat \CI_M$, invariant. The cyclic ordering on tetrahedra is induced from the orientation of $M$.

Invariance under the remainder of the choices above is a consequence of the following observation. Suppose that we have an affine symplectic (or ``canonical'') map
\be \CT:\;(\hat v,\hat U,\hat\Gamma_1,\hat C_1,...,\hat\Gamma_{N-1},\hat C_{N-1}) \mapsto (\hat v',\hat U',\hat\Gamma_1',\hat C_1',...,\hat\Gamma_{N-1}',\hat C_{N-1}')\,, \ee
that changes the linear basis in the product operator algebra. By affine symplectic, we mean a composition of a linear symplectic transformation and a constant shift, so that the primed operators satisfy the same commutation relations as the unprimed ones, namely,
\be [\hat v',\hat U']=\hbar\,,\qquad [\hat \Gamma_j',\hat C_{j'}']={\delta_{jj'}}\hbar\,, \ee
with all other pairs commuting. Moreover, suppose that $\CT$ is restricted so that $\hat v'=\CT_v(\hat v,\hat U,\{\hat C_j\})$ and $\hat U'=\CT_U(\hat v,\hat U,\{\hat C_j\})$ do not depend on any of the $\Gamma_j$'s, and for any $k$ $\hat C_k'=\CT_{C_k}(\hat C_1,...,\hat C_{N-1})$ does not depend on $\hat v$, $\hat U$, or any of the $\hat\Gamma_j$'s. Then the inverse affine symplectic transformation
\be \CT^{-1}:\; (\hat v',\hat U',\hat\Gamma_1',\hat C_1',...,\hat\Gamma_{N-1}',\hat C_{N-1}')\mapsto
(\hat v,\hat U,\hat\Gamma_1,\hat C_1,...,\hat\Gamma_{N-1},\hat C_{N-1}) \label{invAS} \ee
has this same property, namely $\hat v=\CT^{-1}_v(\hat v',\hat U',\{\hat C_j'\})$ and $\hat U=\CT^{-1}_U(\hat v',\hat U',\{\hat C_j'\})$ with no dependence on $\hat \Gamma_j'$'s, and $\hat C_k=\CT^{-1}_{C_k}(\{\hat C_j'\})$ with no dependence on $\hat v',\,\hat U',\,\hat \Gamma_j'$. Let $\hat \CI_M$ be the ideal obtained by reducing $\hat \CI_{(M,\Delta)}$ in unprimed variables, \ie\ by eliminating the $\hat \gamma_j$'s and setting $\hat C_j\to 2\pi i+\kappa_j\hbar$; and let $\hat \CI_M'$ be the corresponding ideal obtained by writing $\hat \CI_{(M,\Delta)}$ in primed variables and subsequently eliminating $\hat \Gamma_j'$'s and setting $\hat C_j'=2\pi i+\kappa_j'\hbar$, with
\be \kappa_k'\hbar := \CT_{C_k}(\hat C_1,...,\hat C_{N-1})\Big|_{\textstyle \big(\hat C_j=2\pi i+\kappa_j\hbar\;\;\forall\;j\big)}-2\pi i\,,\qquad k=1,...,N-1\,. \label{kkp} \ee
Then
\begin{proposition} \label{prop1} With $\CT$, $\CT^{-1}$, and ideals $\hat \CI_M$ and $\hat \CI_M'$ as above, if $\hat \CI_M$ is generated by $G$ polynomials $\big(\hat A_M^{(g)}(\hat\ell,\hat m^2;q)\big)_{g=1}^G$, then $\hat \CI_M'$ is generated by $G$ polynomials $\big(\hat A_M^{(g)}{}'(\hat \ell',\hat m^2{}';q)\big)_{g=1}^G$ with
\be \hat A_M^{(g)}{}'(\hat \ell',\hat m^2{}';q) := \hat A_M^{(g)}\big(\exp(\CT^{-1}_v(\hat v',\hat U',\{\hat C_j'\})),\,
 \exp(\CT^{-1}_U(\hat v',\hat U',\{\hat C_j'\}))\,;q\big)\Big|_{\hat C_j'\to2\pi i+\kappa_j'\hbar\;\;\forall\;j}\,. \label{AAp} \ee
\end{proposition}

We allowed the freedom to have more than one generator $\hat A^{(g)}$ of $\hat \CI_M$ because a priori it is not completely obvious that the result of eliminating variables in a $q$-commutative ring will yield just a single equation, and because the result with multiple generators extends immediately to link complements, or cases where even classically more than one ``A-polynomial'' is needed.

Although Proposition \ref{prop1} may have been difficult to state, it is quite easy to prove. One simply has to notice that due to the form of $\CT$ and $\CT^{-1}$ the operations of
\begin{itemize}
\item[a)] eliminating the $\hat \gamma_j$ from $\hat\CI_{(M,\Delta)}$ and then using $\CT^{-1}$ to rewrite the elimination ideal in terms of primed variables; and
\item[b)] rewriting $\hat\CI_{(M,\Delta)}$ in primed variables and then eliminating the $\hat \gamma_j'$
\end{itemize}
are interchangeable. The definition of $\kappa_j'$ in \eqref{kkp} then ensures that \eqref{AAp} holds. \\

Now, let's use Proposition \ref{prop1}. First suppose that in the process of calculating the ideal $\hat \CI_M$ for a triangulated manifold $M$, we fix a set of $N-1$ $\hat C_j$'s, and choose $N-1$ conjugate operators $\hat \Gamma_j$, $j=1,...,N-1$. Any two such choices $\{\hat \Gamma_j\}$ and $\{\hat \Gamma_j'\}$ are related by a transformation
\be \hat \Gamma_j  = \sum_{j'=1}^N B_{jk}\hat \Gamma_k' + t_j\,, \label{sympTG} \ee
where $B_{jk}$ is a constant $(N-1)\times(N-1)$ symmetric matrix, and $t_j$ are $(N-1)$ constant shifts. Relation \eqref{sympTG} extends to an affine symplectic transformation $\CT^{-1}$ that acts trivially on the rest of the operators. In particular, we can have $\hat v=\hat v'$ and $\hat U=\hat U'$. Then Proposition \ref{prop1} immediately implies that the generator(s) of $\CI_M$, \ie\ the $\hat A$-polynomial(s) of $M$, are independent of the choice of $\hat \Gamma_j$'s.

Similarly, we can tackle the issue of invariance under the choice of $N-1$ $\hat C_j$'s. Suppose, for example, that instead of the first $N-1$ $\hat C_j$'s, we want to choose the last $N-1$. Recalling that $\sum_{j=1}^N \hat C_j =2N\pi i+2Na\hbar$, and setting $(\hat C_1',\hat C_2',...,\hat C_{N-1}')=(\hat C_2,\hat C_3,...,\hat C_N)$, we see that
\be \begin{pmatrix} \hat C_1 \\ \hat C_2 \\ \vdots \\ \hat C_{N-1} \end{pmatrix} =
\begin{pmatrix} -1 & -1 & \cdots & -1 & -1\\
  1 & 0 & \cdots & 0 & 0 \\
  & & \ddots & & \\
  0 & 0 & \cdots & 1 & 0
\end{pmatrix}
\begin{pmatrix}\hat C_1' \\ \hat C_2' \\ \vdots \\ \hat C_{N-1}' \end{pmatrix} +
\begin{pmatrix} 2N\pi i+2Na\hbar \\ 0 \\ \vdots \\ 0\end{pmatrix}\,.
\label{sympTC}
\ee
Writing this in vector notation as $\hat{\mb{C}} = \mb A\,\hat{\mb{C}}'+\mb{t}$, we find that \eqref{sympTC} extends to an affine symplectic transformation $\CT^{-1}$ if it is accompanied by the transformation $\hat{\mb \Gamma} = \mb A^{-1}\,\hat{\mb \Gamma}'$ of the $N-1$ conjugate operators $\hat \Gamma_j$'s. The complete transformation $\CT$ again acts trivially on $\hat v$ and $\hat U$, and the natural specialization $\hat C_j'\to 2\pi i+\kappa_j'\hbar$ with
\be \kappa_j' = \kappa_{j+1}\,,\qquad j=1,...,N-1 \ee
agrees with the definition \eqref{kkp}. Then Proposition \ref{prop1} ensures that the results of calculating $\hat \CI_M$ with old gluing functions $\hat C_j$, or with the new gluing functions $\hat C_j'$, are identical. It is also clear that we could have repeated this argument with the $\hat C_j'$ being \emph{any} size-$(N-1)$ subset of $\{\hat C_j\}_{j=1}^N$, rather than $(\hat C_2,...,\hat C_{N})$, guaranteeing full quantum invariance under a choice of gluing operators.

\begin{figure}[htb]
\centering
\includegraphics[width=4in]{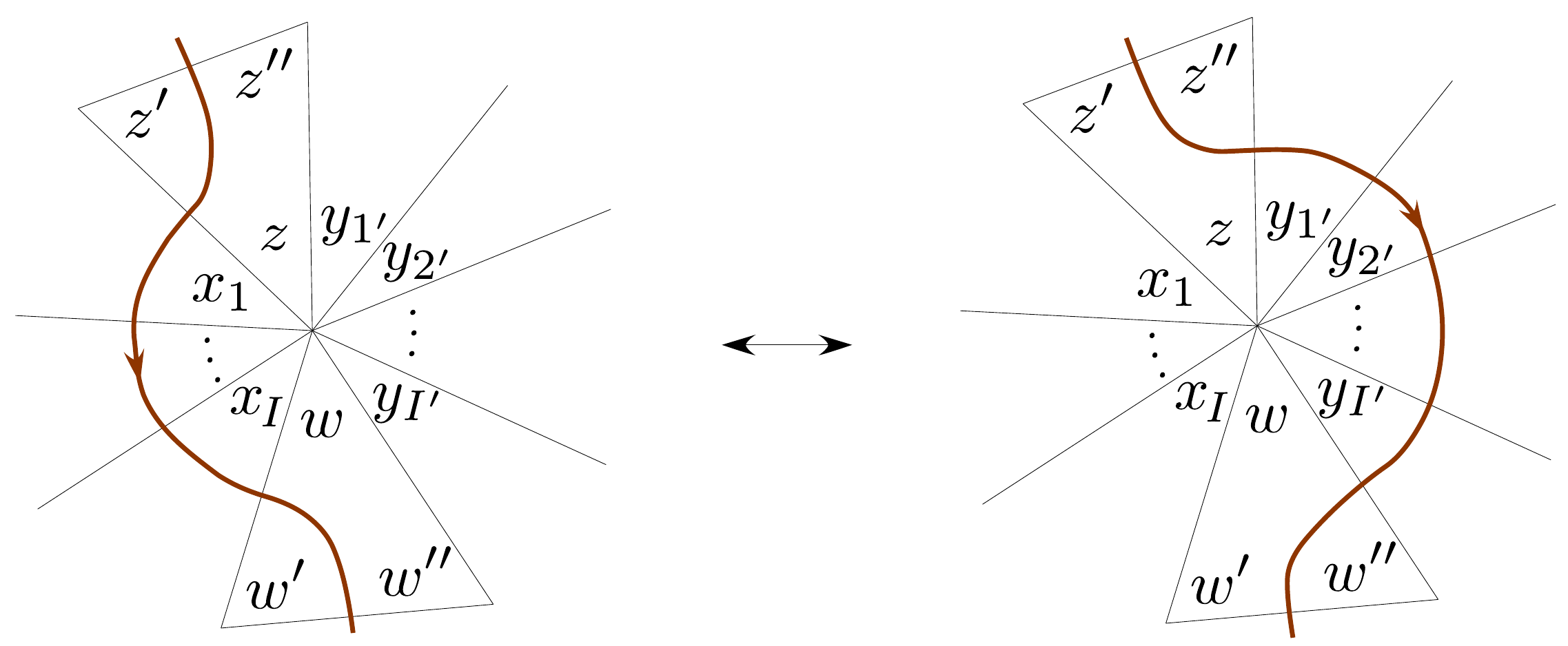}
\caption{The elementary move for a non-self-intersecting path}
\label{fig:pathelem2}
\end{figure}

For independence of meridian and longitude paths, we must be a little more careful. Let us assume that paths do not self-intersect, so that the relevant ``elementary move'' under which we need to check invariance is that of Figure \ref{fig:pathelem}(a), reproduced in Figure \ref{fig:pathelem2}.

Suppose that the path in question is the meridian. (It should be obvious how to repeat the following argument for the longitude.) Then, on the left side of Figure \ref{fig:pathelem2} the meridian operator is
\be \hat U = (\cdots)-\hat Z'+\sum_{i=1}^I\hat X_I-\hat W'\,, \ee
whereas on the right side the meridian is
\be \hat U' = (\cdots) +\hat Z''-\sum_{i'=1}^{I'}\hat Y_i+\hat W''\,, \ee
where $(\cdots)$ indicates the part of the logarithmic holonomy that is unchanged.
Using the vertex equations $\hat Z+\hat Z'+\hat Z''=i\pi+a\hbar$ and the definition of the gluing operator $\hat C_e = \hat Z+\hat W+\sum_i\hat X_i+\sum_{i'}\hat Y_{i'}$ at the central edge `$e$', we have
\be \hat U = \hat U'+\hat C_e-2\pi i-2a\hbar\,. \label{sympTU} \ee
Relation \eqref{sympTU} can be completed to an (inverse) affine symplectic transformation $\CT^{-1}$ as in \eqref{invAS} that acts trivially on $\hat v$ and all the $\hat C_j$'s (of which $\hat C_e$ is one), and defines
\be \hat \Gamma_e=\hat \Gamma_e'-\hat v\,,\qquad \text{and}
\qquad \hat \Gamma_j=\hat \Gamma_j'\quad (j\neq e)\,.\ee
Proposition \ref{prop1} then implies that the calculation of $\hat \CI_M$ is identical before or after the path-changing move as long as
\be \big(\hat U'+\hat C_e-2\pi i-2a\hbar\big)\big|_{\hat C=2\pi i+\kappa_e\hbar} = \hat U'\,.\ee
This requires that $\kappa_e = 2a$.

Since a path could be deformed through \emph{any} edge $e$ in the triangulation of $M$, we see that complete invariance of $\hat \CI_M$ under a choice of path is ensured if and only if
\be \boxed{\kappa_j = 2a\qquad \forall\;j=1,...N}\,.\label{qpathcond}\ee
Note that this is compatible with the constraint $\sum_{j=1}^N\kappa_j=2Na$ in \eqref{qKcons}.

\subsubsection{Cycles on $T^2$ and symmetries of $\hat A$.}
\label{sec:Asymm}

Proposition \ref{prop1} is also useful for understanding what happens when the homological class of ``longitude'' and ``meridian'' cycles on the boundary torus is changed. This is particularly relevant when $M=\ol{M}\bs K$ is the complement of a knot in a three-manifold $\ol{M}$ other than the three-sphere, since the ``longitude'' and ``meridian'' may not be canonically defined there. If we perform a classical $Sp(2,\Z)$ transformation on the boundary cycles $\lambda$ and $\mu$, so that
\be \begin{pmatrix} \lambda'\\ \mu' \end{pmatrix} = 
 \begin{pmatrix} a & b\\ c& d\end{pmatrix}
 \begin{pmatrix} \lambda \\ \mu \end{pmatrix}\,,\ee
then it follows that the corresponding quantum operators satisfy
\be \begin{pmatrix} \hat v'\\ \hat U' \end{pmatrix} = 
 \begin{pmatrix} a & b/2\\ 2c& d\end{pmatrix}
 \begin{pmatrix} \hat v \\ \hat U \end{pmatrix}\,. \ee
Proposition \ref{prop1} then implies that the generator(s) $\hat A^{(g)}(\hat \ell,\hat m^2;q)$ transform as
\begin{align} \hat A^{(g)}(\hat \ell,\hat m^2;q) \;\;\mapsto\;\; &\hat A^{(g)}(e^{d\hat v'-\frac b2\hat U'},e^{-2c\hat v'+a U'};q) \\ &= \hat A^{(g)}(q^{-\frac{bd}{4}}\hat m'{}^{-b}\hat\ell'{}^d,q^{-ac}\hat m'{}^{2a}\hat \ell'{}^{-2c};q)\,.
\nno \end{align}
Had we written the generators of $\hat\CI_M$ as functions of $\hat m$ rather than $\hat m^2$, this would have taken a more symmetric form
\be \hat A^{(g)}(\hat \ell,\hat m;q) \;\;\mapsto\;\;\hat A^{(g)}(q^{-\frac{bd}{4}}\hat m'{}^{-b}\hat\ell'{}^d,q^{-\frac{ac}{4}}\hat m'{}^{a}\hat \ell'{}^{-c};q)\,. \ee

In a similar way, modifications of Proposition \ref{prop1} can help us understand how symmetries of the classical A-polynomial extend to the quantum $\hat A$-polynomial(s). First, on any knot complement $M$ the A-polynomial (or rather, the ideal generated by it) is invariant if one replaces $(\ell,m)\mapsto (\ell^{-1},m^{-1})$.
Geometrically, this corresponds to reversing the orientations of both longitude and meridian cycles --- while preserving the overall orientation of the boundary $T^2$ --- and coincides with the action of the Weyl group $\Z_2$ on $\CP_{T^2}$. Since $(\hat v,\hat U)\mapsto(-\hat v,-\hat U)$ is a symplectic transformation, it is immediate from Proposition \ref{prop1} that the quantum $\hat A$-polynomial(s) transform as
\be \hat A^{(g)}(\hat\ell,\hat m^2;q) \mapsto \hat A^{(g)}(\hat\ell^{-1},\hat m^{-2};q)\,. \ee
We claim that in fact, as ideals,
\be \big(\hat A^{(g)}(\hat\ell,\hat m^2;q)\big)_{g=1}^G = \big(\hat A^{(g)}(\hat\ell^{-1},\hat m^{-2};q)\big)_{g=1}^G\,. \label{symML} \ee
One can show that this is indeed the case by examining the effect of the transformation $(\hat Z_i,\hat Z_i')\to(-\hat Z_i,-\hat Z_i')$ on the operator algebra for tetrahedra, though we omit further details for now. The quantum symmetry \eqref{symML} was discussed in \cite{DGLZ} from a more general point of view.

One can also consider a classical transformation $(\ell,m)\mapsto(\ell^{-1},m)$ or $(\ell,m)\to (\ell,m^{-1})$, which corresponds to reversing the orientation of the boundary $T^2$. For a knot complement $M=S^3\bs K$ in the three-sphere, this is the same as reversing the orientation of $M$.  Quantum mechanically, $(\hat v,\hat U)\mapsto(-\hat v,\hat U)$ or $(\hat v,-\hat U)$ are not quite symplectic transformations and must be accompanied by a change of sign $\hbar\to -\hbar$. Then, one can argue that the generators of ideals $\hat \CI_M,\, \hat \CI_{-M}$ for a knot complement $M$ in $S^3$ and its mirror image $-M$, respectively, are related as
\be \hat A_{-M}{}^{(g)}(\hat \ell,\hat m^2;q) = \hat A_M{}^{(g)}(\hat \ell^{-1},\hat m^2;q^{-1}) = \hat A_M{}^{(g)}(\hat \ell,\hat m^{-2};q^{-1})\,. \label{Amirror} \ee

\subsection{Independence of triangulation}
\label{sec:qPach}

Showing that the ideal ideal $\hat \CI_M$ in \eqref{finalqI} is independent of specific choices of ideal triangulation is the final ingredient in making $\hat \CI_M$, generated by the $\hat A$-polynomial(s) of $M$, a well-defined topological invariant. 
We will argue for this independence by quantizing the classical framework of Section \ref{sec:ind}, and arguing for local invariance under 2-3 Pachner moves. Just as in Section \ref{sec:ind} (and in particular Section \ref{sec:missing}), some subtle issues come into play when the triangulations involved in a Pachner move are especially degenerate. We will briefly discuss these issues at the end.

\begin{figure}[htb]
\centering
\includegraphics[width=5in]{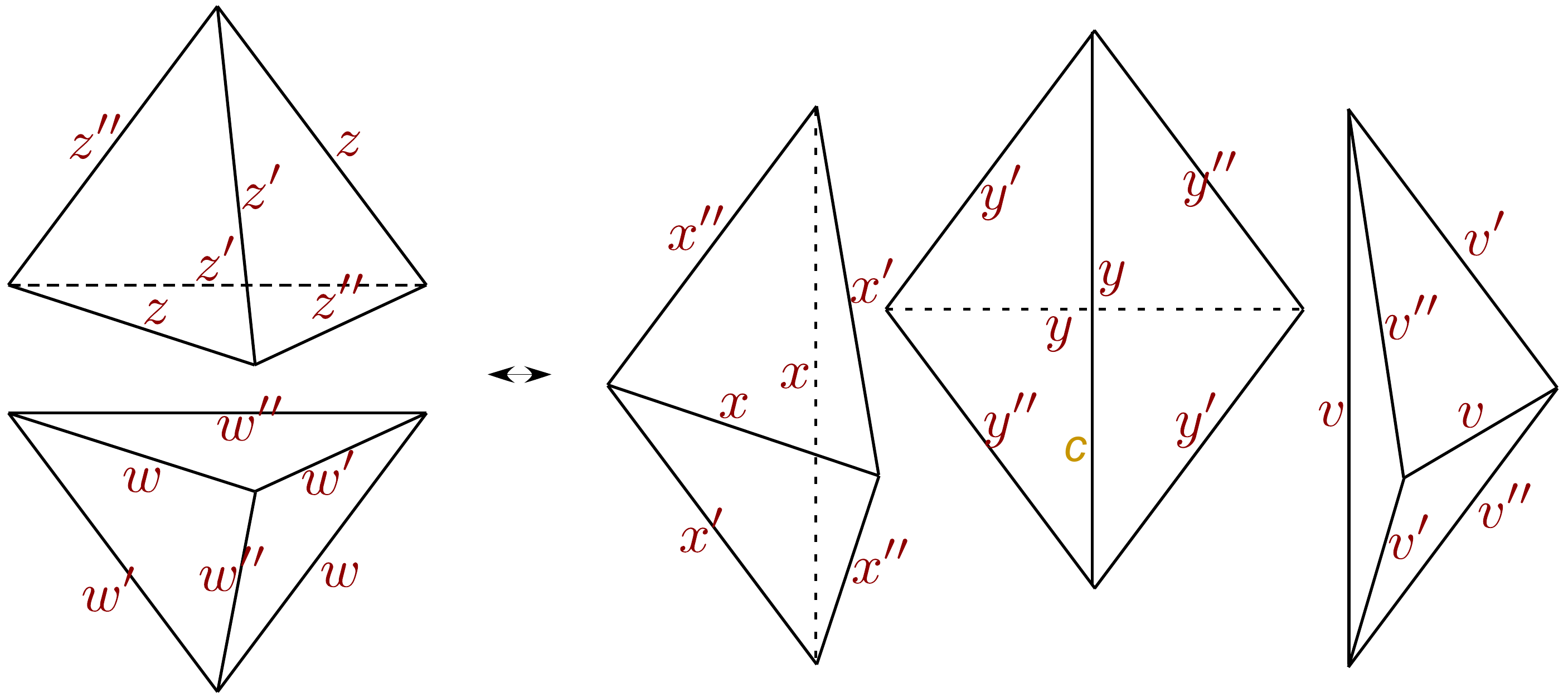}
\caption{Shape parameters for the 2-3 Pachner move}
\label{fig:Pachnerhyp2}
\end{figure}

We suppose then that $M$ has two different triangulations $\{\Delta_i\}_{i=1}^N$ and $\{\tilde\Delta_i\}_{i=1}^{N-1}$ that differ by a 2-3 Pachner move applied to tetrahedra
\be \Delta_{N-1}\,,\; \Delta_N\quad\leftrightarrow\quad \tilde\Delta_{N-1}\,,\; \tilde\Delta_{N}\,,\; \tilde\Delta_{N+1}\,. \ee
The shape parameters involved in this move are assigned as in Figure \ref{fig:Pachnerhyp}, repeated here in Figure \ref{fig:Pachnerhyp2}. The one additional edge in the $\{\tilde\Delta_i\}$ triangulation now corresponds to an operator
\be \hat C:=\hat{\tilde C}_{N+1}=\hat X+\hat Y+\hat V\,. \label{qCPach} \ee
We want to show that the product ideal $\hat \CI_{(M,\tilde\Delta)}$ for the $\{\tilde\Delta_i\}$ triangulation reduces to the product ideal $\hat \CI_{(M,\Delta)}$ (with gluing functions as appropriate for the $\{\Delta_i\}$ triangulation) after eliminating all elements that do not commute with $\hat C$, and setting
\be \hat C \to 2\pi i+\tilde\kappa_{N+1}\hbar\,, \ee
for a $\kappa_{N+1}$ that will be determined. We do \emph{not} immediately assume condition \eqref{qpathcond}, that is $\kappa_j\equiv 2a$, on either the $\kappa_j$ of the $\{\Delta_i\}$ triangulation or the $\tilde \kappa_j$ of the $\{\tilde\Delta_i\}$, so that we can find the most general ``quantum-correction'' structure allowed by the Pachner move.

For any edge except the (9+1) edges involved in the 2-3 move, it is clear that $\hat C_j=\hat{\tilde C}_j$, and that $\kappa_j$ should equal $\tilde \kappa_j$. In order for the remaining gluing operators of the $\{\Delta_i\}$ triangulation to agree with those obtained from the $\{\tilde\Delta_i\}$ triangulation after setting $\hat C \to 2\pi i +\tilde\kappa_{N+1}\hbar$, we require the six relations (\cf\ \eqref{eqs6})
\bse \label{qeqs6}
\begin{align} \hat Z = \hat V'+ \hat Y''+b\hbar\,,\quad \hat Z'= \hat X'+ \hat V''+b\hbar\,,\quad \hat Z''= \hat Y'+ \hat X''+b\hbar\,,\;\;\\ \hat W= \hat Y'+ \hat V''+b\hbar\,,\quad \hat W'= \hat X'+ \hat Y''+b\hbar\,,\quad \hat W''= \hat V'+ \hat X''+b\hbar\,.\end{align}
\ese
These can be taken as the definition of the 2-3 move at the level of operators. We can introduce a quantum correction $b\hbar$ to each of these linear equations, but must introduce the same correction to every equation since, a priori, there is nothing to distinguish one from another.%
\footnote{Even if one does try to introduce six different $b$'s here, invariance under the quantum Pachner move will force them to be equal.} %

Equations \eqref{qeqs6}, combined with the quantum vertex equations for $\hat X$, $\hat Y$, and $\hat Z$, imply the three additional relations (\cf\ \eqref{eqs3})
\bse \label{qeqs3}
\begin{align}
 \hat X &= \hat C-2\pi i-2a\hbar-2b\hbar+ \hat W+ \hat Z\,, \label{qXeq3}\\
 \hat Y &= \hat C-2\pi i-2a\hbar-2b\hbar+ \hat Z'+ \hat W''\,,\\
 \hat V &= \hat C-2\pi i-2a\hbar-2b\hbar+ \hat Z''+ \hat W'\,.
\end{align}
\ese
Just as in the semi-classical case, it is also clear from \eqref{qeqs6} that $\hat Z,\hat Z',\hat Z''$ and $\hat W,\hat W',\hat W''$ have the correct commutation relations with each other, and that all these operators commute with $\hat C$. Moreover, we find that
\be \hat Z+\hat Z'+\hat Z''=\hat W+\hat W'+\hat W''=3\pi i+3a\hbar +3b\hbar -\hat C\,. \ee
Upon setting $\hat C\to 2\pi i+\tilde\kappa_{N+1}\hbar$, these must reduce to the vertex equations for $\hat Z$ and $\hat W$, which fixes $\boxed{\tilde\kappa_{N+1}=2a+3b}$.

Together, Equations \eqref{qeqs6} and \eqref{qeqs3} determine the relation between the correction factors $\kappa_j$ and $\tilde\kappa_j$ on edges involved in the 2-3 Pachner move. A little thought shows that in going from $3\to 2$ tetrahedra, the quantum corrections on the six ``diagonal'' external edges of the 2-3 hexahedron \emph{increase} by $b\hbar$ while the corrections on the three ``equatorial'' external edges \emph{decrease} by $b\hbar$. The only possibility for assigning quantum edge corrections systematically to general triangulations in a way that is consistent with these $b$-shifts is to make $\kappa_j$ proportional to \emph{minus} the number of dihedral angles meeting edge $j$ ($\tilde j$). However, we saw above that a 2-3 Pachner move is only possible if the internal edge involved in the move has $\tilde\kappa_{N+1}=2a+3b$. Combining these restrictions, we conclude that for a 2-3 Pachner move with relations \eqref{qeqs6} to be possible at any place in any given triangulation the quantum edge corrections must be
\be \boxed{\kappa_j = 2a+(6-n_j)b\qquad \text{if $n_j$ dihedral angles meet edge $j$}}\,. \label{qPachedge} \ee

Now consider the wavefunction equations in $\hat \CI_{(M,\tilde\Delta)}$ corresponding to quantized Lagrangians for the three tetrahedra $\tilde\Delta_{N-1},\tilde\Delta_N,\tilde\Delta_{N+1}$. They can be written as
\bse \label{qLxyv}
\begin{align}
 \hat \CL_x=\alpha\hat x'+\alpha^{-1}\hat x''{}^{-1}-1 &\simeq 0\,, \\
 \hat \CL_y=\alpha\hat y'+\alpha^{-1}\hat y''{}^{-1}-1 &\simeq 0\,, \\
 \hat \CL_v=\alpha\hat v'+\alpha^{-1}\hat v''{}^{-1}-1 &\simeq 0\,.
\end{align}
\ese
We want to change variables from $(\hat X',\hat X'',\hat Y',\hat Y'',\hat V',\hat V'')$ to $(\hat Z'',\hat Z,\hat W'',\hat W,\hat\Gamma,\hat C)$, with
\be \hat \Gamma := \hat X''\,. \ee
In exponentiated variables, we want to replace $(\hat x',\hat x'',\hat y',\hat y'',\hat v',\hat v'')$ with monomials in the new exponentiated variables $(\hat z'',\hat z,\hat w'',\hat w,\hat\gamma,\hat c)$. This is easily done by inverting equations \eqref{qCPach} and \eqref{qeqs6} and subsequently exponentiating (\cf\ \eqref{xyvsols}) with the result that
\bse \label{qPachxyv}
\begin{align}
\hat{\CL}_x &= -\alpha q^{3a+2b-\frac12}\hat\gamma{}^{-1}\hat c{}^{-1}\hat z{}^{-1}\hat w{}^{-1}+\alpha^{-1}\hat\gamma{}^{-1}-1\simeq 0 \\
\hat{\CL}_y &=\alpha q^{-b}\hat\gamma{}^{-1}\hat z''+\alpha^{-1}\hat{\gamma}{}^{-1}\hat{z}{}^{-1}\hat w''-1\simeq 0 \\
\hat\CL_z &= \alpha q^{-b}\hat\gamma{}^{-1}\hat w''+\alpha^{-1}\hat\gamma{}^{-1}\hat z''\hat w{}^{-1}-1\simeq 0\,.
\end{align}
\ese

From the three equations \eqref{qPachxyv}, one can eliminate $\hat\gamma$ by first multiplying by $\hat\gamma$ on the left and then taking the difference of any two equations. After some slight manipulation and setting $\hat c\to q^{\tilde\kappa_{N+1}}= q^{2a+3b}$ (which we are allowed to do after eliminating $\hat\gamma$), the two independent equations that remain are
\bse
\begin{align} \hat E_z &:= \alpha^{-1}\hat z\hat w-\alpha^{-1}\hat z\hat z''-\alpha q^{-\frac12+a-b}-\alpha q^{-b}\hat w\hat z{}^2+\alpha^3 q^{-\frac12+a-2b}\hat z+\alpha^3q^{-2b}\hat z{}^2\hat z''\hat w \simeq 0\,, \nno \\
\hat E_w &:= \alpha^{-1}\hat z\hat w-\alpha^{-1}\hat w\hat w''-\alpha q^{-\frac12+a-b}-\alpha q^{-b}\hat z\hat w{}^2+\alpha^3q^{-\frac12+a-2b}\hat w+\alpha^3q^{-2b}\hat z\hat w{}^2\hat w''\simeq 0\,. \nno
\end{align}
\ese
In the semi-classical setting, we saw that these equations had to factor in order to yield the Lagrangian equations for $z$ and $w$. Here, the operators $\hat E_z$ and $\hat E_w$ factor in the noncommutative ring $\C(q)[\hat z''{}^{\pm 1},\hat z{}^{\pm 1},\hat w''{}^{\pm 1},\hat w{}^{\pm 1},...]$ if and only if
\be \boxed{b=\frac{1-2a}{6}}\,. \label{qPachb} \ee
Recalling that $\alpha=q^{\frac{1-2a}{6}}$, this means that $\boxed{q^b=\alpha}$\,. Then we find
\bse \label{Ezw23}
\begin{align}
 \hat E_z &= -(\alpha^{-2}-\hat z\hat w)\hat z\, (\alpha \hat z''+\alpha^{-1}\hat z^{-1}-1)\simeq 0\,,\\
 \hat E_w &= -(\alpha^{-2}-\hat z\hat w)\hat w\, (\alpha \hat w''+\alpha^{-1}\hat w^{-1}-1) \simeq 0\,.
\end{align}
\ese
Just as we were not allowed to have $zw=1$ classically, the operator $(\alpha^{-2}-\hat z\hat w)$ must be invertible. The reasoning is that the ideal $\hat \CI_{(M,\tilde\Delta)}$ was defined in the ring
\begin{align} \hat \CR_{xyv} &= \C(q)[\hat x'{}^({}'{}^){}^{\pm 1},\hat y'{}^({}'{}^){}^{\pm 1},\hat v'{}^({}'{}^){}^{\pm 1},(1-\hat x'{}^({}'{}^))^{- 1},(1-\hat y'{}^({}'{}^))^{- 1},(1-\hat v'{}^({}'{}^))^{- 1}, \\
&\hspace{2in} (\hat x'\hat x''+1)^{-1},\,(\hat y'\hat y''+1)^{-1},\,(\hat v'\hat v''+1)^{-1}]\,, \nno
\end{align}
in which the element $(1-\hat x)=(1+q^{a+1/2}\hat x''{}^{-1}\hat x'{}^{-1})$ is invertible. Due to relation \eqref{qXeq3}, $(q^{{\rm any}\,\#}-\hat z\hat w)$ must therefore be invertible in the reduced ring $\hat \CR_{zw}^{3\to2}$ containing the ideal generated by \eqref{Ezw23}.

In the end, after removing the factors $(\alpha^{-2}-\hat z\hat w)$ from $\hat E_z$ and $\hat E_w$ in \eqref{Ezw23}, we find that the ideal $\hat \CI_{(M,\tilde\Delta)}$ on the right side of the 2-3 Pachner move has reduced to the ideal $\hat \CI_{(M,\Delta)}$ on the left.
While conditions \eqref{qPachedge} and \eqref{qPachb} are the only ones necessary for the quantum version of the 2-3 Pachner move to hold, we know from \eqref{qpathcond} of Section \ref{sec:qpath} that invariance of $\hat\CI_M$ under longitude and meridian path deformations also requires $\kappa_j\equiv 2a$. Combining these restrictions together, we would like to claim that

\begin{conjecture} \label{thm1} The construction of the ideal $\hat\CI_M$ as defined in Section \ref{sec:qglue} for any triangulation $\{\Delta_i\}_{i=1}^N$ of a knot complement $M$, with
\be \boxed{a=\frac12\qquad\text{and}\qquad \kappa_j=1\,\quad j=1,...,N}\,, \label{akfixed} \ee
produces a topological invariant of $M$, independent of the actual triangulation used or any other choices made.
\end{conjecture}

Unfortunately, just as in the classical scenario of Section \ref{sec:missing}, this Conjecture may not be entirely true for certain ``bad'' triangulations of $M$. The subtle problem is that, while the ideal $\hat\CI_{(M,\tilde\Delta)}$ naively reduces to 
the ideal $\hat\CI_{(M,\Delta)}$ as desired under the $3\to2$ Pachner move, the reduced ring
\be \hat\CR_{zw}^{3\to2}=\C(q)[\hat z{}^({}''{}^){}^{\pm 1},\,\hat w{}^({}''{}^){}^{\pm 1},\,
(1-\hat z\hat w)^{-1},\,(\hat w\hat w''+\hat z'')^{-1},\,(\hat z\hat z''+\hat w'')^{-1}]  \ee
is not quite the same as the expected ring
\be \hat\CR_{zw} = \C(q)[\hat z{}^({}''{}^){}^{\pm 1},\,\hat w{}^({}''{}^){}^{\pm 1},\,(1-\hat z{}^({}''{}^))^{\pm 1},\,(1-\hat w{}^({}''{}^))^{\pm 1},\,(\hat z\hat z''+1)^{-1},\,(\hat w\hat w''+1)^{-1}] \,.\ee
In particular, the two rings have very slightly different invertible elements. We certainly do expect that, at the special values \eqref{akfixed} for $a$ and $\kappa_j$, the ideal $\hat\CI_M$ containing the $\hat A$-polynomial(s) of $M$ is invariant under Pachner moves whenever the corresponding classical gluing variety $A(\ell,m^2)$ is invariant. In the exceptional cases where the classical $A(\ell,m^2)$ gains or loses factors, it is possible that \emph{left} factors could appear or disappear from the generator(s) of $\hat\CI_M$.
We would then say, as at the end of Section \ref{sec:missing}, that the ideal $\hat\CI_M$ is really an invariant of a pair $(M,[\Delta])$, consisting of a three-manifold and a triangulation class. \\

We note that all the arguments and constructions of this section and previous ones extend fairly trivially from knot to link complements. In the case of a link complement with $\nu$ components, one would construct an ideal $\hat\CI_M$ that has at least $G\geq\nu$ generators $\hat A_{M}^{(g)}(\hat\ell_1,\hat m_1,...,\hat\ell_\nu,\hat m_\nu)$. This ideal is a topological invariant to the same extent that it would be for a knot complement --- \ie\ perhaps an invariant of some finite number of pairs $(M,[\Delta])$ rather than just three-manifolds $M$.

\subsection{Examples}
\label{sec:ex}

We end this section with several explicit computations of the ideal $\hat \CI_M$ for knot complements $M=S^3\bs K$, where $K$ is the figure-eight knot, the trefoil, and the $\mb{5_2}$ knot. In all these examples, we find that the ideal $\hat\CI_M$ has a unique generator $\hat A(\hat\ell,\hat m^2;q)$, which is identical to the quantized nonabelian $\hat A$-polynomial known previously to produce recursion relations for colored Jones polynomials.

\subsubsection{Figure-eight knot $\mb{4_1}$}
\label{sec:ex41}

For the complement of the figure-eight knot in $S^3$, the generators the generators $\hat\CL_{\Delta_i}$ of the ideal $\hat \CI_{(\mb{4_1},\Delta)}$, for a two-tetrahedron triangulation, were determined in \eqref{ideal41}. Let us define
\be \boxed{\hat M := \hat m^2}\,. \ee
After multiplying $\hat \CL_z$ and $\hat \CL_w$ from \eqref{ideal41} by appropriate factors to clear negative exponents, and using $a=1/2$ as determined in Theorem \ref{thm1}, we then find
\bse \label{ideal412}
\begin{align}
\hat E_{z} &:= \hat M\hat \gamma \hat\CL_{\Delta_z} = 1-\hat M\hat\gamma-q^{\frac12}\hat M{}^2\hat\ell\hat\gamma{}^2 \simeq 0\,,\\
\hat E_{w} &:= \hat c\,\hat\ell\hat \gamma \hat\CL_{\Delta_w}= -q^{\frac12}-\hat c\hat L\hat \gamma+\hat c \hat \ell\hat \gamma{}^2 \simeq 0\,.
\end{align}
\ese

To eliminate $\hat\gamma^2$ from these equations, we form the combination
\be \hat E_{1,1} := \hat c\,\hat\ell\hat E_z+q^{\frac52}\hat M{}^2\hat\ell \hat E_w = -\hat c+q\hat M{}^2+\hat c\hat M\hat \gamma+q^{\frac12}\hat c\hat M{}^2\hat \ell\hat\gamma \simeq 0\,.\ee
In order to eliminate the linear term in $\hat\gamma$, we need a second independent linear equation. We can take it to be
\begin{align} &\hat E_{1,2} := \hat M\hat \ell\hat\gamma \hat E_{1,2}+q\hat c(\hat M\hat \ell+q^{\frac12})\hat E_z \\
&\qquad = -\hat c\hat M\hat\ell-q^{\frac12}\hat c+\hat c\hat M\hat \ell\hat\gamma+q^{\frac 12}\hat c\hat M\hat \gamma+q\hat c\hat M{}^2\hat \ell\hat \gamma-q^2\hat M{}^3\hat\ell\hat\gamma\simeq 0\,. \nno
\end{align}

Now, if these were classical equations, we could simply solve for $\hat\gamma$ using $\hat E_{1,1}= 0$, and substitute the result into $\hat E_{1,2}=0$ to get the A-polynomial. The quantum story is more complicated, but it turns out that it is always possible to find \emph{some} polynomial functions $\hat{f}_1(\hat\ell,\hat M,\hat c)$ and $\hat{f}_2(\hat \ell,\hat M,\hat c)$ so that the linear term in $\hat\gamma$ is eliminated from $\hat f_1\hat E_{1,1}-\hat f_2\hat E_{1,2}$. The fact that we are working in a noncommutative ring causes the degrees of $\hat{f}_1$ and $\hat{f}_2$ in $\hat M$ to be higher than expected, and ultimately leads to extra factors of the form $(1-q^{\#}\hat M^2)$ multiplying each term of the $\hat A$-polynomial. (Classically, these factors could be removed completely.) Here, we find that
\bse
\begin{align}
\hat f_1 &= -\hat c^2\hat\ell-q^{\frac32}\hat c^2-q^2\hat c^2\hat M\hat\ell+q^2\hat c\hat M^2\hat\ell+q^4\hat c\hat M^2\hat \ell+q^4\hat c\hat M^3\hat \ell+q^{\frac{11}{2}}\hat c\hat M^2-q^6\hat M^4\hat\ell\,,\\
\hat f_2 &= -q\hat c^2-q^{\frac32}\hat c^2\hat M\hat \ell+q^{\frac72}\hat c\hat M^3\hat \ell+q^5\hat c\hat M^2
\end{align}
\ese
will do the job.
Then
\begin{align}
\hat \CL_{\mb{4_1}}&:= q^{-3}(\hat f_1\hat E_{1,1}-\hat f_2\hat E_{1,2}) \\
&= q^{-\frac12}\hat c(\hat c-q^4\hat M^2)\hat M^2 \nno \\
&\qquad -(\hat c-q^3\hat M^2)(\hat c^2-q \hat c^2\hat M-(q^2+q^4)\hat c\hat M^2-q^4\hat c\hat M^3+q^6\hat M^4)\hat\ell \nno \\
&\qquad +q^{\frac12}\hat c^2(\hat c-q^2\hat M^2)\hat M^2\hat \ell^2 \nno \\
&\simeq 0\nno
\end{align}
is the generator of the elimination ideal $\hat \CJ_{\mb{4_1}}$.
Setting $\hat c\to q^{\kappa}=q$ now that everything commutes with $\hat c$, we obtain the actual $\hat A$-polynomial that generates $\hat \CI_{\mb{4_1}}$, namely
\begin{align} \hat A_{\mb{4_1}}(\hat \ell,\hat M;q) &= \hat \CL_{\mb{4_1}}\big|_{\hat c\to q} \nno \\
&= q^{\frac32}(1-q^3\hat M^2)\hat M^2-(1-q^2\hat M^2)(1-q\hat M-(q+q^3)\hat M^2-q^3\hat M^3+q^4\hat M^4)\hat\ell \nno \\
 &\quad +q^{\frac52}(1-q\hat M^2)\hat M^2\hat\ell^2\,. 
\label{Ah41}
\end{align}
With the substitution $\hat M=\hat m^2$, this is the $\hat A$-polynomial \eqref{Ah41eg} that we quoted in the introduction.

Note that \eqref{Ah41} is the quantum $\hat A$-polynomial that corresponds to a natural physical normalization for Chern-Simons partition functions \cite{DGLZ}. In particular, $\hat A_{\mb {4_1}}$ leads to a recursion relation for the colored Jones polynomials $J_N(K;q)$, defined as the $SU(2)$ Chern-Simons expectation values of a Wilson loop in the three-sphere,
\be J_N(K;q) := \frac{1}{Z_{CS}^{SU(2)}(S^3)}Z_{CS}^{SU(2)}(K_N\subset S^3)\,. \label{physJ} \ee
In contrast, the colored Jones polynomials $V_N(K;q)$ appearing in the mathematical literature are usually normalized by the expectation value of the unknot $\mb{U}$,
\be V_N(K;q) = \frac{ J_N(K;q)}{J_N(\mb{U};q)}\,. \label{mathJ} \ee
Following the dictionary of Section \ref{sec:recursion}, this means that the operator actually featuring on the LHS of the recursion relation for $V_N(K;q)$ for the figure-eight knot (or any other knot) is
\be \hat A_{V,\mb{4_1}}(\hat \ell,\hat M;q) = \hat A_{\mb{4_1}}(\hat \ell,\hat M;q)(\hat M{}^{1/2}-\hat M{}^{-1/2})\,. \label{Anormd} \ee
After multiplying through on the left by $\hat M{}^{1/2}$ to cancel negative powers of $\hat M$, and commuting all $\hat M$'s to the left of all $\hat\ell$'s, it is easy to see that $\hat A_{V,\mb{4_1}}(\hat\ell,\hat M;q)$ is the same operator appearing in (\eg) \cite{Gar-Le, Gar-twist} on the left-hand side of recursion relations.

Note that since the figure-eight knot is amphicheiral, \ie\ it is homeomorphic to its mirror image, the $\hat A$-polynomial \eqref{Ah41} is invariant under the orientation-reversing map \eqref{Amirror}.

\subsubsection{The trefoil $\mb{3_1}$}
\label{sec:ex31}

Now consider the complement of the trefoil knot in $S^3$. This is a very simple example, but it illustrates several important features which persist generically for more complicated knots.

From the point of view of hyperbolic geometry the trefoil is often treated specially because, like all torus knots, it does not admit a complete hyperbolic metric. As discussed in Section \ref{sec:classrem}, this makes no difference whatsoever for us.

An ideal triangulation for the trefoil is readily computed in the program \texttt{SnapPy} \cite{SnapPy} or \texttt{snap} \cite{snap}.\footnote{It is also readily computable by hand. We heartily thank Bus Jaco for first showing this to us.} The minimal triangulation has two tetrahedra, which we will call $\Delta_z$ and $\Delta_w$. The (quantum) gluing functions and boundary holonomies are
\bse \label{UC31}
\begin{align}
\hat C_1 &= \hat Z'+\hat W'\,, \\
\hat C_2 &= 2\hat Z+\hat Z'+2\hat Z''+2\hat W+\hat W'+2\hat W'' = 4\pi i+2\hbar-\hat Z'-\hat W'\,, \\
\hat U &= -\hat Z+\hat W \,, \\
\hat V &= -4\hat Z+\hat Z'+4\hat W-\hat W' \,,
\end{align}
\ese
where we have used the quantum vertex equation $\hat W+\hat W'+\hat W''=i\pi+\hbar/2$ to eliminate $W''$ from $C_2$.
We lift from a $PSL(2,\C)$ connection to a $SL(2,\C)$ connection by choosing the square root
\be \hat v = -2\hat Z+\frac12 \hat Z'+2\hat W-\frac12 \hat W'+i\pi\,. \ee
Factors of $1/2$ are unavoidable here, as they are for many more complicated knots.

As expected, the gluing functions are not independent because $\hat C_1+\hat C_2=4\pi i+2\hbar$. Throwing away $\hat C_2$, we can choose the operator conjugate to $\hat C:=\hat C_1$ to be
\be \hat \Gamma = \frac12 \hat Z+\frac12 \hat W\,. \label{G31} \ee
Solving \eqref{UC31} and \eqref{G31} for the old operators $\hat Z,\hat Z',\hat W,\hat W'$, we find
\bse
\begin{align}
\hat Z = -\frac{\hat U}{2}+\Gamma\,,\qquad \hat Z'=\hat v-2\hat U+\frac{C}{2}+i\pi\,,\\
\hat W = \frac{\hat U}{2}+\Gamma\,,\qquad \hat W' = -\hat v+2\hat U+\frac{C}{2}-i\pi\,.
\end{align}
\ese
Quantizing and exponentiating, the wavefunction equations generating the left ideal $\hat\CI_{(\mb{3_1},\Delta)}$ then become
\bse
\begin{align}
\hat\CL_z = \hat m^{-1}\hat\gamma-q^{-1}\hat c^{-\frac12}\hat m^4\hat\ell^{-1}-1 \simeq 0\,,\\
\hat\CL_w = \hat m\hat\gamma -q^{-1}\hat c^{-\frac12}\hat m^{-4}\hat\ell-1\simeq 0\,.
\end{align}
\ese

Eliminating $\hat\gamma$ from these equations is trivial. We find
\begin{align} \hat\CL_{\mb{3_1}}& := q^3\hat c^{\frac12}\hat m^4\hat\ell(\hat m^2 \hat\CL_z-\hat\CL_w) \nno \\
&= q^5\hat m^{10}-q^3\hat c^{\frac12}\hat m^4(1-q\hat m^2)\hat\ell-\hat\ell^2 \simeq 0\,. \label{Lc31}
\end{align}
The usual specialization of the logarithmic gluing constraint is $\hat C\to 2\pi i+\hbar$. While this means that $\hat c\to q$, it also implies that we should set $\hat c^{\frac12}\to e^{i\pi+\hbar/2}=-q^{\frac12}$ rather than $\hat c^{\frac12}\to +q^{\frac12}$. Implementing this substitution, then, we obtain
\be \hat \CL_{\mb{3_1}}\big|_{\hat c^{\frac12}\to -q^{\frac12}} = q^5\hat m^{10}+q^{\frac72}\hat m^4(1-q\hat m^2)\hat\ell-\hat\ell^2\,. \label{preA31} \ee
The actual $\hat A$-polynomial for the trefoil, known from recursion relations for the colored Jones (\cf\ \cite{Gar-Le}), is first order in $\hat \ell$, so this is evidently not it. However, the operator \eqref{preA31} factors as
\be \hat\CL_{\mb{3_1}}\big|_{\hat c^{\frac12}\to -q^{\frac12}} = (q^{\frac72}\hat m^4-\hat\ell)(\hat \ell+q^{\frac32}\hat m^6)\,, \label{fact31} \ee
and the factor on the right is precisely the desired quantum $\hat A$-polynomial.%
\footnote{See the comments surrounding \eqref{Anormd} in order to properly compare \eqref{fact31} to the operator $\hat A_{V,\mb{3_1}}$ appearing in the mathematical recursion relation \cite{Gar-Le} for the colored Jones polynomials of the trefoil. The $\hat A$-polynomial for the trefoil knot complement with the opposite orientation can also be computed, as described in Section \ref{sec:Asymm}.} %

In the classical limit $q\to 1$, the factor on the left of \eqref{fact31} becomes $(m^4-\ell)$, and if this vanished it would imply that $z'=w'=1$, which is not an acceptable point in the classical moduli space of tetrahedra. Quantum mechanically, this statement translated to the fact that due to the invertible elements in the original ring $\C(q)[\hat z{}^({}'{}^){}^{\pm 1},(1-\hat z{}^({}'{}^))^{-1},\hat w{}^({}'{}^){}^{\pm 1},(1-\hat w{}^({}'{}^))^{-1},(1+\hat z\hat z')^{-1},(1+\hat w,\hat w')^{-1}]$ containing $\hat \CI_{(\mb{3_1},\Delta)}$, any quantum operator $(q^{\#}\hat m^4-q^{\#}\hat\ell)$ is invertible in the reduced ring containing $\hat \CI_{\mb{3_1}}$. Therefore, the generator of $\hat \CI_{\mb{3_1}}$ is really
\be \hat{A}_{\mb{3_1}}(\hat\ell,\hat m^2;q) = (q^{\frac72}\hat m^4-\hat\ell)^{-1}\hat \CL_{\mb{3_1}}\big|_{\hat c\to q} = \hat \ell+q^{\frac32}\hat m^6 = \hat \ell+q^{\frac32}\hat M^3\,.\ee

\subsubsection{The knot $\mb{5_2}$}

The minimal ideal triangulation of the $\mb{5_2}$ knot complement in $S^3$ requires three tetrahedra. From \texttt{SnapPy} \cite{SnapPy}, we find
\bse \label{uC52}
\begin{align}
\hat C_1 &= 2\hat Z_1+\hat Z_2+\hat Z_2''+\hat Z_3'+\hat Z_3'' = 2\pi i+\hbar +2\hat Z_1-\hat Z_2'-\hat Z_3\,, \\
\hat  C_2 &= \hat Z_1'+\hat Z_2'+\hat Z_2''+\hat Z_3+\hat Z_3' = i\pi +\frac12\hbar+\hat Z_1'-\hat Z_2+\hat Z_3+\hat Z_3'\,, \\
\hat  C_3 &= \hat Z_1'+2\hat Z_1''+\hat Z_2+\hat Z_2'+\hat Z_3+\hat Z_3''=3\pi i+\frac{3}{2}\hbar-2\hat Z_1-\hat Z_1'+\hat Z_2+\hat Z_2'-\hat Z_3'\,, \\
\hat  U &= \hat Z_1-\hat Z_1'+\hat Z_2-\hat Z_2'+\hat Z_3'' = i\pi+\frac12\hbar + \hat Z_1-\hat Z_1'+\hat Z_2-\hat Z_2'-\hat Z_3-\hat Z_3'\,, \\
\hat V &= -2\hat Z_1+2\hat Z_1'-3\hat Z_2+3\hat Z_2'+\hat Z_2''-\hat Z_3-\hat Z_3'-\hat Z_3''=2(-\hat Z_1+\hat Z_1'-2\hat Z_2+\hat Z_2')\,.
\end{align}
\ese
After the second equal sign in each of these expressions, we have used the semi-classical vertex equations to eliminate $Z_1'',\,Z_2'',$ and $Z_3''$. We also take a square root of $\ell^2$ by defining
\be \hat v = i\pi -\hat Z_1+\hat Z_1'-2\hat Z_2+\hat Z_2'\,, \ee
with an extra $i\pi$. One can check that these functions on $\CP_{(\mb{5_2},\Delta)}$ all have the appropriate Poisson brackets.

Since the three gluing functions have a linear dependency $\hat C_1+\hat C_2+\hat C_3=6\pi i+3\hbar$, we can ignore $\hat C_3$. We want to use $C_1$ and $C_2$ as semi-classical moment maps, eventually setting them both equal to $2\pi i+\hbar$ quantum mechanically. Conjugate operators $\hat \Gamma_j$ are given by
\bse \label{G52}
\begin{align}
\hat \Gamma_1 = \hat Z_3+\hat Z_3' \\
\hat \Gamma_2 = \hat Z_1-\hat Z_2-\hat Z_3-\hat Z_3'\,.
\end{align}
\ese
We can now invert equations \eqref{uC52} and \eqref{G52} to express $(\hat U,\hat v,\hat C_j,\hat \Gamma_j)$ (for $j=1,2$) in terms of the $\hat Z$'s. We find
\bse
\begin{align}
\hat Z_1 &= 2\pi i+\frac12\hbar -\hat U-\hat v+\hat \Gamma_2\,,\qquad
\hat Z_1' = i\pi-\hat U-\hat v+C_2-2\hat \Gamma_1\,,\\
\hat Z_2 &= 2\pi i+\frac12\hbar-\hat U-\hat v-\hat \Gamma_1\,,\qquad
\hat Z_2' = 4\pi i+\frac{3}{2}\hbar-2\hat U-\hat v-\hat C_2+\hat \Gamma_2 \\
\hat Z_3 &= 2\pi+\frac12\hbar -\hat v-\hat C_1+\hat C_2+\hat \Gamma_2\,,\qquad
\hat Z_3' = -2\pi i-\frac12\hbar +\hat v+\hat C_1-\hat C_2+\hat \Gamma_1-\hat \Gamma_2\,.
\end{align}
\ese
Exponentiating these expressions and plugging them in to the three equations $\hat \CL_{\Delta_i}\simeq 0$, we arrive at the ideal $\hat \CI_{(\mb{5_2},\Delta)}$, written as
\bse
\begin{align}
\hat\CL_{\Delta_1}&=q\hat M^{-1} \hat\ell^{-1}\hat\gamma_2-q^{\frac12}\hat c_2{}^{-1}\hat M\hat\ell\hat\gamma_1{}^2-1\simeq 0\,,\\
\hat\CL_{\Delta_2}&=q\hat M^{-1}\hat\ell^{-1}\hat\gamma_1{}^{-1}+q^{-1}\hat c_2\hat M^2\hat\ell\hat\gamma_2{}^{-1}-1\simeq 0\,,\\
\hat\CL_{\Delta_3}&=q\hat c_1{}^{-1}\hat c_2\hat\ell^{-1}\hat\gamma_2+q^{\frac32}\hat c_1{}^{-1}\hat c_2\hat\ell^{-1}\hat \gamma_1{}^{-1}\hat\gamma_2-1\simeq 0\,,
\end{align}
\ese
where again
\be \hat M = \hat m^2\,.\ee

Eliminating $\hat\gamma_2$ from $\hat \CI_{(\mb{5_2},\Delta)}$ is easy. Upon multiplying $\hat\CL_{\Delta_1}\simeq 0$ by $\hat M\hat\ell$, we get
\be \hat\gamma_2-\hat M\hat\ell-q^{\frac32}\hat c_2{}^{-1}\hat M^2\hat\ell^2\hat\gamma_1{}^2 \simeq 0\,, \ee
and we can use this to substitute for $\hat\gamma_2$ in the other two equations. Note, however that since we are working in a left ideal such substitutions must always happen on the \emph{right}. In other words, we must bring all factors of $\hat\gamma_2$ to the right before setting
\be \hat\gamma_2 \to \hat M\hat\ell+q^{\frac32}\hat c_2{}^{-1}\hat M^2\hat\ell^2\hat\gamma_1{}^2\,.\ee
Doing this substitution in $\hat\CL_{\Delta_2}$ and $\hat\CL_{\Delta_3}$ and clearing denominators (by multiplying on the left) results in the respective equations
\bse
\begin{align}
\hat E_a &:= \hat c_2-\hat c_2\hat M\hat\ell\hat\gamma_1+\hat c_2^2\hat M^2\hat\ell\hat\gamma_1+q^{\frac12}\hat M\hat\ell\hat\gamma_1{}^2-q^{\frac32}\hat M^2\hat\ell^2\hat\gamma_1{}^3 \simeq 0 \\
\hat E_b &:= q^{\frac12}\hat c_2\hat M+\hat c_2\hat M\hat\gamma_1-q\hat c_1 \hat\gamma_1+q\hat M^2\hat\ell\hat\gamma_1{}^2+q^{\frac12}\hat M^2\hat\ell\hat\gamma_1{}^3\simeq 0\,.
\end{align}
\ese
Since $\hat\gamma_2$ no longer appears, we could set $\hat c_2\to q$ at this stage.

From here, the process of eliminating $\hat\gamma_1$ is essentially the same as in the case of the figure-eight knot. We proceed to eliminate first $\hat\gamma_1{}^3$ terms, then $\hat\gamma_1{}^2$ terms, then linear $\hat\gamma_1$ terms, making sure to generate enough independent equations at each step to do this. In the end, we find a minimal generator $\hat\CL_{\mb{5_2}}$ of the elimination ideal $\hat \CJ_{(\mb{5_2})}$ and set $\hat c_1\to q$ to get the generator of $\hat \CI_{(\mb{5_2})}$. The result is
\begin{align}
\hat{A}_{\mb{5_2}}(\hat\ell,\hat M;q) &= q^{\frac12}(1-q^4\hat M^2)(1-q^5\hat M^2) \\
&\quad - (1-q^2\hat M^2)(1-q^5\hat M^2)(1-2q\hat M-q(q+q^3)\hat M^2 \nno \\&\hspace{2in} + q^2(1-q)(1-q^2)\hat M^3+q^5\hat M^4-q^6\hat M^5)\hat\ell \nno \\
&\quad q^{\frac52}(1-q\hat M^2)(1-q^4\hat M^2)\hat M^2(1-q^2\hat M-q^2(1-q)(1-q^2)\hat M^2 \nno \\&\hspace{2.5in}+q^4(1+q^3)\hat M^3+2q^7\hat M^4-q^9\hat M^5)\hat\ell^2 \nno \\
&\quad q^{14}(1-q\hat M^2)(1-q^2\hat M^2)\hat M^7\hat\ell^3\,.\nno
\end{align}
In the classical commutative limit $q\to 1$, this becomes $(1-M^2)^2$ times the classical nonabelian A-polynomial,
\be A_{\mb{5_2}}(\ell,M) = 1-(1-2M-2M^2+M^4+M^5)\ell+M^2(1-M+2M^3+2M^4-M^5)\ell^2+M^7\ell^3\,,\ee
where $M=m^2$. Note how, in addition to the extra factors $(1-q^{\#}\hat M^2)(1-q^{\#}\hat M^2)$, the noncommutative $\hat A$-polynomial actually contains extra terms --- such as $\hat M^3\hat\ell$ and $\hat M^4\hat\ell^2$ --- whose coefficients completely vanish in the classical limit.

To compare this result with recursion relations in the knot theory literature \cite{Gar-twist}, it is again necessary to renormalize $\hat A_{\mb {5_2}}\to \hat A_{V,\mb{5_2}}$ as in \eqref{Anormd}. Since the $\mb{5_2}$ knot is not amphicheiral, it may also be necessary to apply the transformation \eqref{Amirror} to reverse the orientation of $S^3\bs\mb{5_2}$, depending on the conventions being used.

\section{The wavefunction}
\label{sec:wf}

Having quantized $\hat{A}(\hat \ell,\hat m^2;q)$, we turn to our final task: constructing the holomorphic blocks themselves for rank-one Chern-Simons theory on a knot complement. Recall from Section \ref{sec:CS} that the analytically continued Chern-Simons partition function for a knot complement $M=\ol{M}\bs K$ is a locally holomorphic function of the boundary parameter $u$ (or $m=e^u$) and $\hbar$. For any chosen integration cycle in the Chern-Simons path integral, we can expand \cite{gukov-2003, Wit-anal}
\be Z_{\rm CS}(M;u;\hbar) = \sum_\alpha n_\alpha Z^\alpha_{\rm CS}(M;u;\hbar)\,. \ee
Each block $Z^\alpha(u;\hbar)$ corresponds to a critical point of the Chern-Simons action, hence to a flat connection on $M$. Thus, when $M$ is (\eg) a nontrivial knot complement in $S^3$, there are finitely many blocks.

We argued in Section \ref{sec:Apoly} that (at least for a knot complement in $S^3$) the nonabelian holomorphic blocks constitute a basis of solutions to the homogeneous difference equation $\hat A\cdot Z=0$. The gluing methods of Section \ref{sec:opglue} tell us explicitly how to construct these solutions. We decompose a three-manifold $M$ into tetrahedra; multiply together a tetrahedral block $\psi(z_i')$ (satisfying $(\hat z_i+\hat z_i'{}^{-1}-1)\psi(z_i')=0$) for each tetrahedron; transform the resulting ``product wavefunction'' into a basis where it depends explicitly on $u$ and the gluing functions $C_j$; and finally set $C_j\to 2\pi i+\hbar$. The change of basis and subsequent specialization $C_j\to 2\pi i+\hbar$ are a generalized version of ``integrating out boundary conditions.''

If we were dealing with an honest, physical TQFT, the gluing construction of Section \ref{sec:opglue} would produce an unambiguous result for the wavefunction of a knot complement. In our case we are not working with wavefunctions in actual Hilbert spaces, but rather with analytic continuations of them. The notion of changing basis for a product wavefunction is, therefore, not quite well defined. Nevertheless, we will see that we can still change basis formally in the space of analytically continued wavefunctions, without keeping precise track (\eg) of the integration contours used in Fourier transforms.

Given a knot complement $M$, the formal change of basis leads to a result for holomorphic blocks of the form
\be Z^{\rm gen}(M;u;\hbar) = \int\!\!\!\raisebox{.1cm}{...}\!\!\!\int dp_1\cdots dp_I\, \psi(...)\cdots \psi(...)\,, \label{Zgen}\ee
\ie\ a multiple integral in some complex variables $p_i$ of a product of tetrahedron blocks $\psi$.
Expression \eqref{Zgen} satisfies
\be \hat A(\hat \ell,\hat m^2;q)\,Z^{\rm gen}(u;\hbar) = 0 \label{wfeq} \ee
by construction, in the sense that formal manipulations under the integral sign can validate this equality. Given \emph{any} well-defined integration cycle for \eqref{Zgen} --- \ie\ a cycle such that the integrand approaches zero sufficiently fast near its boundary --- the evaluation of $Z^{\rm gen}(u;\hbar)$ produces an actual well-defined solution to \eqref{wfeq}.
 We might expect, then, that a basis of good integration cycles $C^\alpha$ for \eqref{Zgen} corresponds to a basis for holomorphic blocks, and this is precisely true. Each nonabelian%
\footnote{We are assuming in this discussion that the operator $\hat A(\hat \ell,\hat m^2;q)$ whose nullspace we want to generate knows about all the nonabelian flat connections on a knot complement. As discussed in Sections \ref{sec:classrem} and \ref{sec:missing}, this may not be entirely true if we construct $\hat A(\hat \ell,\hat m^2;q)$ from nongeneric triangulations of $M$, and thereby lose some factors. The correct general statement is that every flat connection appearing as a classical solution to $\hat A(\hat \ell,\hat m^2;q\to 1)=0$ leads to a critical point of the state integral \eqref{Zgen}.} %
flat connection $\CA^\alpha$ leads to a critical point for the integrand of \eqref{Zgen}, which in turn defines a cycle $C^\alpha$ via downward flow, or stationary phase.

The idea of generating solutions to a difference or differential equation via different cycles of a single integral is by no means novel. One encounters similar phenomena often in mathematics. A very simple example is the complex integral
\be f^{\rm gen}(a) = \int dx\, e^{-x^3+ax}\,, \label{faint} \ee
which generates solutions to the second-order (Airy) differential equation
\be (3\,\pd_a^{\,2}-a)\,f(a)=0\,. \label{faex} \ee
The integrand of \eqref{faint} has two critical points, at $x=\pm\sqrt{a/3}$, and each extends by downward flow to a well-defined integration cycle --- giving the two independent functional solutions to \eqref{faex}. This particular example was analyzed in detail in \cite{Wit-anal}.

As noted in the introduction, a state integral model very similar to the one we develop here was presented in \cite{DGLZ}, following \cite{hikami-2006, hikami-2001-16}. The formal expression for $Z^{\rm gen}$ in \cite{DGLZ} takes the same general form as \eqref{Zgen}; and there, just as here, various saddle point contours corresponded to flat connections or holomorphic blocks. Precise equivalence of the two state integral models is expected, but has yet to be established. Both state integral models can be viewed as noncompact, or analytically continued analogues of Kashaev's invariant \cite{Kashaev-invt, kashaev-1997}.

Unfortunately, there are several normalization ambiguities in our present method of finding $Z^{\rm gen}$. They stem from 1) a somewhat ill-defined normalization for the individual tetrahedron blocks $\psi(z_i')$; 2) the lack of a notion of unitarity in our holomorphic picture; and 3) the fact that the Weil representation of the (affine) symplectic group, even when defined so that it is unitary, is still only a projective representation. These issues can all be remedied to some extent, but we will not really attempt to do so here. We will argue that, at worst, holomorphic blocks from \eqref{Zgen} are determined up to a multiplicative ambiguity of the form 
\be \textstyle \exp\left(\frac{\pi^2}{\hbar}\Q+\C+\hbar\, \Q\right)\,,\label{Qform}\ee
their functional dependence on $u$ still being completely fixed.

We will begin in Section \ref{sec:duality} with a discussion of tetrahedral wavefunctions $\psi(z')$, then review the Weil representation of the (affine) symplectic group in Section \ref{sec:Weil}, and finally put everything together in examples of holomorphic blocks for the trefoil and figure-eight knots in Section \ref{sec:wfex}. It is worth noting that the actual quantum dilogarithm we use for $\psi(z')$ is not quite that of Equations \eqref{QDLi} or \eqref{quant-qdl}. Rather, we find it more appropriate to consider a nonperturbatively completed quantum dilogarithm, sometimes known as the ``noncompact'' quantum dilogarithm in the literature \cite{Fad-modular}. This function has some very interesting almost-modular properties, hence the title, and indeed the content, of Section \ref{sec:duality}. \\

With further work, it should be possible to define real Hilbert spaces and to extend the present gluing construction to actual physical wavefunctions in $SU(2)$, or $SL(2,\R)$, or $SL(2,\C)$ Chern-Simons theory. A first and necessary step in this direction would be to understand how to include abelian flat connections into the construction of holomorphic blocks --- or, alternatively, how to introduce inhomogeneity into the operator equation $\hat A\cdot Z=0$. It is likely that using integration cycles with finite boundaries for \eqref{Zgen} will play a key role. We hope to clarify such matters in the future.

\subsection{S-duality}
\label{sec:duality}

Let us take another look at Equation \eqref{qLangeq}, the wavefunction equation for the holomorphic block of a tetrahedron:
\be \hat\CL_{\Delta} \cdot \psi(Z) = \big(\hat z+\hat z'{}^{-1}-1\big)\,\psi(Z') = 0\,. \label{wfLe2} \ee
The operators here act as
\be \hat z\,\psi(Z') = \psi(Z'+\hbar)\,,\qquad \hat z'\psi(Z')=e^{Z'}\psi(Z')\,,\ee
and it is easy to see that the formal solution to \eqref{wfLe2} can be written as an infinite product,
\be \psi(Z') \overset{?}{=} \prod_{r=1}^\infty\big(1-q^re^{-Z'}\big)\,. \label{psiold} \ee

This solution, however, has quite a lot of ambiguity. First, it is clear that we could multiply $\psi(Z')$ by any function of $\hbar$ and still get a solution to \eqref{wfLe2}, since $\hbar$ is just a constant. But we can actually do much more. If we allow $\psi(Z')$ to be an honest function of the logarithmic variable $Z'$ rather than a function of $z'=e^{Z'}$ (in other words, breaking its periodicity under $Z'\to Z'+2\pi i$), we can multiply it by \emph{any other function} $f\left(\exp \frac{2\pi i}{\hbar} Z'\right)$ and still solve \eqref{wfLe2}. This is because
\be e^{\hat Z}f\big(e^{\frac{2\pi i}{\hbar}Z'} \big) = f \big(e^{\frac{2\pi i}{\hbar}Z'+2\pi i} \big) = f \big(e^{\frac{2\pi i}{\hbar}Z'} \big)\,. \ee

Since it is the exponentiated variables $z$, $z'$, $\ell$, $m^2$, etc. that correspond to $SL(2,\C)$ holonomy eigenvalues, and therefore are the natural classical coordinates on phase spaces $\CP_\Delta$, $\CP_{(M,\Delta)}$, or $\CP_{T^2}$, one might argue that the wavefunction $\psi(Z')$ should be periodic in $Z'\to Z'+2\pi i$. There are several excellent reasons, however, to break the classical $Z'\to Z'+2\pi i$ periodicity of phase space quantum mechanically, at least in an analytically continued context, and to use a modified solution to \eqref{wfLe2} that has a very specific set of $e^{\frac{2\pi i}{\hbar}Z'}$ corrections.

The function we would like to use is
\be \boxed{\,\psi(Z') = \Phi_{\hbar/2}(-Z'+i\pi+\hbar/2)\,}\,, \label{psinew} \ee
where $\Phi_{\hbar/2}$ is the ``noncompact'' quantum dilogarithm of \cite{Fad-modular}.%
\footnote{The subscript of $\Phi$ here is written as $\hbar/2$ in order to agree with the definition of $\Phi_\hbar(p)$ in \cite{DGLZ}. Recall that $\hbar_{\rm here}=2\hbar_{\text{ref \cite{DGLZ}}}$.} %
This function is meromorphic in $Z'$ on the entire complex plane (with essential singularity at $Z'=\infty$), and can be defined in various regimes as either a ratio of infinite products
\be \Phi_{\hbar/2}(p) = \left\{\begin{array}{@{\;}l@{\quad}l}
\ds \prod_{r=1}^\infty\frac{1+q^{r-1/2}e^p}{1+\mb{q}^{-r+1/2}e^{\mb p}} & \Re(\hbar)<0\,, \vspace{.2cm}\\
\ds \prod_{r=1}^\infty\frac{1+\mb{q}^{r-1/2}e^{\mb p}}{1+q^{-r+1/2}e^p} & \Re(\hbar)>0\,,
\end{array}\right. \label{QDLprod} \ee
where
\be \boxed{\, \mb p := \frac{2\pi i}{\hbar}p\,,\qquad \bm \hbar := -\frac{4\pi^2}{\hbar}\,,\qquad\text{and}\qquad \mb q:= e^{\bm \hbar}\,}\,;\ee
or via an integral formula
\be \Phi_{\hbar/2} = \exp\left(\frac14 \int_{\R+i\epsilon} \frac{dx}{x} \frac{e^{-ipx}}{\sinh(\pi x)\sinh(\hbar x/(2i))}\, \right)\,. \label{QDLint} \ee
Properties of this remarkable function have been discussed in many places, including \cite{Barnes-QDL, Fad-modular, kashaev-2000, FKV, volkov-2003, Bytsko-Teschner-noncpt, FG-qdl-cluster, DGLZ}. It was central to the state integral model constructed by \cite{DGLZ, hikami-2006, hikami-2001-16}.

Rather than being periodic in $Z\to Z'+2\pi i$, our desired function $\psi(Z')$ in \eqref{psinew} satisfies
\be \psi(Z'+2\pi i) = \big(1-e^{-\frac{2\pi i}{\hbar}Z'}\big)\psi(Z')\,. \label{qperiod2} \ee
This is the aforementioned breaking of periodicity, also anticipated in Section \ref{sec:log}. We emphasize, however, that in the classical limit $\hbar \to 0$, the difference between the two functions \eqref{psiold} and \eqref{psinew} becomes invisible. The functions have identical asymptotic expansions in $\hbar$. We can think of the noncompact quantum dilogarithm as a nonperturbative completion of its compact cousin \eqref{psiold}.

Our strongest motivation for using \eqref{psinew} as the analytically continued wavefunction of a tetrahedron stems from the observation that this function alone is invariant under cyclic permutations
\be Z'\mapsto Z''\mapsto Z \mapsto Z\,. \label{wfperm0} \ee
In terms of wavefunctions, such permutations are implemented via a Fourier transform, the details of which will be explained in Section \ref{sec:wfcyclic}. The property that allows invariance under permutations is the amazing fact that the Fourier transform of the noncompact quantum dilogarithm is essentially itself (\cf\ \cite{FKV, volkov-2003, Woronowicz}).

Another consequence of the noncompact quantum dilogarithm's Fourier self-trans-form, combined with its famous pentagon identity \cite{Fad-modular}, is that the state integral model we construct for holomorphic blocks on triangulated knot complements $(M,\Delta)$ is invariant under 2--3 Pachner moves.%
\footnote{As usual, we ignore the subtleties of the ``bad'' triangulations discussed in Sections \ref{sec:classrem}, \ref{sec:missing} when making such statements.} %
The proof of this fact is very similar to arguments given in (\eg) \cite{FockChekhov, FG-qdl-cluster, Kash-Teich} in the context of quantum Teichm\"uller theory, so we will not explain it further here. (Once we have a well-defined tetrahedron wavefunction $\psi(Z')$ that is invariant under permutations \eqref{wfperm0}, invariance of a state integral model under 2--3 Pachner moves and other combinatorial choices is actually guaranteed for us. This is because we explicitly construct the state integral model as a wavefunction dual to the operator gluings in Section \ref{sec:quant}, and we have already checked in Section \ref{sec:quant} that the ``gluing together'' of $\hat A_M$ is independent of combinatorial choices.)

Mathematically, we might note that just as the noncompact quantum dilogarithm \eqref{psinew} is a rather nicer function of $Z'$ than \eqref{psiold} (and is more suitable for a state integral model), it is also a much nicer function of $\hbar$. Namely, while \eqref{psiold} is only defined for $\Re\hbar<0$ (or $|q|=1$) and has a natural boundary at the real line, the function \eqref{psinew} can be analytically continued to a holomorphic function on the entire complex plane minus a half-line. This is reminiscent of the properties of similar functions discussed in \cite{LewisZag}, in relation to the Eichler integrals of Maass waveforms.

Physically, we have one additional motivation for using the noncompact quantum dilogarithm \eqref{psinew} rather than its ``compact'' cousin \eqref{psiold}, and this has to do with S-duality. In several recent works \cite{Witten-path, Witten-HK, gukov-2008, Gukov-quantMS}, it has been shown that various ingredients of Chern-Simons theory on a three-manifold $M$ --- such as its Hilbert space, or its holomorphic blocks --- can be constructed in four-dimensional $\CN=4$ super Yang-Mills theory on a manifold $M\times[0,1]$ or $M\times\R_+$. The relation between the complexified coupling $\tau$ of super Yang-Mills theory and the Chern-Simons coupling $\hbar$ in these pictures is roughly
\be \hbar \sim 2\pi i\tau\,. \ee
As super Yang-Mills theory is self-dual under the interchange $\tau\leftrightarrow -1/\tau$ (together with the exchange of the gauge group with its Langlands dual), one might wonder whether a similar symmetry might manifest itself in Chern-Simons theory, under
\be S\,:\;\hbar \to -\bm\hbar\,. \ee

An alternative hint of S-duality (in fact, of modularity) in certain colored Jones polynomials was also found in \cite{QMF}. Unfortunately, the precise physical interpretation of S-duality in Chern-Simons theory is far from clear at the moment. Nevertheless, if one uses the noncompact quantum dilogarithm \eqref{psinew} as the holomorphic block of a tetrahedron, S-duality, at least in our present analytically continued context, suddenly appears surprisingly simple. Namely, due to the properties $\Phi_{\hbar/2}(p)=\Phi_{\bm\hbar/2}(\mb p)$ and $\Phi_{-\hbar/2}(p)=\Phi_{\hbar/2}(p)^{-1}$, the first an elementary consequence of \eqref{QDLprod} and the second a formal consequence of \eqref{QDLint}, we have
\be S:\;\psi(Z') \overset{Z'\to \mb Z',\;\hbar\to-\bm\hbar}{\longmapsto} \psi(Z'-\hbar)^{-1}\,,\ee
where we have accompanied the inversion of $\hbar$ with a Jacobi-type transformation $Z'\to\mb Z'=\frac{2\pi i}{\hbar}Z'$. Better yet, if we just send $\hbar \to \bm\hbar=-4\pi^2/\hbar$ (preserving the upper half-$\hbar$-plane) rather than $\hbar \to -\bm\hbar$, we find an exact invariance
\be \sigma:\; \psi(Z')\overset{Z'\to \mb Z',\;\hbar\to\bm\hbar}{\longmapsto} \psi(Z')\,. \label{psidual} \ee

The requirement that $\psi(Z')$ be self-dual under the transformation $\sigma:\,(Z',\hbar)\mapsto(\mb Z',\bm\hbar)$ as in \eqref{psidual} can be used as a fairly natural way to narrow down its overall normalization. In particular, we see that if we multiply $\psi(Z')$ by a ``constant'' function $f(\hbar)$, this function must be symmetric under $\hbar\leftrightarrow\bm\hbar$ in order to preserve \eqref{psidual}. Combining this with the observation at the end of Section \ref{Volrat} that the leading asymptotics of $\psi(Z')$ should match the analytically continued hyperbolic volume of an ideal tetrahedron, we can estimate that any normalization factor $f(\hbar)$ should be of the general form
\be f(\hbar)\;\in\;\exp\left(\frac{\pi^2}{\hbar}\,\Q+\C+\hbar\,\Q\right)\,. \label{QDLnormest} \ee
This is a very rough, conservative estimate, though it will suffice here. We will see in Section \ref{sec:Weil} that it coincides with the form of ambiguities introduced into the state integral model through the projectivity of the Weil representation.

\subsubsection{Modular double and the dual $\hat{\mb A}$}
\label{sec:Heis}

The analytically continued wavefunction $\psi(Z')= \Phi_{\hbar/2}(-Z'+i\pi+\hbar/2)$ is quasi-periodic under shifts of $Z'$ by $\hbar$, $\psi(Z'+\hbar) = (1-e^{-Z'})\psi(Z')$, and this quasi-periodicity is embodied in the functional equation
\bse \label{ddiff}
\be \hat\CL_\Delta\cdot\psi(Z')= (\hat z+\hat z'{}^{-1}-1)\psi(Z')=0\, \label{ddiff1} \ee
as in \eqref{wfLe2}. The noncompact quantum dilogarithm, however, is also quasi-periodic under shifts by $2\pi i$, as in \eqref{qperiod2}, and therefore satisfies a second dual equation
\be \hat{\bm\CL}_\Delta\cdot\psi(Z') = (\hat{\mb z}+\hat {\mb z}'{}^{-1}-1)\psi(Z') = 0\,,  
\label{ddiff2}\ee
\ese
where
\be \hat{\mb Z}=\frac{2\pi i}{\hbar}\hat Z\,,\qquad \hat{\mb Z}'=\frac{2\pi i}{\hbar}\hat Z'\,,\qquad\text{and}\qquad
\hat{\mb z}=\exp({\hat{\mb Z}})\,,\qquad \hat{\mb z}'=\exp({\hat{\mb Z}'})\,, \label{defbold2} \ee
so that, in particular,
\be \hat{\mb z}\,\psi(Z') = \psi(Z'+2\pi i)\,\qquad\text{and}\qquad
\hat{\mb z}'\,\psi(Z') = e^{\frac{2\pi i}{\hbar}Z'}\psi(Z')\,.\ee

The combination of equations (\ref{ddiff}a-b) completely fixes the functional dependence of $\psi(Z')$ on the logarithmic variable $Z'$, determining $\psi(Z')= \Phi_{\hbar/2}(-Z'+i\pi+\hbar/2)$ up to overall normalization. This follows from an analogous statement in the operator algebra. Namely, the full algebra of exponential operators generated by $\hat{\mb z}$ and $\hat{\mb z}'$ in addition to $\hat z$ and $\hat z'$ is isomorphic to the logarithmic algebra generated by $\hat Z$ and $\hat Z'$\, \cite{Fad-modular}:
\be \C(q,\mb{q})[\hat z{}^{\pm 1},\hat z'{}^{\pm 1},\hat{\mb z}{}^{\pm 1},\hat{\mb z}'{}^{\pm 1}] \simeq \C(\hbar)[\hat Z,\hat Z']\,. \ee
Thus, once $\hat{\mb z}$ and $\hat{\mb z}'$ are included, pairs of operator equations can determine wavefunctions that depend on $Z'$ rather than just the exponentiated variable $z'=e^{Z'}$.

Enlarging a $q$-commuting algebra of exponential operators by adding their duals in this manner is sometimes referred to as forming its \emph{modular double} \cite{Fad-modular} (also \cf\ \cite{Kash-Heis, Kharchev-L-STS}). It is important to observe that original and dual variables always mutually commute. For example,
\be \hat{z}\,\hat{\mb z}' = \exp\big([\hat Z,\hat{\mb Z}']\big) \hat{\mb z}'\hat z = \exp\left(\frac{2\pi i}{\hbar}\cdot\hbar\right)\hat{\mb z}'\hat z= \hat{\mb z}'\hat z\,, \ee
and similarly, $\hat{\mb z}\,\hat z'=\hat z'\hat{\mb z}$. Therefore, the modular double of a $q$-commutative algebra really just contains two commuting copies of the algebra itself. There is an involution $\sigma_*$ sending operators to their duals, \eg,
\be \sigma_*\,\hat Z = \hat{\mb Z}\,,\qquad \sigma_*\,\hat{\mb z}'=\hat z'\,, \ee
and also
\be \sigma_*\,\hbar = \bm{\hbar}\qquad\imp\qquad \sigma_*q = \mb q\,,\ee
so that $\sigma_*$ interchanges the two copies of an algebra of exponential operators in a modular double. Note that the commutation relations for dual variables are
\be [\hat{\mb Z},\hat{\mb Z}'] = \left(\frac{2\pi i}{\hbar}\right)^2\hbar = \bm\hbar \qquad\imp\qquad \hat{\mb z}\,\hat{\mb z}'=\mb{q}\,\hat{\mb z}'\hat{\mb z}\,,
\ee
so that $\sigma_*$ preserves the commutator.

The quantum vertex equations have a particularly nice transformation under $\sigma_*$. In Section \ref{sec:qtet}, we promoted the classical constraint $Z+Z'+Z''=i\pi$ on the phase space of a tetrahedron to a quantum relation \eqref{qvertex},
\be \hat Z+\hat Z'+\hat Z''= i \pi+a\hbar\,.\label{qvertex2} \ee
We used many ingredients in Section \ref{sec:quant} to argue that $a=1/2$ was the only consistent value for $a$. Acting on the left-hand side of \eqref{qvertex2} with $\sigma_*$ we find
\begin{align} \sigma_*(\hat Z+\hat Z'+\hat Z'') &=
 \hat {\mb Z}+\hat {\mb Z}'+\hat{\mb Z}'' \nno \\
 &= \frac{2\pi i}{\hbar}(\hat Z+\hat Z'+\hat Z'') \nno \\
 &= \frac{2\pi i}{\hbar}(i\pi + a \hbar) \nno \\
 &= 2a\pi i + \frac{\bm\hbar}{2}\,,
\end{align}
which is equal to $\sigma_*(i\pi+a\hbar)=i\pi+a\bm{\hbar}$ if and only if $\boxed{a=1/2}$\,. Thus, simple compatibility of the vertex equation with the involution $\sigma_*$ immediately fixes the quantum correction $a$. (We have assumed here that in addition to \eqref{defbold2} we have $\hat{\mb Z}''=\frac{2\pi i}\hbar \hat Z''$.)

Since a tetrahedron wavefunction $\psi(Z')$ satisfies both $\hat\CL_\Delta\,\psi(Z')=0$ and $\hat{\bm\CL}_\Delta\,\psi(Z')=0$, one might wonder whether the holomorphic blocks $Z^\alpha(M;u)$ of an entire knot complement $M$ are also annihilated by a dual operator, in addition to $\hat A$. The answer is easily found to be yes. The general TQFT gluing methods of Section \ref{sec:opglue} imply that if the holomorphic blocks $\psi(Z_i')$ for every tetrahedron $\Delta_i$ in a triangulation of $M$ satisfy $\hat{\bm\CL}_{\Delta_i}\,\psi(Z_i')=0$ as well as $\hat\CL_{\Delta_i}\,\psi(Z_i')=0$, then there is an element $\hat{\mb A}(\hat{\bm\ell},\hat{\mb m}^{2};\mb q)$ in the left ideal generated by all the $\hat{\bm\CL}_{\Delta_i}$'s (and specialized to $\hat C_j=\hbar$) that annihilates the glued holomorphic blocks $Z^\alpha(M;u)$.

Explicitly, the element $\hat{\mb A}(\hat{\bm\ell},\hat{\mb m}^{2};\mb q)$ is constructed by applying the involution $\sigma_*$ to all the algebras and ideals of Section \ref{sec:quant}, sending
\be \hat Z_i \mapsto \hat{\mb Z}_i=\frac{2\pi i}{\hbar}\hat Z_i\,,\qquad \hat Z_i' \mapsto \hat{\mb Z}_i'=\frac{2\pi i}{\hbar}\hat Z_i'\,,\qquad\hat Z_i'' \mapsto \hat{\mb Z}_i''=\frac{2\pi i}{\hbar}\hat Z_i''\,, \ee \be \hbar \mapsto \bm\hbar\,,\qquad q\mapsto \mb q\,, \nno\ee
as well as
\be \hat C_j \mapsto \hat{\mb C}_j=\frac{2\pi i}\hbar \hat C_j\,,\qquad \Gamma_j \mapsto \hat{\bm \Gamma}_j=\frac{2\pi i}\hbar \hat \Gamma_j\,,\qquad
 U \mapsto \hat{\mb U}=\frac{2\pi i}\hbar \hat U\,,\qquad
 v \mapsto \hat{\mb v}=\frac{2\pi i}\hbar \hat v\,, \nno\ee
\be \hat c_j\mapsto \hat{\mb c}_j=\exp{\hat{\mb C}_j}\,, \qquad
\hat \gamma_j \mapsto \hat{\mb \gamma}_j=\exp \hat{\mb \Gamma}_j\,,\ee
\be \hat m^2 \mapsto \hat{\mb m}^2=\exp\hat{\mb U}\,,\qquad
\hat\ell \mapsto \hat{\bm \ell} = \exp\hat{\mb v}\,,\qquad \text{etc.}\,\nno\ee
and then repeating every argument in Section \ref{sec:quant} word for word. The final result is that $\hat{\mb A}(\hat{\bm\ell},\hat{\mb m}^{2};\mb q)$ is equivalent to the operator obtained by just applying $\sigma_*$ directly to $\hat A(\hat\ell,\hat m^2;q)$, with one exception. Namely, if the negative square root of $\hat\ell$ was taken by setting $\hat v=\hat V/2+i\pi$ (\cf\ the factor of $i\pi$ in \eqref{qUv41}, or the discussion of these roots around \eqref{vVclass}), then the correct effective transformation of $\hat \ell$ from $\hat A$ to $\hat{\mb A}$ is
\be -\hat \ell=\exp\big(\hat V/2\big)\quad\mapsto\quad \exp\big(\hat{\mb V}/2\big)=\exp\big(\frac{2\pi i}{\hbar}(\hat{\mb v}-i\pi)\big)=\mb q^{-1/2}\hat{\bm\ell}\,. \label{Lexc} \ee
The fundamental reason for the exception is that the choice of a root is the only place where the symmetry between the quasi-periods $2\pi i$ and $\hbar$ of the noncompact quantum dilogarithm is broken. Everywhere else, they occur in the $\sigma_*$--covariant combination $2\pi i+\hbar$. Thus, for example, the specialization $\hat c_j\to q$ (in the final step of calculating $\hat A$) is mirrored by the dual specialization $\hat{\mb c}_j = \exp\big(\frac{2\pi i}\hbar\hat C_j\big) \to \exp(2\pi i+\bm\hbar) = \mb q\,.$ To summarize,
\be \hat{\mb A}(\hat{\bm\ell},\hat{\mb m}^{2};\mb q) =\left\{\begin{array}{c@{\quad}l} \sigma_*\hat A(\hat \ell,\hat m^2;q) &\text{if}\; \sqrt{\ell^2}=\ell \vspace{.2cm}\\
\sigma_*\hat A(-q^{-1/2}\hat \ell,\hat m^2;q) & \text{if}\;\sqrt{\ell^2}=-\ell\,.
\end{array}\right. \label{ADrule}
\ee

It is pleasing to note that requiring $\sigma_*$--invariance of the quantum-corrected specialization of gluing functions $\hat C_j\to 2\pi i+\kappa_j\hbar$ immediately fixes the constants $\boxed{\kappa_j\equiv1}$, the same way that $\sigma_*$--invariance of the vertex equations fixes $a=1/2$. Duality effectively allows us to bypass the painstaking analysis of topological invariance in Sections \ref{sec:qpath}--\ref{sec:qPach}.

Just as the semi-classical phase spaces and Lagrangians for tetrahedra were closely related to constructions in semi-classical Teichm\"uller theory (Section \ref{sec:triang}), quantized tetrahedra are related to quantum Teichm\"uller theory \cite{FockChekhov, Kash-Teich}, and in turn to quantum Liouville theory \cite{Verlinde-TeichLiouv, Teschner-TeichLiouv}. The appearance of $\sigma$-duality in Chern-Simons theory can be considered a 3d lift of the well-known 2d S-duality, often written $b\leftrightarrow b^{-1}$ in quantum Teichm\"uller \cite{FockChekhov, Kash-Teich, FG-qdl-cluster} and Liouville \cite{Tesch-Liouv} theories. This connection will be more fully explored elsewhere \cite{DG-S}.

\subsection{Wavefunctions and the Weil representation}
\label{sec:Weil}

The gluing methods of Section \ref{sec:opglue} show that the construction of $\hat A$-polynomials from tetrahedra in Section \ref{sec:quant} should be mirrored by a construction of holomorphic blocks --- \ie\ of the solutions to $\hat A(\hat \ell,\hat m^2;q)\,Z(U)=0$ as well as (now) the dual equation $\hat{\mb A}(\hat{\bm\ell},\hat{\mb m}^{2};\mb q)\,Z(U)=0$. Explicitly, let us assume that we have a triangulated knot complement $M=\bigcup_{i=1}^N\Delta_i$. To build the holomorphic blocks, we should
\begin{enumerate}
\item Assign a wavefunction $\psi(Z_i')=\Phi_{\hbar/2}(-Z_i'+i\pi+\hbar/2)$ to each tetrahedron $\Delta_i$.
\item Multiply these wavefunctions together to form a product $\Psi(Z_1',...,Z_N') = \psi(Z_1')\cdots\psi(Z_N')$.
\item Change the basis, or representation, or polarization, of this wavefunction to obtain $\widetilde\Psi(U,C_1,...,C_{N-1})$, with an explicit dependence on the gluing functions $C_j$.
\item Set $C_j\to 2\pi i+\hbar$ in $\widetilde\Psi(U,C_1,...,C_{N-1})$.
\end{enumerate}
This procedure should produce the holomorphic block integral \eqref{Zgen} described in the introduction to this section.

The only nontrivial step above is the third one. It mirrors the change of symplectic basis (or canonical transformation) that we performed in the algebra of operators in Sections \ref{sec:glue} and \ref{sec:qglue}, rewriting
\be (\hat Z_1',...,\hat Z_N',\hat Z_1,...,\hat Z_N)\overset{\varphi_*}{\longmapsto}(\hat U,\hat C_1,...,\hat C_{N-1},\hat v,\hat\Gamma_1,...,\hat\Gamma_{N-1})\,. \label{pZtoC}
\ee
This transformation is implemented by an element of the affine symplectic group $ISp(2N,\C)$, or more precisely a combination of $Sp(2N,\Q)$ transformations and translations by rational multiples of $i\pi$ and $\hbar$. The action of the affine symplectic group on the operator algebra changes the representation of operators on wavefunctions. For example, corresponding to \eqref{pZtoC}, we would expect to go from a representation on functions $f(Z_1',...,Z_N')$ (with $\hat Z_i=\hbar\,\pd_{Z_i'}$) to a representation on functions $f(U,C_1,...,C_{N-1})$ (with $\hat v=\hbar\,\pd_U$ and $\hat \Gamma_j=\hbar\,\pd_{C_j}$). The map between one representation and another, which intertwines the action $\varphi_*$ of the affine symplectic group in the operator algebra, is precisely what is needed in Step 3 above to send
\be \Psi(Z_1',...,Z_N')\overset{\varphi}{\longmapsto}\widetilde\Psi(U,C_1,...,C_{N-1})\,. \label{pwf} \ee

When wavefunctions live in a Hilbert space $L^2(\R^N)$, the desired intertwining action is known (mathematically) as the Weil representation of the affine symplectic group $ISp(2N,\R)$ \cite{Shale-rep, Weil-rep}. It is a unitary but projective representation of $ISp(2N,\R)$, so its action on wavefunctions in only defined up to a phase. In our case, we don't quite have a Hilbert space, but we can still apply Weil transformations in a formal manner. To do so, we can imagine our locally holomorphic wavefunction $\Psi$ to be the analytic continuation of an actual element in $L^2(\R^N)$.%
\footnote{One could now make this statement much more precise. For example, at pure imaginary $\hbar$, the quantum dilogarithm $\psi(Z')$ is not square integrable, but it \emph{is} integrable and has a well-defined Fourier transform along a distinguished contour. In principle, this allows for the definition of a distinguished ``real'' integration cycle in a final expression for the wavefunction of $M$, such as \eqref{Zgen}. If we were interested in the actual partition function of physical Chern-Simons theory, this is precisely the type of procedure that we should go through. However, since we actually want holomorphic blocks (\ie\ a complete basis for the vector space of solutions to $\hat A\,Z^\alpha=0$), we would end up analytically continuing the final integral \eqref{Zgen} anyway, replacing any distinguished integration cycle by well-defined critical-point cycles $C^\alpha$. Thus, there is presently no reason to be any more careful about the actual integrations being performed.} %
Note also that if all the $Z_i'$ and also $\hbar$ are taken to be pure imaginary, then our symplectic transformation $\varphi_*$ becomes an element of $ISp(2N,\Q)\subset ISp(2N,\R)$ rather than $ISp(2N,\C)$. In this case, the classic Weil representation is fully well-defined --- and can thereafter be analytically continued.

Let us assume, then, that a formal Weil transformation of our holomorphic wavefunctions does exist. Given an element $\varphi_*$ of $ISp(2N,\C)$ as above, the easiest way to find the corresponding intertwining action $\varphi$ on wavefunctions is to write $\varphi_*$ as a product of generators. The generators of $ISp(2N,\C)$ can be taken to be of four basic types. Acting on column vectors $\big(\vec{\hat Z}{}',\vec{\hat Z}\,\big)^T=(\hat Z_1',...,\hat Z_N',\hat Z_1,...,\hat Z_N)^T$, the first three types of generators are $Sp(2N,\C)$ matrices
\be \label{spgens}
\begin{pmatrix} I-J & -J \\ J & I-J \end{pmatrix}\,,\qquad
\begin{pmatrix} A & 0 \\ 0 & A^{-1\,T} \end{pmatrix}\,,\qquad
\text{and}\qquad
\begin{pmatrix} I & 0 \\ B & I \end{pmatrix}\,,
\ee
written in terms of $N\times N$ blocks, where $J$ is diagonal with entries $0$ and $1$, $A$ is nonsingular, and $B$ is symmetric ($B^T=B$). The fourth type of generators contains the translations
\be \label{tgens}
\begin{pmatrix} \vec{\hat Z}{}' \\ \vec{\hat Z}\end{pmatrix}
\overset{\varphi_*}{\mapsto} \begin{pmatrix} \vec{\hat Z}'+\vec s \\ \vec{\hat Z}\end{pmatrix}\qquad\text{and}\qquad
\begin{pmatrix} \vec{\hat Z}{}' \\ \vec{\hat Z}\end{pmatrix}
\overset{\varphi_*}{\mapsto} \begin{pmatrix} \vec{\hat Z}' \\ \vec{\hat Z}+\vec t\end{pmatrix}\,.
\ee

The intertwiner corresponding to each generator is fairly simple (\cf\ \cite{Guillemin-Sternberg}). Let's first consider the three types of $Sp(2N,\C)$ elements, and suppose that
\be \begin{pmatrix} \vec X\\\vec Y\end{pmatrix} = \varphi_*\,\begin{pmatrix} \vec Z'\\\vec Z \end{pmatrix}\,,\ee
with $\varphi_*\in Sp(2N,\C)$, such that $Z_i=\hbar\,\pd_{Z_i'}$ on functions $f(\vec Z')$ and $Y_i=\hbar\,\pd_{X_i}$ on transformed functions $\widetilde f(\vec{X})$. (To avoid cluttering notation, we omit the hats `\;$\hat{}$\;' on operators here.) Then:
\bse \label{Wactions}
\begin{itemize}

\item If $\varphi_* = \begin{pmatrix} 0 & -I \\ I & 0\end{pmatrix}$, which is the first matrix in \eqref{spgens} specialized to $J=I$, then
\be f(\vec Z')\,\overset{\varphi}{\longmapsto}\,\widetilde f(\vec{X})=\frac{1}{(2\pi i\hbar)^{N/2}}\int d\vec Z'\,f(\vec Z')\,e^{\frac 1\hbar\vec X\cdot \vec Z'}\,. \label{wFT} \ee

\item More generally, if $\varphi_* = \begin{pmatrix} I-J & -J \\ J & I-J \end{pmatrix}$, then a one-dimensional Fourier transform of the type \eqref{wFT} should be performed for every coordinate corresponding to a `1' on the diagonal of $J$.

\item If $\varphi_* = \begin{pmatrix} A & 0 \\ 0 & A^{-1\,T} \end{pmatrix}$, then
\be f(\vec Z') \,\overset{\varphi}{\longmapsto}\,\widetilde f(\vec{X})=
 \frac{1}{\sqrt{\det A}}\,f(A^{-1}\vec X)\,.\ee

\item If $\varphi_* = \begin{pmatrix} I & 0 \\ B & I \end{pmatrix}$, then
\be f(\vec Z') \,\overset{\varphi}{\longmapsto}\,\widetilde f(\vec{X})=
f(\vec X)\,e^{\frac1{2\hbar} \vec X^T B\vec X}\,. \label{sB} \ee

\end{itemize}
Similarly, for translations we find:
\begin{itemize}
\item If\; $\begin{pmatrix} \vec{Z}{}' \\ \vec{Z}\end{pmatrix}
\overset{\varphi_*}{\mapsto} \begin{pmatrix} \vec X\\\vec Y\end{pmatrix}= \begin{pmatrix} \vec{Z}'+\vec s \\ \vec{Z}\end{pmatrix}$\,, then
\be f(\vec Z') \,\overset{\varphi}{\longmapsto}\,\widetilde f(\vec{X})= f(\vec X-\vec s)\,; \label{ts} \ee

\item and if\; $\begin{pmatrix} \vec{Z}{}' \\ \vec{Z}\end{pmatrix}
\overset{\varphi_*}{\mapsto} \begin{pmatrix} \vec X\\\vec Y\end{pmatrix}= \begin{pmatrix} \vec{Z}' \\ \vec{Z}+\vec t\end{pmatrix}$\,, then
\be f(\vec Z') \,\overset{\varphi}{\longmapsto}\, \widetilde f(\vec{X})= f(\vec X)\,e^{\frac1\hbar \vec t\cdot\vec X}\,. \label{tt} \ee
\end{itemize}
\ese

The wavefunction transformations \eqref{Wactions} can be composed to create the formal intertwiner corresponding to any $\varphi_*\in ISp(2N,\C)$. We shall see some examples momentarily. The reason this Weil transformation is only formal is due to the Fourier transform \eqref{wFT}. For the types of functions we are considering (locally holomorphic functions of $\vec Z'$), the Fourier integral may not be defined on a fixed, canonical integration cycle. Rather, as we have mentioned several times, such an integral must ultimately be interpreted as taken on \emph{any} cycle that makes the integral converge; and such cycles are generally constructed by downward flow from critical points.

The normalization constants $1/\sqrt{\det A}$ and $(2\pi i\hbar)^{-N/2}$ are included because they \emph{would} make transformations \eqref{Wactions} unitary if $\hbar$ were pure imaginary and if we were acting on $L^2(\R^N)$. These normalizations continue to be natural in an analytically continued context. \\

In \eqref{ts}--\eqref{tt}, we have written the two different types of translations separately. Despite the fact that these translations commute as elements of $ISp(2N,\C)$, the intertwining actions on functions $f(\vec Z')$ do not! Indeed, note that if we map $(\vec Z',\vec Z)\mapsto (\vec Z'+\vec s,\vec Z)\mapsto (\vec Z'+\vec s,\vec Z+\vec t)=(\vec X,\vec Y)$, then
\be f(\vec Z')\mapsto f(\vec X-\vec s)\,e^{\frac1\hbar \vec t\cdot\vec X}\,, \label{tst} \ee
whereas if we map $(\vec Z',\vec Z)\mapsto (\vec Z',\vec Z+\vec t)\mapsto (\vec Z'+\vec s,\vec Z+\vec t)=(\vec X,\vec Y)$ in the opposite order, then
\be f(\vec Z')\mapsto e^{-\frac1\hbar\vec t\cdot\vec s}f(\vec X-\vec s)\,e^{\frac1\hbar \vec t\cdot\vec X}\,.\label{tts}\ee
The two transformations \eqref{tst} and \eqref{tts} differ by the multiplicative constant $e^{-\frac1\hbar\vec t\cdot\vec s}$, which would simply be a phase if we were acting with $ISp(2N,\R)$, at imaginary $\hbar$. This is one sign that the Weil representation, formal or otherwise, is only projective.

Since all of the translations we consider are by rational multiples of either $i\pi$ or $\hbar$, the extra projective factor in \eqref{tts} must be of the general form
\be \exp\Big(\frac{\pi^2}{\hbar}\Q+\C+\hbar\,\Q\Big)\,. \label{projform} \ee
One can similarly check whether other expected commutation relations for the generators of $ISp(2N,\C)$ are modified by projective factors in the Weil representation. Aside from a mild, $\hbar$--independent ambiguity arising from commutation of $Sp(2N,\C)$ generators among themselves, the only other notable projective factor comes from commutation of a lower-diagonal symplectic transformation \eqref{sB} with a translation \eqref{ts}. This factor is
\be e^{\frac1{\hbar} \vec s^T B\vec s}\,, \ee
which again must be of the form \eqref{projform} since $B$ is always rational for us.

In the introduction to this section, we claimed that our construction of holomorphic blocks for a knot complement was naturally well-defined up to a multiplicative factor precisely of the form \eqref{projform}. By combining the self-dual normalization of $\psi(Z')$ discussed on page \pageref{QDLnormest} with the current observations about projective factors and Weil generator normalizations, we have substantiated this claim.

We note that, mathematically, the projective Weil representation of $ISp(2N,\R)$ could be lifted to a true representation of an extension of the affine metaplectic group $IMp(2N,\R)$ by the Weyl group of $Sp(2N,\R)$ (\cf\ \cite{Burdet}). Unfortunately, we see no physical motivation or meaning of this lift in the present context.

\subsubsection{Cyclic permutations}
\label{sec:wfcyclic}

As our first toy example of transforming wavefunctions in the Weil representation, we can consider the effect of cyclic permutations $Z\to Z'\to Z''\to Z$ on the tetrahedron wavefunctions $\psi(Z')$.

The operator algebra corresponding to a tetrahedron is really generated by three logarithmic elements $\hat Z,\,\hat Z',\,\hat Z''$, which satisfy the quantum vertex equation
\be \hat Z+\hat Z'+\hat Z'' = i\pi+\hbar/2\,. \ee
We explained in Section \ref{sec:Heis} that (ignoring subtle details related to invertible elements) this algebra is equivalent to the modular double of an algebra in exponentiated operators. Explicitly including the vertex equations, we have
\begin{align} &\C(\hbar)[\hat Z,\hat Z',\hat Z'']/\raisebox{-.1cm}{\small$(\hat Z+\hat Z'+\hat Z'' = i\pi+\hbar/2)$} \nno \\&\hspace{1.5in}\,\simeq\, \C(q,\mb{q})[\hat z,\hat z',\hat z'',\hat{\mb z},\hat{\mb z}',\hat{\mb z}'']/\raisebox{-.1cm}{\small$(\hat z\hat z'\hat z''=q,\;\hat{\mb z}\hat{\mb z}'\hat{\mb z}''=\mb q)$}\,.
\label{vxHeis}\end{align}

The operator algebras \eqref{vxHeis} are manifestly invariant under cyclic permutations. Once we impose the vertex equations and eliminate one of the $\hat Z$'s in favor of the other two (eliminating the center of the operator algebra), cyclic invariance becomes the statement that the canonical transformations
\be \rho_*\,:\;\begin{pmatrix} \hat Z'\\\hat Z\end{pmatrix}
\longmapsto \begin{pmatrix} \hat Z''\\\hat Z'\end{pmatrix}
 = \begin{pmatrix} -\hat Z-\hat Z'+i\pi+\hbar/2 \\ \hat Z'
 \end{pmatrix}\qquad\text{(+ cyclic)}
\ee
%
%(or rather their inverses $\rho^*=\rho_*^{-1}$)
are isomorphisms of the algebras
\be
\begin{diagram}
 & & \C[\hat Z,\hat Z'] & & \\
 & \ruTo^{\rho_*} & & \rdTo^{\rho_*} & \\
 \C[\hat Z'',\hat Z] & & \lTo^{\rho_*} & & \C[\hat Z',\hat Z'']
\end{diagram} \label{Ctri}
\ee
that preserve the Lagrangian operator $\hat\CL_\Delta$ and its dual $\hat{\bm \CL}_\Delta$. Accordingly, the formal Weil transformations $\rho$ of wavefunctions should be isomorphisms, in a suitable sense, of the analytically continued functional spaces
\be
\begin{diagram}
 & & \CH_{\pd\Delta}'\sim\{f(Z')\} & & \\
 & \ruTo^{\rho} & & \rdTo^{\rho} & \\
 \CH_{\pd\Delta}\sim\{f(Z)\} & & \lTo^{\rho} & & \CH_{\pd\Delta}''\sim\{f(Z'')\}
\end{diagram} \label{Htri}
\ee
In particular, these maps $\rho$ must preserve the wavefunction of a tetrahedron.

Here it should be understood that the operators $(\hat Z,\hat Z'),\,(\hat Z',\hat Z''),\,(\hat Z'',\hat Z)$, respectively, act on the spaces $\CH_{\pd\Delta}',\CH_{\pd\Delta}'',\CH_{\pd\Delta}$ as $(\hbar\pd_{Z'},Z'\cdot),$ $(\hbar\pd_{Z''},Z''\cdot),$ $(\hbar\pd_Z,Z\cdot)$. The desired preservation of wavefunctions means that, in the Weil representation,
\be \psi(Z') \overset{\rho}{\longmapsto} \psi(Z'')\,, \label{psiZ3}\ee
up to a projective factor.

As an element of $ISp(2,\C)$, $\rho_*$ is a composition of
\be ST^T := \mtt 0{-1}10 \mtt 1011 \ee
with a translation by $i\pi+\hbar/2$. The easily verified identity $\rho_*^3=1$ is an extension of the standard $Sp(2,\Z)$ identity $(ST^T)^3=I$ to the affine symplectic group. In terms of wavefunctions, one can check formally that in the Weil representation
\be f(Z) \overset{\rho^3}\longmapsto e^{-\frac1{2\hbar}(i\pi+\hbar/2)^2}\,f(Z)\, \ee
for any $f(Z)$, so that $\rho^3=\text{id.}$ up to an expected projective factor.

To check invariance of the tetrahedron wavefunction under a single $\rho$, we need to use the fact that the Fourier transform of the noncompact quantum dilogarithm is (\cf\ \cite{FKV})
\be \frac{1}{\sqrt{2\pi i\hbar}}\int dY\,\Phi_{\hbar/2}(Y)\,e^{\frac1\hbar XY} = e^{-\frac{i\pi}{4}-\frac{4\pi^2-\hbar^2}{24\hbar}}\,\Phi_{\hbar/2}(-X+i\pi+\hbar/2)\,, \label{FT} \ee
for a suitably defined contour.%
\footnote{Note that factors like $\exp\big(\frac{i\pi}{4}\big)$ are sensitive to the precise rotation required to define a good contour. Since we are only interested in transformations up to projective factors, we do not keep careful track of them here.} %
Inverting this relation gives
\be \int dX\,\Phi_{\hbar/2}(-X+i\pi+\hbar/2)\,e^{\frac1{2\hbar}X^2-\frac1\hbar XY} = e^{\frac{i\pi}{4}+\frac{4\pi^2-\hbar^2}{24\hbar}}\,\Phi_{\hbar/2}(Y)\,. \label{iFT}
\ee
By decomposing the transformation $\rho$ into generators and using the Weil prescriptions in (\ref{Wactions}a-e), we then find that
\begin{align}
\psi(Z') & \overset{T^T}{\longmapsto} \psi(Z')\,e^{\frac1{2\hbar}Z'{}^2} \nno\\
&\overset{S}{\longmapsto} \frac{1}{\sqrt{2\pi i\hbar}}\int dZ'\,\psi(Z')\,e^{\frac{1}{2\hbar}\big(Z'{}^2+2Z''Z'\big)} \nno\\
&\overset{{\rm translate}}{\longmapsto}
 \frac{1}{\sqrt{2\pi i\hbar}}\int dZ'\,\psi(Z')\,e^{\frac{1}{2\hbar}\big(Z'{}^2+2Z''Z'-2Z'(i\pi+\hbar/2)\big)} \nno \\
 &\overset{\eqref{iFT}}{=}\, e^{\frac{5}{12}i\pi-\frac{1}{6\hbar}(i\pi+\hbar/2)^2}\,\psi(Z'')\,.
\end{align}
The wavefunction maps to itself modulo projective factors, precisely as needed. This identity would not have worked had we used the ``compact'' quantum dilogarithm \eqref{psiold} as the definition of $\psi(Z')$.

For this to work, however, it is crucial that the noncompact quantum dilogarithm \eqref{psinew} is used as the definition of $\psi(Z')$, rather than the compact version \eqref{psiold}; the compact quantum dilogarithm \eqref{psiold} does \emph{not} satisfy the identity \eqref{iFT}.

\subsection{Examples}
\label{sec:wfex}

We finish with two illustrative examples of holomorphic blocks, tying together all the ideas and methods of this paper. The first is the complement of the trefoil knot, $M_{\mb{3_1}}=S^3\bs\mb{3_1}$. For fixed boundary condition $u$ (or $U=2u$), there is only one nonabelian flat connection on $M_{\mb{3_1}}$, and the corresponding holomorphic block can be computed exactly and nonperturbatively. The second example is the complement of the figure-eight knot, $M_{\mb{4_1}}=S^3\bs\mb{4_1}$, for which we obtain an abstract generating integral of holomorphic blocks, written as a one-dimensional integral of two quantum dilogarithms. The integrand has two critical points, each corresponding to a distinct nonabelian flat connection on $M_{\mb{4_1}}$. By performing stationary phase approximations around these critical points, we obtain the asymptotic expansions of the holomorphic blocks to (in principle) all perturbative orders in $\hbar$, and find complete agreement with the state integral model of \cite{DGLZ}.

In general, it is \emph{guaranteed} that the number of distinct critical points of an integral for holomorphic blocks agrees with the number of classical solutions to $A(\ell,m^2)=0$ at fixed $m$. More precisely, when we construct an integral to generate solutions to $\hat A(\hat \ell,\hat m^2;q)\,Z(u)=0$, its critical points must correspond to solutions of $\hat A(\ell,m^2;1)=0$. The formal argument for this is essentially identical to the discussion in Section 3.4 of \cite{DGLZ}. Thus, in the end, every flat connection $\CA^\alpha$ that appears as a solution to a classical equation $\hat A(\ell,m^2;q\to 1)=0$ must be uniquely identified with a critical point $\alpha$ (and presumably an entire contour $C^\alpha$) of the state integral model.

\subsubsection{Trefoil $\mb{3_1}$}
\label{sec:wfex31}

The complement of the trefoil knot $M_{\mb{3_1}}=S^3\bs\mb{3_1}$ can be decomposed into two ideal tetrahedra, with (say) shape parameters $Z,Z',Z''$ and $W,W',W''$. According to Section \ref{sec:ex31}, the symplectic transformation in the operator algebra from $(\hat Z',\hat Z,\hat W',\hat W)$ to $(\hat U,\hat C,\hat v,\hat \Gamma)$ takes the form
\be \varphi_*\,:\;
\begin{pmatrix} \hat Z'\\\hat Z\\\hat W'\\\hat W\end{pmatrix}\mapsto \begin{pmatrix}\hat U\\\hat C\\\hat v\\\hat \Gamma\end{pmatrix} = 
\begin{pmatrix} 0 & 0 & -1 & 1 \\ 1 & 1 & 0 & 0 \\
 {\small \frac12} & {\small-\frac12} & -2 & 2 \\ 0 & 0 & {\small \frac12} & {\small \frac12} \end{pmatrix}
\begin{pmatrix} \hat Z'\\\hat Z\\\hat W'\\\hat W\end{pmatrix} + 
\begin{pmatrix} 0 \\ 0 \\ i\pi \\ 0 \end{pmatrix}\,.
\label{S31}
\ee
The $Sp(4,\Q)$ matrix in \eqref{S31} can be factored into generators%
\footnote{A useful algorithm for decomposing such matrices into generators is described in \cite{HuaReiner}.} %
\be \begin{pmatrix} 0 & 0 & -1 & 1 \\ 1 & 1 & 0 & 0 \\
 {\small \frac12} & {\small-\frac12} & -2 & 2 \\ 0 & 0 & {\small \frac12} & {\small \frac12} \end{pmatrix} = 
\begin{pmatrix} 1&0&0&0\\0&0&0&-1\\0&0&1&0\\0&1&0&0\end{pmatrix}
\begin{pmatrix} 1&-1&0&0\\{\small -\frac12}&{\small -\frac12}&0&0\\ 0&0&{\small\frac12}&{\small-\frac12}\\0&0&-1&-1\end{pmatrix}
\begin{pmatrix} 1&0&0&0\\0&1&0&0\\2&-2&1&0\\-2&2&0&1\end{pmatrix}
\begin{pmatrix} 0&0&-1&0\\0&0&0&-1\\1&0&0&0\\0&1&0&0\end{pmatrix}\,,
\ee
and it is helpful to label the product appearing on the right-hand side here as $M_1M_2M_3M_4$.

The Weil transformations rules described in (\ref{Wactions}a-e) immediately determine how the product wavefunction of two tetrahedra should transform. We start with $\Psi(Z',W')=\psi(Z')\psi(W')$, and calculate
\begin{align} \psi(Z')\psi(W') &\overset{M_4}{\longmapsto}
 \frac{1}{2\pi i\hbar}\int dZ'dW'\,\psi(Z')\psi(W')\,e^{\frac1\hbar(Z'U+W'X)} \nno \\
 &\overset{M_3}{\longmapsto} \frac{1}{2\pi i \hbar}\int dZ'dW'\, \psi(Z')\psi(W')\,e^{\frac1\hbar(Z'U+W'X)+\frac{1}{\hbar}(U^2+X^2-2UX)} \nno \\
 &\overset{M_2}{\longmapsto} \frac{i}{2\pi i\hbar} \int dZ'dW'\,
 \psi(Z')\psi(W')\,e^{\frac1\hbar\big[-X(W'+Z')+\frac12U(2U+Z'-W')\big]} \nno \\
 &\overset{M_1}{\longmapsto} \frac{i}{(2\pi i\hbar)^{3/2}}\int dZ'dW'dX\,\psi(Z')\psi(W')\,e^{\frac1\hbar\big[-X(W'+Z')+\frac12U(2U+Z'-W')+C X\big]} \nno \\
 &= \frac{-i}{\sqrt{2\pi i\hbar}}\int dZ'\,\psi(Z')\psi(C-Z')\,e^{\frac1\hbar UZ'+\frac1\hbar U(U-C/2)} \nno \\
 &\overset{\rm translate}{\longmapsto} \frac{-i}{\sqrt{2\pi i\hbar}}\int dZ'\,\psi(Z')\psi(C-Z')\,e^{\frac1\hbar UZ'+\frac1\hbar U(U-C/2)+\frac{i\pi U}{\hbar}} \nno \\
 &= \frac{-i}{\sqrt{2\pi i\hbar}}\int dp\, \psi(p+C/2)\psi(-p+C/2) e^{\frac1\hbar U(U+p)+\frac{i\pi U}{\hbar}} \label{311} \\
 &\qquad =:\widetilde\Psi(U,C)\,.\nno
\end{align}
This is the product wavefunction in the transformed basis. In order to find the holomorphic block, we should specialize to $C=2\pi i+\hbar$. The product of $\psi$'s in \eqref{311} then becomes
\be \psi(p+C/2)\psi(-p+C/2) \longrightarrow \Phi_{\hbar/2}(p)\Phi_{\hbar/2}(-p) = e^{-\frac{3p^2+\pi^2-\hbar^2/4}{6\hbar}}\,,\ee
where the last equality is due to a standard (noncompact!) quantum dilogarithm identity. Thus, at $C=2\pi i+\hbar$, the integral in \eqref{311} is just a Gaussian! There is a unique critical point, and exact integration on a downward-flow contour constructed from it yields
\be Z_{\mb{3_1}}(U;\hbar) = \widetilde\Psi(U,2\pi i+\hbar) = -\big(\sqrt{i}\,e^{\frac{4\pi^2-\hbar^2}{24\hbar}}\big)\, \exp\left(\frac{3U^2}{2\hbar}+\frac{i\pi U}{\hbar}\right)\,.
\label{Z31}
\ee

It is very easy to check that the $\hat A$-polynomial derived in Section \ref{sec:ex31} annihilates this unique nonabelian holomorphic block. Namely,
\be (\hat\ell+q^{3/2}\hat M^3)\,Z_{\mb{3_1}}(U;\hbar) = 0\,,\ee
where as usual $q=e^\hbar$, and the operators $\hat M=\hat m^2$ and $\hat \ell$ act as
\be \hat M\,Z(U)=e^U Z(U)\,,\qquad \hat\ell\, Z(U) = Z(U+\hbar)\,.
\ee
As predicted in Section \ref{sec:Heis}, there is also a dual $\hat{\mb A}$ polynomial that annihilates the same holomorphic block, given by $\hat{\mb A}(\hat{\bm \ell},\hat{\mb m}^2;\mb q) = \hat A(-\mb q^{-1/2}\hat{\bm \ell},\hat{\mb m}^2;\mb q) = -\mb q^{-1/2}\hat{\bm\ell}+\mb q^{3/2}\hat{\mb M}^3$, with $\hat{\mb M}=\hat{\mb m}^2$. Indeed, it is also easy to check explicitly that
\be (-\mb q^{-1/2}\hat{\bm\ell}+\mb q^{3/2}\hat{\mb M}^3)\,Z_{\mb{3_1}}(U;\hbar) = 0\,, \ee
where now
\be \hat{\mb M}\,Z(U) = e^{\frac{2\pi i}{\hbar}U}\,Z(U)\,,\qquad
 \hat{\bm\ell}\,Z(U) = Z(U+2\pi i)\,.\ee
The wavefunction itself has an explicit duality
\be \sigma\!\cdot\!Z_{\mb{3_1}}(U;\hbar) = Z_{\mb{3_1}}\Big(\frac{2\pi i}{\hbar}U;-\frac{4\pi^2}{\hbar}\Big) = (M\mb M^{-1})^{1/2}\,Z_{\mb{3_1}}(U;\hbar)\,, \ee
where $\mb M=e^{\frac{2\pi i}{\hbar}U}$.

The holomorphic block \eqref{Z31} has appeared before in the literature, in several different guises. In particular, it agrees --- up to our standard projective factors $\exp\Big(\frac{\pi^2}{\hbar}\,\Q+\C+\hbar\,\Q\Big)$ --- with the direct analytic continuation of the trefoil's colored Jones polynomial, computed by \cite{Mur-Hik}. \\

As a final amusing and illustrative exercise, we can check directly that the operator
\be \hat\CL_{\mb{3_1}} = q^5\hat M^5-q^3\hat c^{1/2}\hat M^2\hat\ell+q^4\hat c^{1/2}\hat M^3\hat\ell-\hat\ell^2\, \label{31cop} \ee
from \eqref{Lc31} actually annihilates the full product wavefunction $\widetilde\Psi(U,C)$ \emph{before} setting $C\to 2\pi i+\hbar$. Recall that this operator is a $\hat\gamma$--independent element of the left ideal that annihilates the trefoil's product wavefunction in any representation, so in particular it must annihilate $\widetilde\Psi(U,C)$. Recall also that on $\widetilde\Psi(U,C)$, the operators $\hat c,\hat M,\hat \gamma$ act as
\be \hat c \,\widetilde\Psi(U,C)=e^C\,\widetilde\Psi(U,C)\,,\qquad
\hat M\,\widetilde\Psi(U,C)=M\,\widetilde\Psi(U,C)=e^U\,\widetilde\Psi(U,C) \ee\be \hat\ell\,\widetilde\Psi(U,C)=\widetilde\Psi(U+\hbar,C)\,.\nno\ee

The proof proceeds by formal manipulations under the integral sign. Using the functional equation
\be \psi(Z'+\hbar)=(1-e^{-Z'})\psi(Z') \ee
for the tetrahedron wavefunction, we find
\bse \label{31v}
\begin{align} \widetilde\Psi(U,C) &= \frac{-i}{\sqrt{2\pi i\hbar}}\int dp\, \psi(p+C/2)\psi(-p+C/2) e^{\frac1\hbar U(U+p)+\frac{i\pi U}{\hbar}} \nno \\
&\overset{p\to p+\hbar}{=}
 \frac{-i}{\sqrt{2\pi i\hbar}}\int dp\, \psi(p+C/2+\hbar)\psi(-p+C/2-\hbar) e^{\frac1\hbar U(U+p)+\frac{i\pi U}{\hbar}+U} \nno \\
 &= \frac{-i}{\sqrt{2\pi i\hbar}}\,M\int dp\,\psi(p+C/2)\psi(-p+C/2) \frac{1-e^{-p-\frac C2}}{1-e^{p-\frac C2+\hbar}}\,e^{\frac1\hbar U(U+p)+\frac{i\pi U}{\hbar}}\,. \label{31v1}
\end{align}
Similarly,
\begin{align} q^{-1}\hat c^{-1/2}\hat M^{-2}\hat\ell\,\widetilde\Psi(U,C) &= 
 \frac{+i}{\sqrt{2\pi i\hbar}}\int dp\, \psi(p+C/2)\psi(-p+C/2) e^{\frac1\hbar U(U+p)+\frac{i\pi U}{\hbar}+p-\frac C2} \nno\\
 &\hspace{-1in}\overset{p\to p+\hbar}{=} \frac{+i}{\sqrt{2\pi i\hbar}}\,M\int dp\, \psi(p+C/2)\psi(-p+C/2)\frac{1-e^{-p-\frac C2}}{1-e^{p-\frac C2+\hbar}} e^{\frac1\hbar U(U+p)+\frac{i\pi U}{\hbar}}e^{p-\frac C2+\hbar}\,, \label{31v2}
\end{align}
and
\begin{align} q^{-1}\hat c^{-1/2}\hat M^3\hat\ell^{-1}\,\widetilde\Psi(U,C)
 = \frac{+i}{\sqrt{2\pi i\hbar}}\,M\int dp\, \psi(p+C/2)\psi(-p+C/2) e^{\frac1\hbar U(U+p)+\frac{i\pi U}{\hbar}}e^{-p-\frac C2}\,.
\label{31v3}
\end{align}
\ese
Combining together the right-hand sides of (\ref{31v}a-c), we obtain
\begin{align} & \widetilde\Psi(U,C) + q^{-1}\hat c^{-1/2}\hat M^{-2}\hat\ell\,\widetilde\Psi(U,C)- q^{-1}\hat c^{-1/2}\hat M^3\hat\ell^{-1}\,\widetilde\Psi(U,C) \nno\\
 &\qquad = \frac{-i}{\sqrt{2\pi i\hbar}}\,M\int dp\, \psi(p+C/2)\psi(-p+C/2) e^{\frac1\hbar U(U+p)+\frac{i\pi U}{\hbar}} \nno \\
 &\qquad = \hat M\,\widetilde\Psi(U,C)\,,
\end{align}
which is equivalent to the desired relation $\hat\CL_{\mb{3_1}}\widetilde{\Psi}(U,C)=0$.

More generally, one can use the same basic methods illustrated here to prove that much more complicated integral formulas for holomorphic blocks are annihilated by linear difference operators. In essence, one shifts both the integration variable(s) and the functional variable $U$ by $\pm\hbar$ over and over again, until sufficiently many expressions are generated to satisfy a nontrivial linear relation.

\subsubsection{Figure-eight knot $\mb{4_1}$}
\label{sec:wfex41}

The next simplest holomorphic blocks are those for the complement of the figure-eight knot in the three-sphere, $M=S^3\bs \mb{4_1}$. As discussed in Sections \ref{sec:hyp}--\ref{sec:quant}, the figure-eight knot complement can be composed from two tetrahedra, which again are assigned shape parameters $Z$ and $W$. According to Section \ref{sec:qglue}, the transformation from the original set of generators $(\hat Z',\hat Z,\hat W',\hat W)$ in the operator algebra to $(\hat U,\hat C,\hat v,\hat \Gamma)$ is given by
\be \varphi_*\,:\; \begin{pmatrix} \hat Z'\\\hat Z\\\hat W'\\\hat W\end{pmatrix}\mapsto \begin{pmatrix}\hat U\\\hat C\\\hat v\\\hat \Gamma\end{pmatrix} = 
\begin{pmatrix} 1&0&0&-1\\1&1&-1&-1\\-1&0&1&0\\0&0&0&1 \end{pmatrix}
\begin{pmatrix} \hat Z'\\\hat Z\\\hat W'\\\hat W\end{pmatrix} + 
\begin{pmatrix} 0 \\ 2\pi i+\hbar \\ i\pi \\ 0 \end{pmatrix}\,.
\label{S41}
\ee
Let's denote the $Sp(4,\Z)$ matrix appearing in \eqref{S41} as $M$. It can be decomposed as
\be M = M_1M_2M_3M_4M_5:=\mtt I 0 {B_1} I \mtt 0 {-I} I 0 \mtt {A_3} 0 0 {A_3^T{}^{-1}} \mtt I 0 {B_4} I \mtt 0 {-I} I 0\,,
\ee
where the $2\times2$ blocks here involve the identity $I$ as well as
\be B_1 = \mtt {-1}000\,,\qquad A_3=\mtt{-1}10{-1}\,,\qquad\text{and}\qquad B_4=\mtt0110\,.\ee

The corresponding Weil transformation $\varphi$ of the product wavefunction $\Psi(Z',W') = \psi(Z')\psi(W')$ can be constructed in steps, following rules (\ref{Wactions}a--e). We find
\begin{align} &\psi(Z')\psi(W') \overset{M_5}{\longmapsto}
 \frac{1}{2\pi i\hbar}\int dZ'dW'\,\psi(Z')\psi(W')\,e^{\frac{1}{\hbar}(XZ'+YW')} \nno \\
 &\qquad\overset{M_4}{\longmapsto}
 \frac{1}{2\pi i\hbar}\int dZ'dW'\,\psi(Z')\psi(W')\,e^{\frac{1}{\hbar}(XZ'+YW'+XY)} \nno \\
 &\qquad\overset{M_3}{\longmapsto}
 \frac{1}{2\pi i\hbar}\int dZ'dW'\,\psi(Z')\psi(W')\,e^{\frac{1}{\hbar}(-(X+Y)Z'-Y W'+(X+Y)Y)} \nno \\
  &\qquad\overset{M_2}{\longmapsto}
 \frac{1}{(2\pi i\hbar)^2}\int dZ'dW'dXdY\,\psi(Z')\psi(W')\,e^{\frac{1}{\hbar}(-(X+Y)Z'-Y W'+(X+Y)Y+UX+CY)} \nno \\
 &\qquad\overset{M_1}{\longmapsto}
 \frac{1}{(2\pi i\hbar)^2}\int dZ'dW'dXdY\,\psi(Z')\psi(W')\,e^{\frac{1}{\hbar}(-(X+Y)Z'-Y W'+(X+Y)Y+UX+CY-\tfrac12U^2)} \nno \\
 &\qquad = \frac{-1}{2\pi i\hbar}\int dW'dY\,\psi(Y+U)\psi(W')\,e^{\frac1\hbar(-YW'+(C-U)Y-\tfrac12U^2)} \nno \\
 &\qquad \overset{\rm translate}{\longmapsto} \frac{-1}{2\pi i\hbar}\int dW'dY\,\psi(Y+U)\psi(W')\,e^{\frac1\hbar(-YW'+(C-2\pi i-\hbar-U)Y-\tfrac12U^2+i\pi U)} \,. \label{wf411}
\end{align}
The integral over $W'$ in \eqref{wf411} can be performed by using \eqref{FT}, yielding
\begin{align} \eqref{wf411} &=
 \frac{i}{\sqrt{2\pi\hbar}}\int dY\, \psi(U+Y)\psi(Y)\,e^{-\frac{\pi^2-\hbar^2/4}{6\hbar}+\frac1\hbar\big(i\pi U-\frac{U^2}{2}-i\pi Y-UY+\frac{Y^2}{2}-\frac{Y\hbar}{2}+(C-2\pi i -\hbar)Y\big)} \nno \\
 &\hspace{-.3in}\overset{Y\to -p+i\pi+\hbar/2}{=}  \frac{1}{\sqrt{2\pi\hbar}}\int dp\, \Phi_{\hbar/2}(p-U)\Phi_{\hbar/2}(p)\,e^{-\frac U2-\frac{U^2}{2\hbar}+\frac{\pi^2-\hbar^2/4}{3\hbar}+\frac1\hbar\big(\frac{p^2}{2}+pU\big)+\frac1\hbar (C-2\pi i-\hbar)\big(-p+i\pi+\frac\hbar2\big)} \nno \\
 &= \frac{1}{\sqrt{2\pi\hbar}}\int dp\,\frac{\Phi_{\hbar/2}(p-U)}{\Phi_{\hbar/2}(-p)}\,e^{\frac{4\pi^2-\hbar^2}{24\hbar}-\frac U2-\frac{U^2}{2\hbar}+\frac{pU}{\hbar}+\frac1\hbar (C-2\pi i-\hbar)\big(-p+i\pi+\frac\hbar2\big)} \label{wf412} \\ 
 &=:\widetilde\Psi(U,C)\,. \nno
\end{align}
Finally, the generating integral of holomorphic blocks becomes
\begin{align} Z^{\rm gen}_{\mb{4_1}}(U;\hbar) &= \widetilde\Psi(U,C=2\pi i+\hbar) \nno \\
 &= \frac{1}{\sqrt{2\pi\hbar}}e^{\frac{4\pi^2-\hbar^2}{24\hbar}-\frac U2-\frac{U^2}{2\hbar}}\int dp\,\frac{\Phi_{\hbar/2}(p-U)}{\Phi_{\hbar/2}(-p)}\,e^{\frac{pU}{\hbar}} \label{wf413}
\end{align}

It can be shown explicitly that the integral \eqref{wf413} is formally a solution to the equation $\hat A_{\mb 4_1}(\hat\ell,\hat m^2;q)Z^{\rm gen}_{\mb{4_1}}(U;\hbar)=0$. The proof uses the methods outlined at the end of Section \ref{sec:wfex31}. After removing a factor of
\be
2^{-1/2}e^{\frac{4\pi^2-\hbar^2}{24\hbar}} \in \exp\Big(\frac{\pi^2}{\hbar}\,\Q+\C+\hbar\,\Q\Big) \label{fac41} \ee
from \eqref{wf413}, the integral can also be shown to be equivalent to the (presumably well-normalized) state integral model of \cite{DGLZ} for the figure-eight knot. The excess factor in \eqref{fac41} is precisely of the form that projective ambiguities are expected to take.

The integrand in \eqref{wf413} has two critical points $\alpha={\rm geom.}$ and $\alpha={\rm conj.}$ in the $\hbar\to 0$ limit, corresponding to the two classical, nonabelian flat connections on the figure-eight knot complement. These are sometimes known as the geometric and conjugate flat connections. The asymptotic expansions of the geometric and conjugate blocks $Z^{\alpha={\rm geom},\,{\rm conj}}_{\mb{4_1}}(U;\hbar)$ can be calculated by performing a saddle-point approximation of \eqref{wf413}. For example, if we set
\be Z^{\alpha}_{\bm{4_1}}(U;\hbar) \sim \exp\left(\frac1\hbar S_0^\alpha(U)+\delta^\alpha\log\hbar+S_1^\alpha(U)+\hbar\, S_2^\alpha(U)+\hbar^2\, S_3^\alpha(U)+\ldots\right)\,, \ee
we find that
\bse
\begin{align} S_0^\alpha(U) &=\frac{\pi^2}{6}+ \Li_2\left(\frac{1+M\ell}{1-M^2}\right)-\Li_2\left(\frac{M(1+M\ell)}{-1+M^2}\right)+\frac12 U^2+U\log\left(\frac{1+M\ell}{-1+M^2}\right)\,,\\
\delta^\alpha &\equiv 0\,,\\
 S_1^\alpha(U) &= \frac12\log\left(\frac{M(-1+M^2)}\Delta\right)\,,\\
 S_2^\alpha(U) &= -\frac1{24}+\frac{(-1+M^2)^3\,(1-M-2M^2+15M^3-2M^4-M^5+M^6)}{24\Delta^3}\,,\\
 S_3^\alpha(U) &= \frac{(-1+M^2)^6\,(1-M-2M^2+5M^3-2M^4-M^5+M^6)}{2\Delta^6}\,, \\
 \quad\text{etc.} \nno
\end{align}
\ese
The notation here is such that $M=m^2=e^U=e^{2u}$ (as usual), and $\Delta = \pd_\ell A_{\mb{4_1}}(\ell,M) = -1+M+2M^2+M^3-M^4+2M^2\ell$. In $\Delta$ and elsewhere, the variable $\ell$ satisfies $A_{\mb{4_1}}(\ell,M)=M^2-(1-M-2M^2-M^3+M^4)\ell+M^2\ell=0$, and so is implicitly a function of $M$. The choice of critical point or flat connection $\alpha$ is encoded in the choice of a solution $\ell(M) = \ell^\alpha(M)$ to the quadratic equation $A_{\mb{4_1}}(\ell,M)=0$. These asymptotics all agree perfectly with those found in Section 4.2 of \cite{DGLZ}, up to the projective factor \eqref{fac41}.

\acknowledgments

I would like to thank Christopher Beem, Abhijit Champanerkar, Nathan Dunfield, Hiroyuki Fuji, Sasha Goncharov, Daniel Halpern-Leistner, Lotte Hollands, Bus Jaco, Rinat Kashaev, Christoph Koutschan, Joerg Teschner, and Edward Witten for many illuminating conversations and insightful comments. I am particularly grateful to Sergei Gukov, Don Zagier, and Stavros Garoufalidis, conversations with whom this paper originated from, for their continual help and support along the way.
I would also like to thank KITP, Santa Barbara for their hospitality during the program on Langlands-type dualities in quantum field theory, August 2010, supported in part by DARPA under Grant No.
HR0011-09-1-0015 and by the National Science Foundation under Grant No. PHY-0551164. I have been partially supported by NSF grant PHY-0969448 during the completion of this work.

\bibliographystyle{JHEP_TD}
\bibliography{Ahat}

\end{document}